\documentclass[prd,aps,twocolumn,epsfig,amsmath,showpacs,nofootinbib]{revtex4}

\usepackage{times}
\usepackage{hyperref}

\usepackage{graphicx}
\usepackage{dcolumn}
\usepackage{bm}

\begin{document}

\title{Baryon magnetic moments in large-$N_c$ chiral perturbation theory: Effects of the decuplet-octet mass difference and flavor symmetry breaking}

\author{
Giovanna Ahuatzin
}
\affiliation{
Academia de Ciencias, Universidad Polit\'ecnica de San Luis Potos{\'\i}, Urbano Villal\'on 500, San Luis Potos{\'\i}, San Luis Potos{\'\i} 78369, M\'exico
}

\author{
Rub\'en Flores-Mendieta\footnote{On sabbatical leave from Instituto de F{\'\i}sica, Universidad Aut\'onoma de San Luis Potos{\'\i}, San Luis Potos{\'\i}, San Luis Potos{\'\i}, M\'exico}
}
\affiliation{
Theory Center, Thomas Jefferson National Accelerator Facility, Newport News, Virginia 23606, USA
}

\author{
Mar{\'\i}a A.\ Hern\'andez-Ruiz
}
\affiliation{
Facultad de Ciencias Qu{\'\i}micas, Universidad Aut\'onoma de Zacatecas, Apartado Postal 585, Zacatecas, Zacatecas 98060, M\'exico
}

\author{Christoph P.\ Hofmann}
\affiliation{
Facultad de Ciencias, Universidad de Colima, Bernal D{\'\i}az del Castillo 340, Colima, Colima 28045, M\'exico
}

\date{\today}

\begin{abstract}
The magnetic and transition magnetic moments of the ground-state baryons are computed in heavy baryon chiral perturbation theory in the large-$N_c$ limit, where $N_c$ is the number of colors. SU(3) symmetry breaking is systematically studied twofold: On the one hand, one-loop nonanalytic corrections of orders $m_q^{1/2}$ and $m_q \ln m_q$ are included, with contributions of baryon intermediate states from both flavor octet and flavor decuplet multiplets, assuming degeneracy between baryon states within a given flavor multiplet but nondegeneracy between baryons of different multiplets. On the other hand, perturbative SU(3) symmetry breaking is also analyzed by including all relevant leading-order operators that explicitly break SU(3) at linear order. The resultant expressions are compared with the available experimental data and with other determinations in the context of conventional heavy baryon chiral perturbation theory for three flavors of light quarks and at the physical value $N_c=3$. The agreement reached is quite impressive.
\end{abstract}


\maketitle

\section{Introduction}\label{sec:intro}

The SU(3) group theoretical approach to deal with baryon magnetic moments was first developed by Coleman and Glashow \cite{cg61}; their analysis led to the celebrated relations---named after them---among octet baryons in terms of two parameters \cite{coleman}, namely,
\begin{eqnarray}
\mu_{\Sigma^+}^{(0)} = \mu_p^{(0)}, \qquad  \mu_{\Sigma^-}^{(0)} + \mu_n^{(0)} = -\mu_p^{(0)}, \quad 2\mu_\Lambda^{(0)} = \mu_n^{(0)}, \nonumber \\[4mm]
\mu_{\Xi^-}^{(0)} = \mu_{\Sigma^-}^{(0)}, \qquad \mu_{\Xi^0}^{(0)} = \mu_n^{(0)}, \qquad  2\mu_{\Lambda\Sigma^0}^{(0)} = -\sqrt{3}\mu_n^{(0)}, \label{eq:cg}
\label{eq:treeval}
\end{eqnarray}
along with the isospin relation
\begin{equation}
\mu_{\Sigma^+}^{(0)} - 2 \mu_{\Sigma^0}^{(0)} + \mu_{\Sigma^-}^{(0)} = 0, \label{eq:isos}
\end{equation}
where the superscript $(0)$ will denote the SU(3) symmetric values hereafter. Soon after the discovery of relations (\ref{eq:cg}), experimental analyses found discrepancies by a few standard deviations from the SU(3) values. Since then, a number of methods have been used in order to improve the numerical predictions of Coleman and Glashow by including SU(3) breaking effects. Among these methods, heavy baryon chiral perturbation theory \cite{jm255,jm259} and the $1/N_c$ expansion of QCD \cite{djm94,djm95}, where $N_c$ is the number of colors, are two schemes to understand the low-energy consequences of hadrons.

Furthermore, the combined use of chiral perturbation theory and the $1/N_c$ expansion is another calculational scheme which constrains the low-energy interactions of baryons with the meson nonet in a more effective way than each method alone \cite{jen96}. Let us recall that in the chiral limit $m_q\to 0$ and mesons become massless Goldstone boson states; as a result, there is an expansion in powers of $m_q/\Lambda_\chi$, where $\Lambda_\chi\sim 1$ GeV is the scale of chiral symmetry breaking. On the other hand, in the large-$N_c$ limit, decuplet and octet baryons become degenerate, namely, $\Delta\equiv M_T-M_B \propto 1/N_c \to 0$, where $M_T$ and $M_B$ denote the SU(3) invariant masses of the decuplet and octet baryon multiplets, respectively. It turns out that decuplet and octet baryon states constitute a single irreducible representation of the contracted spin-flavor symmetry of baryons in large-$N_c$ QCD \cite{djm94,djm95}. Corrections about the large-$N_c$ limit then appear in powers of $1/N_c$. All in all, the combined expansion in $m_q/\Lambda_\chi$ and $1/N_c$ requires us to consider the double limit $m_q\to 0$ and $N_c\to \infty$.

Caldi and Pagels \cite{caldi}, in the framework of chiral perturbation theory, found that corrections to baryon magnetic moments appear in the nonanalytic forms $m_q^{1/2}$ and $m_q\ln m_q$, which can be obtained from meson-loop graphs. In heavy baryon chiral perturbation theory \cite{jm255,jm259}, loop graphs have a calculable dependence on the ratio $m_\Pi/\Delta$, where $m_\Pi$ denotes the mass of meson $\Pi=\pi,K,\eta$. For the theory to be valid, the conditions $m_\Pi \ll \Lambda_\chi$ and $\Delta \ll \Lambda_\chi$ must be met, although the ratio $m_\Pi/\Delta$ can take any value \cite{fmhjm}.

In a previous paper, \cite{rfm09} we computed one-loop corrections to baryon magnetic moments within a combined expansion in $m_q$ and $1/N_c$. We considered contributions of orders $\mathcal{O}(m_q^{1/2})$ and $\mathcal{O}(m_q\ln m_q)$ to relative order $1/N_c^3$ in the $1/N_c$ expansion. The best way of approaching this problem was in the degeneracy limit $\Delta\to 0$. The resultant theoretical expressions agreed, order by order, with others obtained within baryon chiral perturbation theory \cite{jen92,ms97,geng,geng2,tib} for octet and decuplet baryons and also for octet-octet and decuplet-octet transitions. Additionally, a comparison with the current experimental data \cite{part} through a least-squares fit allowed us to get information about the free parameters of the theory. Although the predicted values obtained for all 27 possible magnetic moments were according to expectations, the fit somehow seemed to be not entirely satisfactory in the sense that the SU(3) invariants of chiral perturbation theory are not well reproduced through the analysis.

It would be desirable to relax the restriction $\Delta \to 0$ and consider the more realistic case $\Delta\neq 0$. Indeed, in the present paper, we do so as a second approximation in the contributions arising from loop graphs of order $\mathcal{O}(m_q^{1/2})$ and $\mathcal{O}(m_q\ln m_q)$. Our motivation here is not really to be definitive about the determination of baryon magnetic moments in the combined scheme but rather to explore the effects $\Delta \neq 0$ has on the fit to experimental data. Noticeable improvements should be observed in the best-fit values of the parameters in the fit and also in the value of $\chi^2$ itself.

In  this paper we will consider two sources of SU(3) symmetry breaking. The first source, the implicit one, originates from the loops themselves when using the physical masses of the mesons. Here the corrections are of orders $\mathcal{O}(m_s^{1/2})$ and $\mathcal{O}(m_s\ln m_s)$, depending on the topology of the Feynman diagrams. The second source, the explicit one, is also related to the light quark masses and transforms as a flavor octet. We will loosely refer to this correction as perturbative symmetry breaking (SB).

This paper is organized as follows. In Sec.~\ref{sec:mm}, apart from introducing our notation and conventions, we provide an overview on the determination of baryon magnetic moments in large-$N_c$ chiral perturbation theory. We start our discussion by defining the tree-level values and then, in Sec.~\ref{sec:1l} we continue with computing one loop-corrections. We first concentrate on corrections of order $\mathcal{O}(m_q^{1/2})$ in Sec.~\ref{sec:mq} by constructing the baryon operator which describes such contribution; the dependence on $\Delta$ is explicitly included at this level. We proceed further in order to achieve the reduction of the operators in the two flavor representations involved. Next, in Sec.~\ref{sec:mqlnmq} we compute corrections of order $\mathcal{O}(m_q\ln m_q)$; the analysis is more challenging than the previous one due to the considerable amount of group theory involved. We complement the analysis by including SB in Sec.~\ref{sec:su3}. The theoretical expressions obtained are then compared with other determinations in the framework of chiral perturbation theory and cross-checked with the very well-known sum rules found in the literature in Sec.~\ref{sec:checks}. In Sec.~\ref{sec:num} we carry out several least-squares fits in order to determine the best-fit parameters of the theory which allow us to predict numerical values of the unobserved magnetic moments; we then compare them with other numerical predictions. Finally, in Sec.~\ref{sec:final} we discuss our findings. This work is complemented by three appendices. In Appendices \ref{app:reduc1} and \ref{app:reduc2}, we provide the reduction of the baryon operators for both kinds of one-loop corrections discussed here. In Appendix \ref{app:OpBasis} we list the explicit results that make up the contributions of order $\mathcal{O}(m_q\ln m_q)$.

\section{\label{sec:mm}Baryon magnetic moment in large-$N_c$ chiral perturbation theory}

The present analysis builds on earlier works, particularly on Refs.~\cite{djm94,djm95,jen96,dai}, which established the mathematical groundwork on large-$N_c$ QCD and the $1/N_c$ expansion for baryons, and also on Ref.~\cite{rfm09}, where baryon magnetic moments in large-$N_c$ chiral perturbation theory in the degeneracy limit were discussed. Thus, we only give an outline of some relevant issues here.

The chiral Lagrangian for baryons in the $1/N_c$ expansion was established in Ref.~\cite{jen96}. It takes the form
\begin{eqnarray}
\mathcal{L}_{\text{baryon}} & = & i \mathcal{D}^0 - \mathcal{M}_{\text{hyperfine}} + \text{Tr} \left(\mathcal{A}^k \lambda^c \right) A^{kc} \nonumber \\
&  & \mbox{} + \frac{1}{N_c} \text{Tr} \left(\mathcal{A}^k \frac{2I}{\sqrt 6}\right) A^k + \ldots, \label{eq:ncch}
\end{eqnarray}
with
\begin{equation}
\mathcal{D}^0 = \partial^0 \openone + \text{Tr} \left(\mathcal{V}^0 \lambda^c\right) T^c. \label{eq:kin}
\end{equation}
The ellipses in Eq.~(\ref{eq:ncch}) refer to higher partial wave meson couplings, showing up at subleading orders in the $1/N_c$ expansions for $N_c > 3$. All of these higher partial waves vanish in the large-$N_c$ limit. Accordingly, the meson coupling to baryons is purely $p$ wave. In particular, the ellipses do not mean that we omit terms or make unjustified approximations. Much like in the analogous study related to the baryon axial vector current performed in Ref.~\cite{rfm12}, the terms shown in Eq.~(\ref{eq:ncch}) are the only ones relevant to our analysis.

Meson fields are contained in the vector and axial vector combinations
\begin{eqnarray}
&  & \mathcal{V}^0 = \frac12 \left(\xi \partial^0 \xi^\dagger + \xi^\dagger \partial^0 \xi\right), \quad
\mathcal{A}^k = \frac{i}{2} \left(\xi \nabla^k \xi^\dagger - \xi^\dagger \nabla^k \xi\right), \nonumber \\
&  & \mbox{\hglue2.3truecm} \xi(x)=\exp[i\Pi(x)/f].
\end{eqnarray}
Here the matrix $\Pi(x)$ stands for the nonet of Goldstone boson fields, and $f$ is the pion decay constant which takes the value $f \approx 93$ $\mathrm{MeV}/c^2$.

Each term in the chiral Lagrangian (\ref{eq:ncch}) involves a baryon operator. The baryon kinetic energy term is proportional to the spin-flavor identity; the quantity $\mathcal{M}_{\text{hyperfine}}$ is the hyperfine baryon mass operator, which takes into account the spin splittings of the tower of baryon states with spins $1/2,\ldots, N_c/2$ in the flavor representations. Furthermore, the flavor octet and the flavor singlet meson combinations couple to the flavor octet $V^{0a}$ and flavor singlet $V^{0}$ baryon charges, respectively, given by
\begin{equation}
V^{0c} = \left\langle \mathcal{B}^\prime \left| \left(\overline{q}\gamma^0 \frac{\lambda^c}{2}q\right)_{\mathrm{QCD}}\right|\mathcal{B} \right\rangle \label{eq:v0a}
\end{equation}
and
\begin{equation}
V^{0} = \left\langle \mathcal{B}^\prime \left| \left(\overline{q}\gamma^0 \frac{I}{\sqrt{6}}q\right)_{\mathrm{QCD}}\right|\mathcal{B} \right\rangle. \label{eq:v0}
\end{equation}
In a similar fashion, the $\ell=1$ flavor octet and flavor singlet axial vector meson combinations couple to the flavor octet $A^{kc}$ and flavor singlet $A^{k}$ axial vector currents, respectively, which read
\begin{equation}
A^{kc} = \left\langle \mathcal{B}^\prime \left| \left(\overline{q}\gamma^k\gamma^5 \frac{\lambda^c}{2}q\right)_{\textrm{QCD}}\right|\mathcal{B} \right\rangle \label{eq:makc}
\end{equation}
and
\begin{equation}
A^{k} = \left\langle \mathcal{B}^\prime \left| \left(\overline{q}\gamma^k\gamma^5 \frac{I}{\sqrt{6}}q\right)_{\textrm{QCD}}\right|\mathcal{B} \right\rangle. \label{eq:ma0}
\end{equation}
The subscript $\mathrm{QCD}$ in Eqs.~(\ref{eq:v0a})-(\ref{eq:ma0}) is used as a reminder that the quark fields are QCD quark fields. 

A baryon operator has a well-defined $1/N_c$ expansion, which can be written as
\begin{equation}
\mathcal{O}_{\mathrm{QCD}}=\sum_n c_n \frac{1}{N_c^{n-1}}\mathcal{O}_n,
\end{equation}
where the operator basis $\mathcal{O}_n$ is conformed by polynomials in the SU(6) spin-flavor generators \cite{djm95},
\begin{equation}
J^k = q^\dagger \frac{\sigma^k}{2} q, \qquad T^c = q^\dagger \frac{\lambda^c}{2} q, \qquad G^{kc} = q^\dagger
\frac{\sigma^k}{2}\frac{\lambda^c}{2} q. \label{eq:su6gen}
\end{equation}
Here the quantities $q^\dagger$ and $q$ represent SU(6) operators that create and annihilate states in the fundamental spin-flavor representation of SU(6), and $\sigma^k$ and $\lambda^c$ are the Pauli and Gell-Mann matrices, respectively. The commutation relations obeyed by the SU(6) spin-flavor generators can be found in Ref.~\cite{djm95}.

Specifically, the $1/N_c$ expansion of the baryon mass operator $\mathcal{M}$, which transforms as a $(0,\mathbf{1})$ under $\textrm{SU(2)}\times\textrm{SU(3)}$, can be written as \cite{jen96}
\begin{equation}
\mathcal{M} = m_{(0)}^{0,1}N_c\openone + \sum_{n=2,4}^{N_c-1} m_{(n)}^{0,1} \frac{1}{N_c^{n-1}}J^n, \label{eq:massop}
\end{equation}
where the coefficients $m_{(n)}^{0,1}$ are \textit{a priori} unknown parameters of order $\mathcal{O}(\Lambda_\chi)$, and the superscripts attached to them indicate the spin-flavor representation they belong to. While the first term in Eq.~(\ref{eq:massop}) represents the overall spin-independent mass of the baryon multiplet, the spin-dependent terms define $\mathcal{M}_{\textrm{hyperfine}}$.

On the other hand, the $1/N_c$ expansion of the baryon axial vector operator $A^{kc}$ can be constructed by keeping in mind that only its space components have nonzero matrix elements at zero recoil. Thus, it transforms as a $(1,\mathbf{8})$ under SU(2)$\times$SU(3) and it is $T$ odd. \cite{djm95}. At the physical value $N_c=3$, we have
\begin{equation}
A^{kc} = a_1 G^{kc} + b_2 \frac{1}{N_c} \mathcal{D}_2^{kc} + b_3 \frac{1}{N_c^2} \mathcal{D}_3^{kc} + c_3 \frac{1}{N_c^2} \mathcal{O}_3^{kc}, \label{eq:akc}
\end{equation}
where the coefficients $a_1$, $b_1$, $b_2$, and $c_3$ are of order unity and the operators that come along with them read
\begin{equation}
\mathcal{D}_2^{kc} = J^kT^c, \label{eq:d2kc}
\end{equation}
\begin{equation}
\mathcal{D}_3^{kc} = \{J^k,\{J^r,G^{rc}\}\}, \label{eq:d3kc}
\end{equation}
\begin{equation}
\mathcal{O}_3^{kc} = \{J^2,G^{kc}\} - \frac12 \{J^k,\{J^r,G^{rc}\}\}. \label{eq:o3kc}
\end{equation}
Successive higher-order operators are constructed from the previous ones by anticommuting them with $J^2$. Besides, the operators $\mathcal{D}_n^{kc}$ are diagonal: nonvanishing matrix elements only occur between states with the same spin. The operators $\mathcal{O}_n^{kc}$, in turn, are purely off-diagonal: nonvanishing matrix elements only occur between states with different spin.

Now, the starting point of the analysis of Ref.~\cite{rfm09} relies on the fact that, in the large-$N_c$ limit, the baryon magnetic moments possess the same kinematical properties as the baryon axial-vector couplings, so they are described in terms of the same operators. The magnetic moment operator is also a spin-1 object and transforms as an SU(3) octet. Thus, in a complete analogy to expression (\ref{eq:akc}), the $1/N_c$ expansion of the operator which yields baryon magnetic moments can be written as \cite{rfm09}
\begin{equation}
M^{kc} = m_1 G^{kc} + m_2 \frac{1}{N_c} \mathcal{D}_2^{kc} + m_3 \frac{1}{N_c^2} \mathcal{D}_3^{kc} + m_4 \frac{1}{N_c^2} \mathcal{O}_3^{kc}, \label{eq:mmag}
\end{equation}
where the series has also been truncated at $N_c=3$. By assuming the SU(3) symmetry limit, the unknown coefficients $m_i$ (also of order unity) are independent of $k$, so they are unrelated to the ones of the series (\ref{eq:akc}) at this limit. The magnetic moments are proportional to the light quark electromagnetic charge matrix $\mathcal{Q}=\textrm{diag}(2/3,-1/3,-1/3)$ and can be separated into isovector and isoscalar components, $M^{k3}$ and $M^{k8}$, respectively. Thus, the baryon magnetic moment operator can ultimately be defined as
\begin{equation}
M^k = M^{kQ} \equiv M^{k3} + \frac{1}{\sqrt{3}} M^{k8}. \label{eq:mmsep}
\end{equation}
Hereafter, the spin index $k$ of $M^k$ will be set to 3, whereas the flavor index $Q$ will stand for $Q=3+(1/\sqrt{3})8$, so any operator of the form $X^Q$ should be understood as $X^3+(1/\sqrt{3})X^8$. In the same spirit, $X^{\bar{Q}}$ should be understood as $X^3-(1/\sqrt{3})X^8$. In particular, $T^Q=T^3+(1/\sqrt{3})T^8$ is the SU(3) flavor generator corresponding to $\mathcal{Q}$.

The magnetic moments in conventional heavy baryon chiral perturbation theory (the effective field theory with no $1/N_c$ expansion) are parametrized by four SU(3) invariants $\mu_D$, $\mu_F$, $\mu_C$, and $\mu_T$ \cite{jen92}, while in the present analysis, they are parametrized in terms of $m_i$, with $i=1,\ldots,4$, introduced in Eq.~(\ref{eq:mmag}). At $N_c=3$, they are related by \cite{rfm09}
\begin{subequations}
\label{eq:rel2}
\begin{eqnarray}
&  & \mu_D = \frac12 m_1 + \frac16 m_3, \\
&  & \mu_F = \frac13 m_1 + \frac16 m_2 + \frac19 m_3, \\
&  & \mu_C = \frac12 m_1 + \frac12 m_2 + \frac56 m_3, \\
&  & \mu_T = -2m_1 - m_4.
\end{eqnarray}
\end{subequations}

In a complete parallelism with Ref.~\cite{djm95}, the operator analysis in this work is performed within the quark representation of the spin-flavor symmetry of large-$N_c$ baryons, which uses the algebraic structure of the nonrelativistic quark model to classify baryon operators. This statement does not mean, however, that either the quaks in the baryon are treated as nonrelativistic or that the validity of the quark model is implicitly assumed.

We should stress the fact that the present analysis of baryon magnetic moments is based on large-$N_c$ chiral perturbation theory, i.e., the combination of heavy baryon chiral perturbation theory with the $1/N_c$ expansion. We want to point out that either method is fully systematic and, above all, model independent. Heavy baryon chiral perturbation theory corresponds to a consistent and systematic expansion in powers of momentum and of the light quark masses. In the $1/N_c$ expansion, on the other hand, one systematically evaluates deviations from the exact spin-flavor symmetry by computing $1/N_c$ corrections to the large-$N_c$ limit. The (combined) chiral Lagrangian for baryons in the $1/N_c$ expansion (\ref{eq:ncch}) is the most general expression which respects the symmetries of QCD and is consistent with the $1/N_c$ expansion. It is important to note that, unlike, e.g., the quark model, these expansions in momentum, quark mass and $1/N_c$, do not make use of any model description of the baryons. In particular, the expressions (\ref{eq:mmag}) and (\ref{eq:mmsep}) are model independent: they represent the most general expression (up to order $1/N_c^2$) consistent with the $N_c$ expansion, while the microscopic details of QCD only manifest themselves in the specific values of the coefficients $m_1, \dots, m_4$.

The matrix elements of the baryon operators $V^{0a}=v^0T^a$, $A^{ia}$, $M^i$, or $\mathcal{M}$ between SU(6) symmetric states can thus be connected to physics in a straightforward way. $V^{0a}$ is a spin-0 and a flavor octet, so it transforms as $(0,\mathbf{8})$ under SU(2)$\times$SU(3). The operator $V^{0a}$ at $q^2=0$ is a special $(0,\mathbf{8})$ operator; it is the generator of SU(3) symmetry transformations, and its matrix elements correspond to the vector form factors $f_1(q^2=0)\equiv g_V$ as conventionally defined in baryon semileptonic decays. In a similar manner, $A^{ia}$ is spin 1 and a flavor octet. Its matrix elements between baryon octet states at $q^2=0$ correspond to the axial vector form factors $g_1(q^2=0)\equiv g_A$ also as defined in baryon semileptonic decays, with a normalization such that $g_A/g_V=F+D$ for neutron beta decay.

On the other hand, we have already pointed out that since the magnetic moment is a spin-1 octet operator, it has a $1/N_c$ expansion identical in structure to the axial current. The matrix elements of $M^i$, for $i=3$, thus yield the actual values of the baryon magnetic moments $\mu_B$. To derive a relation between magnetic moments and form factors, one needs to look at the baryon matrix elements of the electromagnetic current $j_\mu^{\textrm{em}}$. Thus, $\mu_B$ corresponds to $F_1(0)+F_2(0)\equiv G_M(0)$, where $F_1(q^2)$ and $F_2(q^2)$ are the Dirac and Pauli form factors, respectively, and $G_M(q^2)$ is the magnetic form factor. In the limit $q^2\to 0$, $F_1$ and $F_2$ are the charge and the anomalous magnetic moments of the baryons, respectively. For electromagnetic transitions analogous form factors can be defined.

At tree level, the baryon magnetic and transition magnetic moments $\mu_B^{(0)}$ can be straightforwardly computed from Eq.~(\ref{eq:mmsep}). The required matrix elements of the operators involved in such an expression are listed in Ref.~\cite{rfm09} and will not be repeated here. Let us now proceed to discuss the one-loop corrections.

\section{\label{sec:1l}One-loop corrections to baryon magnetic moments}

\begin{figure}[ht]
\scalebox{0.32}{\includegraphics{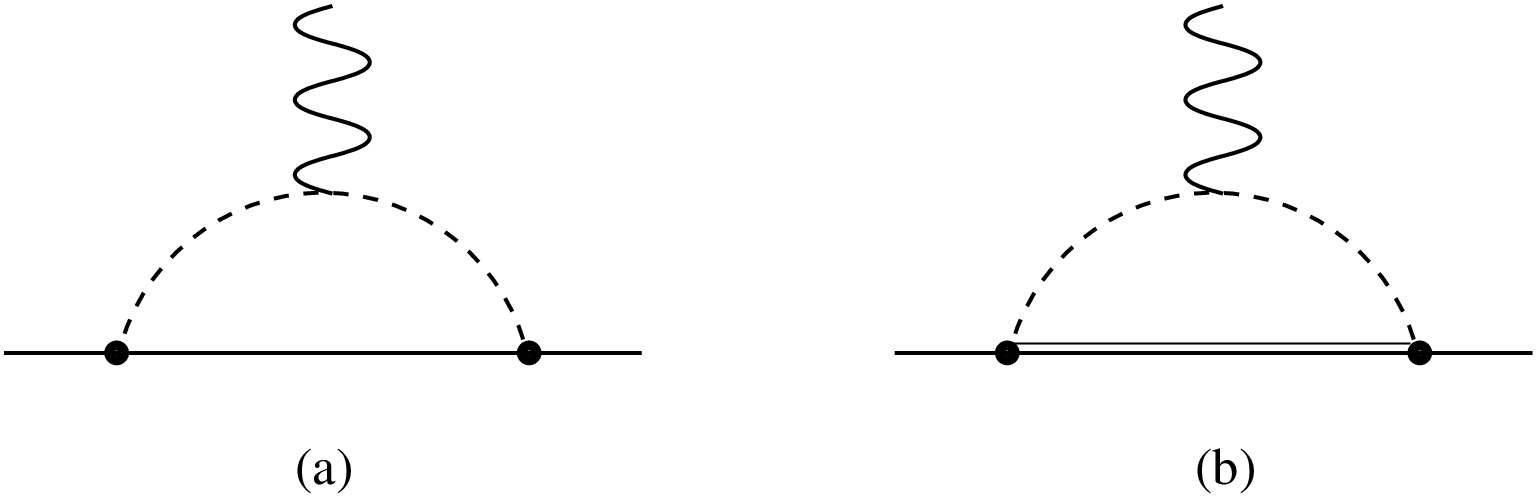}}
\caption{\label{fig:mmloop1}Feynman diagrams which yield nonanalytic $m_q^{1/2}$ corrections to the magnetic moments of octet baryons. Dashed lines denote mesons, and single and double solid lines denote octet and decuplet baryons, respectively. For decuplet baryons and decuplet-octet transitions, the diagrams are similar.}
\end{figure}

\begin{figure}[ht]
\scalebox{0.32}{\includegraphics{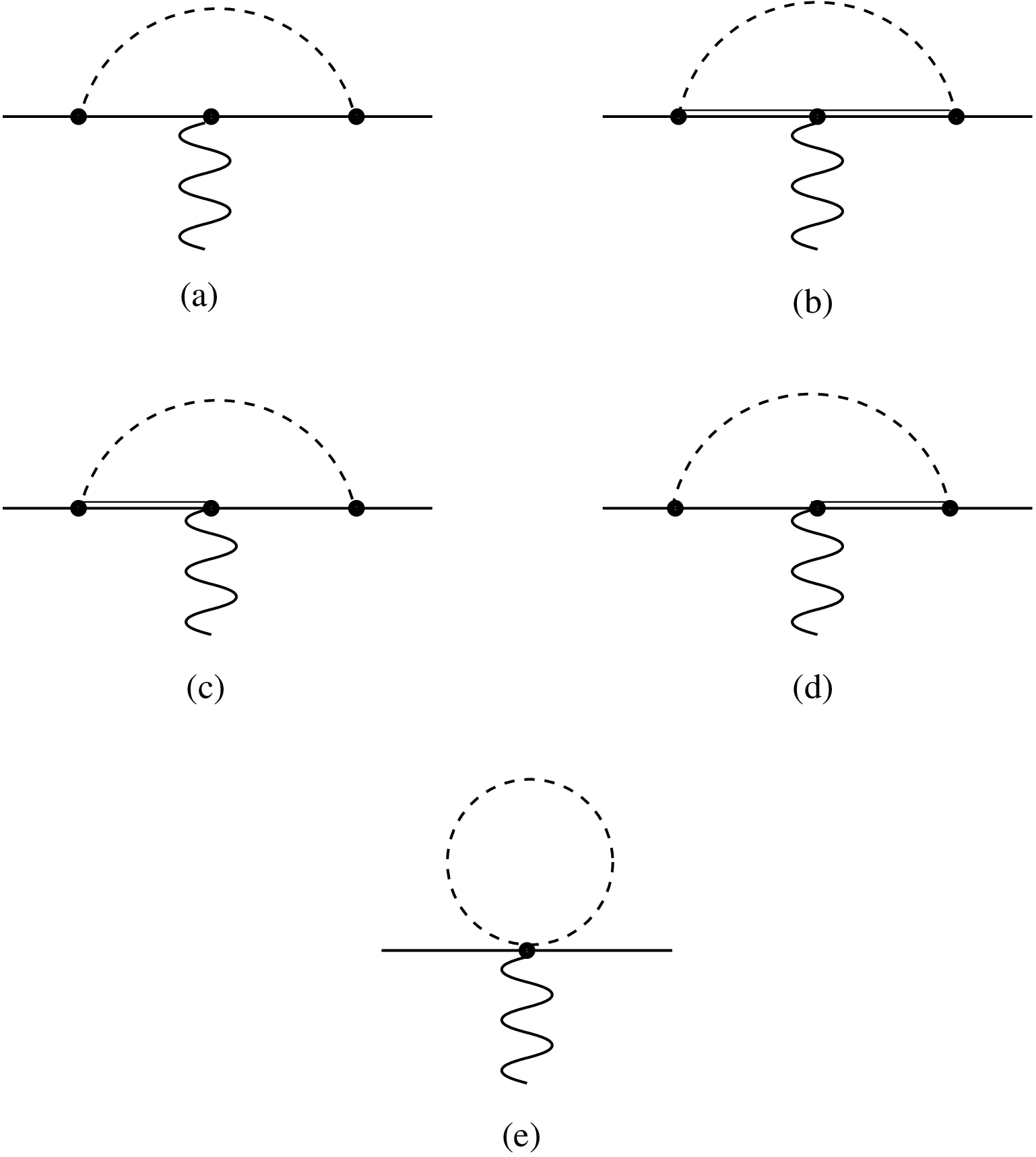}}
\caption{\label{fig:mmloop2}Feynman diagrams which yield nonanalytic $m_q \ln m_q$ corrections to the magnetic moments of octet baryons. Dashed lines denote mesons, and single and double solid lines denote octet and decuplet baryons, respectively. The wavefunction renormalization graphs are omitted in the figure but are nevertheless considered in the analysis. For decuplet baryons and decuplet-octet transitions, the diagrams are similar.}
\end{figure}

The diagrams that contribute to baryon magnetic moments at one-loop order are displayed in Figs.~\ref{fig:mmloop1} and \ref{fig:mmloop2}. These diagrams are given by the product of a group theoretic structure times a loop integral, which depends nonanalytically on the light quark masses $m_q$. The explicit dependence is $\mathcal{O}(m_q^{1/2})$ and $\mathcal{O}(m_q \ln m_q)$ for Figs.~\ref{fig:mmloop1} and \ref{fig:mmloop2}, respectively. Since $m_q\propto m_\Pi^2 \propto p^2$, in the chiral momentum counting scheme, these two types of diagrams are of order $p^3$ and $p^4$, respectively. In this counting, the tree-level values are order $p^2$.

The group theoretical structures that come along with the integrals over the loops have a rather complex dependence on $N_c$. It has been argued that baryons with strangeness of order $N_c^0$ have matrix elements of $T^c$ and $G^{k8}$ $(c=1,2,3)$ of order $N_c^0$, matrix elements of $T^c$ and $G^{kc}$ $(c=4,5,6,7)$ of order $\sqrt{N_c}$, and matrix elements of $T^8$ and $G^{kc}$ $(c=1,2,3)$ of order $N_c$ \cite{djm95}. To overcome this apparent complexity, let us use the fact that the pion-baryon vertex is proportional to $g_A/f$. Thus, in the large-$N_c$ limit, $g_A\propto N_c$ and $f\propto \sqrt{N_c}$, so the pion-baryon vertex scales as $\sqrt{N_c}$. Next, we can assume a naive power counting scheme for baryons with spins of order 1,
\begin{equation}
T^a \sim N_c, \qquad G^{ia} \sim N_c, \qquad J^i \sim 1; \label{eq:crules}
\end{equation}
i.e., factors of $J^i/N_c$ are $1/N_c$ suppressed relative to factors of $T^a/N_c$ and $G^{ia}/N_c$. This $N_c$-counting rule works if we only consider the lowest-lying baryon states, namely, those related to the $\mathbf{56}$ dimensional representation of SU(6).

With these simple tools, we can argue that the one-loop diagrams of Fig.~\ref{fig:mmloop1} are of order $\mathcal{O}(N_c)$. In the limit of small $m_s$, the symmetry breaking part of these diagrams is $\mathcal{O}(m_s^{1/2})$, so their overall contribution to baryon magnetic moments is $\mathcal{O}(m_s^{1/2}N_c)$, whereas the tree-level value is order $N_c$. As for the one-loop diagrams of Fig.~\ref{fig:mmloop2}, the large-$N_c$ dependence has been discussed in detail in Refs.~\cite{fmhjm,rfm06} for the axial vector current. Those conclusions can be extended to the baryon magnetic moment operator. Therefore, diagrams of Fig.~\ref{fig:mmloop2} are at most of order $\mathcal{O}(N_c^0)$, or $1/N_c$ times the tree-level value.

Recent studies that focused on the computation of baryon magnetic moments within covariant chiral perturbation theory, Refs.~\cite{geng,geng2}, raise an important issue here. We need to point out that there is no one-to-one correspondence between the diagrams of covariant baryon chiral perturbation theory and heavy baryon chiral perturbation theory. While the total result for any measurable quantity, of course, must be the same, the contributions from different diagrams can be rearranged. Indeed, in the present case of magnetic moments, there are two types of diagrams that are different from zero in the covariant approach but do not contribute in the heavy baryon version. In the covariant approach, the tadpole diagram (b) as well as the diagrams (f) and (i) in Fig.~1 of Ref.~\citep{geng2}, yield nonzero contributions. The same diagrams in the heavy baryon approach, however, do not contribute to magnetic moments. This is a consequence of the spin symmetry, which emerges at leading order in heavy baryon chiral perturbation theory. More precisely, the tadpole graph (b) corresponds to a vertex which is spin independent and can thus not contribute to magnetic moments. On the other hand, in the diagrams (f) and (i), the momentum $p$ of the external photon only enters through the combination $(k+p) \cdot v$ in the baryon propagator. So again, it does not have the correct structure to lead to a magnetic moment because the magnetic moment depends on the space component of $p^{\mu}$ rather than on its $p^0$ component. In the covariant approach, there are extra $\gamma$ matrices that may produce spin dependence and thus lead to nonzero contributions resulting from the very same diagrams. Note that all $\gamma^{\mu}$ matrices can be eliminated in heavy baryon approach and reduced to expressions involving the 4-velocity $v^{\mu}$ of the heavy baryon field and the velocity-dependent spin operator $S^{\mu}$ only \citep{jm255}. Explicit expressions for type-1 and type-2 diagrams which do contribute in heavy baryon chiral perturbation theory (Figs.~\ref{fig:mmloop1} and \ref{fig:mmloop2} in the present work) can be found in Ref.~\citep{lmw95} [see formulas (22) and (28), respectively].

Coming back to the main goal of the present paper, the analysis of baryon magnetic moments in the framework of large-$N_c$ chiral perturbation theory presented in Ref.~\cite{rfm09} was carried out in the degeneracy limit $\Delta\to 0$. We now intend to find out the effects of a nonvanishing $\Delta$, as well as flavor symmetry breaking. The calculation introduces a number of issues not discussed in Ref.~\cite{rfm09}. Because both types of diagrams involve rather different operator reduction patterns, we proceed to evaluate them separately.

\subsection{\label{sec:mq}Diagrams of order ${\mathcal O}(m_q^{1/2})$}

The analysis of one-loop corrections of order ${\mathcal O}(m_q^{1/2})$ in the degeneracy limit has been discussed in detail in Sec.\ IV.A of Ref.~\cite{rfm09}. Now, for a nonvanishing $\Delta$, one can discern that an immediate modification can be found in the baryon propagator in the loop integral of Fig.~\ref{fig:mmloop1}, which now has an explicit dependence on $\Delta$. To deal with this issue, we can follow the approach implemented in the analysis of flavor $\mathbf{27}$ nonanalytic corrections to the baryon masses presented in Ref.~\cite{jen96}. In this work, is was stated that in the chiral limit the baryon propagator is diagonal in spin, so it can be expressed as
\begin{equation}
\frac{i\mathcal{P}_{\mathsf{j}}}{k^0-\Delta_{\mathsf{j}}}, \label{eq:barprop}
\end{equation}
where $\mathcal{P}_{\mathsf{j}}$ is a spin projector operator for spin $J=\mathsf{j}$, which satisfies by definition
\begin{subequations}
\label{eq:spr}
\begin{eqnarray}
&  & \mathcal{P}_{\mathsf{j}}^2=\mathcal{P}_{\mathsf{j}}, \\
&  & \mathcal{P}_{\mathsf{j}} \mathcal{P}_{\mathsf{j^\prime}} = 0, \qquad \mathsf{j} \neq \mathsf{j^\prime}
\end{eqnarray}
\end{subequations}
and $\Delta_{\mathsf{j}}$ stands for the difference of the hyperfine mass splitting for spin $J=\mathsf{j}$ and the external baryon, namely,
\begin{equation}
\Delta_{\mathsf{j}} = \mathcal{M}_{\textrm{hyperfine}}|_{J^2=\mathsf{j}(\mathsf{j}+1)}-\mathcal{M}_{\textrm{hyperfine}}|_{J^2=\mathsf{j}_{\textrm{ext}}(\mathsf{j}_{\textrm{ext}}+1)}.
\end{equation}

Thus, for $p$-wave meson emission, $\Delta_{\mathsf{j}}$ reduces to \cite{jen96}
\begin{eqnarray}
\Delta_{\mathsf{j}} = \left\{
\begin{array}{ll}
\displaystyle  \frac{1}{N_c}2 \, \mathsf{j} \, m_{(2)}^{0,1}, & \mathsf{j}_{\textrm{ext}}=\mathsf{j}-1, \\[2mm]
0, & \mathsf{j}_{\textrm{ext}}=\mathsf{j}, \\[2mm]
\displaystyle -\frac{1}{N_c}2 \, \mathsf{j} \, m_{(2)}^{0,1}, & \mathsf{j}_{\textrm{ext}}=\mathsf{j}+1,
\end{array}
\right.
\end{eqnarray}
\textit{at leading order} $1/N_c$ in the $1/N_c$ expansion.

A realization of $\mathcal{P}_{\mathsf{j}}$ is given by \cite{jen96}
\begin{equation}
\mathcal{P}_{\mathsf{j}} = \frac{\Pi_{\mathsf{j} \neq \mathsf{j^\prime}}(J^2-J^2_{\mathsf{j^\prime}})}{\Pi_{\mathsf{j} \neq \mathsf{j^\prime}}(J^2_{\mathsf{j}}-J^2_{\mathsf{j^\prime}})}, \label{eq:fullpr}
\end{equation}
i.e., the projection operators for spin $J_{\mathsf{j}}$ is given by the product over all $J_{\mathsf{j}^\prime}=1/2,3/2,\ldots,N_c/2$ not equal to $J_{\mathsf{j}}$. The general form of the spin projector (\ref{eq:fullpr}) for arbitrary $N_c$ can be found in Ref.~\cite{jen96}; however, here we just need the spin-$\frac12$ and spin-$\frac32$ projectors for $N_c=3$, which read
\begin{subequations}
\label{eq:projnc3}
\begin{eqnarray}
\mathcal{P}_\frac12 & = & -\frac13 \left(J^2-\frac{15}{4}\right), \\
\mathcal{P}_\frac32 & = & \frac13 \left(J^2-\frac{3}{4}\right),
\end{eqnarray}
\end{subequations}
where
\begin{subequations}
\begin{equation}
\Delta_\frac12 = \left\{
\begin{array}{ll}
\displaystyle 0, & \mathsf{j}_{\textrm{ext}}=\frac12, \\[2mm]
\displaystyle -\Delta, & \mathsf{j}_{\textrm{ext}}=\frac32,
\end{array}
\right.
\end{equation}
\begin{equation}
\Delta_\frac32 = \left\{
\begin{array}{ll}
\displaystyle \Delta, & \mathsf{j}_{\textrm{ext}}=\frac12, \\[2mm]
\displaystyle 0, & \mathsf{j}_{\textrm{ext}}=\frac32, \\[2mm]
\end{array}
\right.
\end{equation}
\end{subequations}
and
\begin{equation}
\Delta = \frac{3}{N_c}m_{(2)}^{0,1}.
\end{equation}
It is straightforward to check that expressions (\ref{eq:projnc3}) meet conditions (\ref{eq:spr}).

The diagram in Fig.~\ref{fig:mmloop1} is thus given by the product of a baryon operator times a flavor tensor containing information about the loop integrals. Using the baryon propagator (\ref{eq:barprop}), the loop graphs of Fig.~\ref{fig:mmloop1} can be expressed as
\begin{equation}
\delta M_{\textrm{loop 1}}^k = \sum_{\textsf{j}} \epsilon^{ijk} A^{ia} \mathcal{P}_{\textsf{j}}A^{jb} \Gamma^{ab}(\Delta_{\textsf{j}}), \label{eq:corrloop1}
\end{equation}
where the explicit sum over spin $\mathsf{j}$ has been indicated, whereas the sums over spin and flavor indices are understood. Here $A^{ia}$ and $A^{jb}$ are used at the meson-baryon vertices, and $\Gamma^{ab}(\Delta_{\textsf{j}})$ is an antisymmetric tensor which explicitly depends on the difference of the hyperfine mass splitting $\Delta_{\textsf{j}}$. This tensor can be decomposed as
\begin{equation}
\Gamma^{ab}(\Delta_{\textsf{j}}) = A_0(\Delta_{\textsf{j}}) \Gamma_0^{ab} + A_1(\Delta_{\textsf{j}}) \Gamma_1^{ab} + A_2(\Delta_{\textsf{j}}) \Gamma_2^{ab},
\end{equation}
where the tensors $\Gamma_i^{ab}$ are written as \cite{dai}
\begin{subequations}
\begin{eqnarray}
&  & \Gamma_0^{ab} = f^{abQ}, \\
&  & \Gamma_1^{ab} = f^{ab\bar{Q}}, \\
&  & \Gamma_2^{ab} = f^{aeQ}d^{be8} - f^{beQ}d^{ae8} - f^{abe}d^{eQ8}. \label{eq:tens}
\end{eqnarray}
\end{subequations}
Let us recall that $\Gamma_0^{ab}$ and $\Gamma_1^{ab}$ are both SU(3) octets, except that the former transforms as the electric charge, whereas the latter also transforms as the electric charge but is rotated by $\pi$ in isospin space. In turn, $\Gamma_2^{ab}$ breaks SU(3) as $\mathbf{10}+\overline{\mathbf{10}}$ \cite{dai}.

On the other hand, the coefficients $A_i(\Delta_{\textsf{j}})$ are linear combinations of the functions $I(m_\pi,\Delta_{\textsf{j}},\mu)$ and $I(m_K,\Delta_{\textsf{j}},\mu)$, which result from doing the loop integrals; they read
\begin{subequations}
\label{eq:ais}
\begin{eqnarray}
A_0(\Delta_{\textsf{j}}) & = & \frac13 [ I(m_\pi,\Delta_{\textsf{j}},\mu)+2I(m_K,\Delta_{\textsf{j}},\mu) ], \\
A_1(\Delta_{\textsf{j}}) & = & \frac13 [ I(m_\pi,\Delta_{\textsf{j}},\mu)-I(m_K,\Delta_{\textsf{j}},\mu) ], \\
A_2(\Delta_{\textsf{j}}) & = & \frac{1}{\sqrt{3}}[ I(m_\pi,\Delta_{\textsf{j}},\mu)-I(m_K,\Delta_{\textsf{j}},\mu) ],
\end{eqnarray}
\end{subequations}
where the loop integral is \cite{jen92}
\begin{widetext}
\begin{eqnarray}
\left(\frac{8\pi^2 f^2}{M_N} \right) I(m,\Delta,\mu) = -\Delta \ln \frac{m^2}{\mu^2} + \left\{ \begin{array}{ll} \displaystyle 2\sqrt{m^2-\Delta^2}\left[\frac{\pi}{2}-\tan^{-1} \frac{\Delta}{\sqrt{m^2-\Delta^2}} \right], & |\Delta| \leq m, \\
\displaystyle \sqrt{\Delta^2-m^2} \left[-2i\pi + \ln{\frac{\Delta-\sqrt{\Delta^2-m^2}}{\Delta+\sqrt{\Delta^2-m^2}}} \right], & |\Delta| > m, \end{array} \right. \label{eq:loopi}
\end{eqnarray}
\end{widetext}
where $M_N$ and $m$ denote the nucleon and meson masses, respectively, and $\mu$ is the renormalization scale.

Thus, the one-loop correction arising from Fig.~\ref{fig:mmloop1} can be decomposed into the pieces emerging from the flavor $\mathbf{8}$ and flavor $\mathbf{10}+\overline{\mathbf{10}}$ representations as
\begin{eqnarray}
\delta M_{\textrm{loop 1}}^k & = & \sum_{\mathsf{j}} \big[A_0(\Delta_{\mathsf{j}}) M_{\mathbf{8},\textrm{loop 1}}^{kQ}(\mathcal{P}_{\mathsf{j}}) + A_1(\Delta_{\mathsf{j}}) M_{\mathbf{8},\textrm{loop 1}}^{k\bar{Q}}(\mathcal{P}_{\mathsf{j}}) \nonumber \\
&  & \mbox{} + A_2(\Delta_{\mathsf{j}}) M_{\mathbf{10}+\overline{\mathbf{10}},\textrm{loop 1}}^{kQ}(\mathcal{P}_{\mathsf{j}}) \big], \label{eq:loop1}
\end{eqnarray}
where the flavor contributions read
\begin{equation}
M_{\mathbf{8},\textrm{loop 1}}^{kc}(\mathcal{P}_{\mathsf{j}}) = \epsilon^{ijk} f^{abc} A^{ia}\mathcal{P}_{\mathsf{j}}A^{jb}, \label{eq:m8l1}
\end{equation}
and
\begin{eqnarray}
M_{\mathbf{10}+\overline{\mathbf{10}},\textrm{loop 1}}^{kc}(\mathcal{P}_{\mathsf{j}}) & = & \epsilon^{ijk}(f^{aec}d^{be8} - f^{bec}d^{ae8} \nonumber \\
&  & \mbox{} - f^{abe}d^{ec8})A^{ia} \mathcal{P}_{\mathsf{j}} A^{jb}. \label{eq:m10l1}
\end{eqnarray}
For computational purposes, a free flavor index $c$ has been left in Eqs.~(\ref{eq:m8l1}) and (\ref{eq:m10l1}). This free index can be set to $Q=3+(1/\sqrt{3})8$ [or $\bar{Q}=3-(1/\sqrt{3})8$ as the case may be] once the operator reductions on the right-hand sides of such equations have been performed.

\begin{widetext}
The correction $\delta M_{\textrm{loop 1}}^k$, Eq.~(\ref{eq:loop1}), to the SU(3) symmetric value of the baryon magnetic moment can be organized as
\begin{eqnarray}
\delta M_{\textrm{loop 1}}^k & = &
\mathcal{P}_{1/2} \epsilon^{ijk}A^{ia}\mathcal{P}_{1/2}A^{jb}\left[A_0(0)\Gamma_0^{ab}+A_1(0)\Gamma_1^{ab}+A_2(0)\Gamma_2^{ab} \right]\mathcal{P}_{1/2} \nonumber \\
&  & \mbox{} + \mathcal{P}_{1/2} \epsilon^{ijk}A^{ia}\mathcal{P}_{3/2}A^{jb}\left[A_0(\Delta)\Gamma_0^{ab}+A_1(\Delta)\Gamma_1^{ab}+A_2(\Delta)\Gamma_2^{ab} \right]\mathcal{P}_{1/2} \label{eq:mm12}
\end{eqnarray}
for octet baryons,
\begin{eqnarray}
\delta M_{\textrm{loop 1}}^k & = &
\mathcal{P}_{3/2} \epsilon^{ijk}A^{ia}\mathcal{P}_{1/2}A^{jb}\left[A_0(-\Delta)\Gamma_0^{ab}+A_1(-\Delta)\Gamma_1^{ab}+A_2(-\Delta)\Gamma_2^{ab} \right]\mathcal{P}_{3/2} \nonumber \\
&  & \mbox{} + \mathcal{P}_{3/2} \epsilon^{ijk}A^{ia}\mathcal{P}_{3/2}A^{jb}\left[A_0(0)\Gamma_0^{ab}+A_1(0)\Gamma_1^{ab}+A_2(0)\Gamma_2^{ab} \right]\mathcal{P}_{3/2} \label{eq:mm32}
\end{eqnarray}
for decuplet baryons, and
\begin{eqnarray}
\delta M_{\textrm{loop 1}}^k & = &
\mathcal{P}_{3/2} \epsilon^{ijk}A^{ia}\mathcal{P}_{1/2}A^{jb}\left[A_0(0)\Gamma_0^{ab}+A_1(0)\Gamma_1^{ab}+A_2(0)\Gamma_2^{ab} \right]\mathcal{P}_{1/2} \nonumber \\
&  & \mbox{} + \mathcal{P}_{3/2} \epsilon^{ijk}A^{ia}\mathcal{P}_{3/2}A^{jb}\left[A_0(\Delta)\Gamma_0^{ab}+A_1(\Delta)\Gamma_1^{ab}+A_2(\Delta)\Gamma_2^{ab} \right]\mathcal{P}_{1/2} \label{eq:mm3212}
\end{eqnarray}
for decuplet-octet transitions.
\end{widetext}

To proceed further, let us notice that the operator $\epsilon^{ijk} f^{abc} A^{ia}\mathcal{P}_{\mathsf{j}}A^{jb}$ can be decomposed as $\alpha\epsilon^{ijk}f^{abc}A^{ia}A^{jb} + \beta\epsilon^{ijk}f^{abc} A^{ia}J^2 A^{jb}$, where $\alpha$ and $\beta$ are some coefficients. The first summand in the expression mentioned previously corresponds to the degeneracy case $\Delta\to 0$ discussed in Ref.~\cite{rfm09}, whereas the second one is the new contribution to be dealt with in the present analysis. Now, in the product operators such as $\epsilon^{ijk} f^{abc} A^{ia}J^2A^{jb}$, $\epsilon^{ijk} f^{abe}d^{ec8} A^{ia}J^2A^{jb}$, and so on found in Eqs.~(\ref{eq:m8l1}) and (\ref{eq:m10l1}), there will appear up to eight-body operators if we truncate the $1/N_c$ expansion of $A^{kc}$ at the physical value $N_c=3$. The leading order in $1/N_c$ is contained in the product $\epsilon^{ijk} f^{abc} G^{ia}J^2G^{jb}$ and similar terms with two $G$'s, which will be proportional to the square of $a_1$, the leading parameter introduced in Eq.~(\ref{eq:akc}). To perform the current analysis on an equal footing as Ref.~\cite{rfm09}, we work out terms up to relative order $\mathcal{O}(1/N_c^3)$, which implies evaluating products up to seven-body operators in Eqs.~(\ref{eq:m8l1}) and (\ref{eq:m10l1}). The contributions ignored will be proportional to $b_3^2$, $c_3^2$, and $b_3c_3$, which we consider small compared to the ones retained. Because the operator basis is complete \cite{djm95}, the reduction, although long and tedious, is always possible. In Appendix \ref{app:reduc1} we present the relevant reductions of baryon operators up to the order in $1/N_c$ required here.

Gathering together partial results, the \textit{spin-dependent} contributions to be combined with their spin-independent counterparts given in Eqs.~(35) and (36) of Ref.~\cite{rfm09} are as follows:

\begin{widetext}
(1) flavor $\mathbf{8}$ representation:

\begin{eqnarray}
\epsilon^{ijk}f^{abc}A^{ia}J^2A^{jb} & = & - \frac12 (N_c+N_f) a_1^2 \,G^{kc} + \left[\frac12(1+N_f)a_1^2 + \frac{3 N_f}{N_c^2} a_1c_3 \right]\mathcal{D}_2^{kc} \nonumber \\
&  & \mbox{} + \left[-\frac18 (N_c+N_f)a_1^2 - \frac{N_f}{4 N_c} a_1b_2 - \frac{N_c+N_f}{2 N_c^2}a_1b_3 - \frac{3(N_c+N_f)}{2N_c^2} a_1 c_3 \right]\mathcal{D}_3^{kc} \nonumber \\
&  & \mbox{} + \left[-\frac14 (N_c+N_f)a_1^2 - \frac{1+N_f}{N_c}a_1b_2 - \frac{3(N_c+N_f)}{N_c^2}a_1b_3 - \frac{N_c+N_f}{2N_c^2} a_1c_3 \right] \mathcal{O}_3^{kc} \nonumber \\
&  & \mbox{} + \left[\frac14 a_1^2 - \frac{N_f}{4 N_c^2}b_2^2 - \frac{N_f-2}{2N_c^2} a_1b_3 + \frac{7 N_f+12}{4N_c^2} a_1c_3 \right] \mathcal{D}_4^{kc} + \left[- \frac{N_c+N_f}{4N_c^2} a_1c_3 - \frac{N_f}{2N_c^3} b_2b_3 \right] \mathcal{D}_5^{kc} \nonumber \\ 
&  & \mbox{} + \left[-\frac{1}{2 N_c}a_1b_2 - \frac{N_c+N_f}{2N_c^2}a_1b_3 - \frac{N_c+N_f}{4N_c^2} a_1c_3 - \frac{1+N_f}{N_c^3} b_2c_3 \right] \mathcal{O}_5^{kc} \nonumber \\
&  & \mbox{} + \frac{1}{2N_c^2}a_1c_3 \mathcal{D}_6^{kc} -\frac{1}{2N_c^3}b_2c_3 \mathcal{O}_7^{kc} + \mathcal{O}(\mathcal{D}_3J^2\mathcal{D}_3); \label{eq:loop1Ared}
\end{eqnarray}

(2) flavor $\mathbf{10}+\overline{\mathbf{10}}$ representation

\begin{eqnarray}
&  & \epsilon^{ijk}(f^{aec}d^{be8} - f^{bec}d^{ae8} - f^{abe}d^{ec8})A^{ia}J^2A^{jb} \nonumber \\
&  & \mbox{\hglue1.2truecm} = \frac12 a_1^2 \left(\{T^c,G^{k8}\} - \{G^{kc},T^8\}\right) - \frac{1}{N_c}a_1 b_2 \left( \{G^{kc},\{J^r,G^{r8}\}\} - \{G^{k8},\{J^r,G^{rc}\}\} \right) \nonumber \\
&  & \mbox{\hglue1.5truecm} + \frac{1}{2N_c^2}\left(-4 a_1 b_3+5 a_1 c_3\right) \left(\{\mathcal{D}_2^{kc},\{J^r,G^{r8}\}\} -\{\mathcal{D}_2^{k8},\{J^r,G^{rc}\}\} \right) \nonumber \\
&  & \mbox{\hglue1.5truecm} + \left[-\frac14 a_1^2 - \frac{3}{N_c^2}a_1 b_3 - \frac{1}{2N_c^2}a_1c_3\right] \left(\{J^2,\{G^{kc},T^8\}\} - \{J^2,\{G^{k8},T^c\}\} \right) \nonumber \\
&  & \mbox{\hglue1.5truecm} + \left[-\frac{1}{2 N_c}a_1b_2 - \frac{1}{N_c^3}b_2c_3 \right] \left(\{J^2,\{G^{kc},\{J^r,G^{r8}\}\}\} - \{J^2,\{G^{k8},\{J^r,G^{rc}\}\}\} \right) \nonumber \\
&  & \mbox{\hglue1.5truecm} + \left[-\frac{1}{2N_c^2}a_1b_3 - \frac{1}{4N_c^2}a_1c_3 \right] \left(\{J^2,\{J^2,\{G^{kc},T^8\}\}\} - \{J^2,\{J^2,\{G^{k8},T^c\}\}\} \right) \nonumber \\
&  & \mbox{\hglue1.5truecm} + \left[-\frac{1}{2N_c^2}a_1b_3 + \frac{1}{4N_c^2}a_1c_3 \right] \left(\{J^2,\{\mathcal{D}_2^{kc},\{J^r,G^{r8}\}\}\} - \{J^2,\{\mathcal{D}_2^{k8},\{J^r,G^{rc}\}\}\} \right) \nonumber \\
&  & \mbox{\hglue1.5truecm} - \frac{1}{2N_c^3} b_2 c_3 \left(\{J^2,\{J^2,\{G^{kc},\{J^r,G^{r8}\}\}\}\} - \{J^2,\{J^2,\{G^{k8},\{J^r,G^{rc}\}\}\}\}\right) + \mathcal{O}(\mathcal{D}_3J^2\mathcal{D}_3); \label{eq:loop1Bred}
\end{eqnarray}
\end{widetext}
\noindent
where the free flavor index $c$ will be set to $Q=3+(1/\sqrt{3})8$ or $\bar{Q}=3-(1/\sqrt{3})8$ as required in Eq.~(\ref{eq:loop1}). The symbol $\mathcal{O}(\mathcal{D}_3J^2\mathcal{D}_3)$ in Eqs.~(\ref{eq:loop1Ared}) and (\ref{eq:loop1Bred}) means that in the structures such as $\epsilon^{ijk}f^{abc}A^{ia}J^2A^{jb}$, $\epsilon^{ijk} f^{aec}d^{be8}A^{ia}J^2A^{jb}$, and so on we have included all terms up to seven-body operators, such as $\mathcal{D}_2J^2\mathcal{D}_3$, but have neglected contributions which are eight-body operators---like $\mathcal{D}_3J^2\mathcal{D}_3$---or higher. In addition, the operator $[J^2,[T^8,G^{kc}]]$ and its anticommutator with $J^2$ have been omitted in expression (\ref{eq:loop1Bred}) because they do not contribute to any observed magnetic moments.

Notice also that Eqs.~(\ref{eq:loop1Ared}) and (\ref{eq:loop1Bred}) have been rearranged to exhibit explicitly leading and subleading terms in $1/N_c$. It is simple to realize that the one-loop contribution $\delta M_{\textrm{loop 1}}^k$, Eq.~(\ref{eq:loop1}), is order $\mathcal{O}(N_c)$. In the limit of small $m_s$, the symmetry breaking part of $\delta M_{\textrm{loop 1}}^k$ is $\mathcal{O}(m_s^{1/2})$, so the overall contribution of Eq.~(\ref{eq:loop1}) to baryon magnetic moments is $\mathcal{O}(m_s^{1/2}N_c)$; this is the reason why this correction is dominant over the one of Fig.~\ref{fig:mmloop2}.

At this stage, analytical expressions for all 27 possible baryon magnetic and transition magnetic moments can readily be obtained by evaluating the matrix elements of the baryon operators indicated in Eqs.~(\ref{eq:mm12})--(\ref{eq:mm3212}) between baryon SU(6) symmetric states. Most matrix elements are listed in Ref.~\cite{rfm09}, except for a few ones which result from anticommutators of some of the already existing operators with $J^2$, for which the matrix elements can be trivially evaluated. As an example, for $\mu_{\Sigma^-}$ one finds
\begin{widetext}
\begin{eqnarray}
\mu_{\Sigma^-}^{\textrm{(loop 1)}} & = &
\left[\frac{7}{18}a_1^2 +\frac{2}{9}a_1b_2 + \frac{1}{18}b_2^2 + \frac{7}{27}a_1b_3 + \frac{2}{27}b_2b_3 \right]I(m_\pi,0,\mu) \nonumber \\
&  & \mbox{} + \left[\frac{1}{36}a_1^2-\frac{1}{18}a_1 b_2+\frac{1}{36}b_2^2+\frac{1}{54}a_1 b_3-\frac{1}{54}b_2 b_3\right]I(m_K,0,\mu) + \left[ -\frac{1}{18}a_1^2-\frac{1}{18}a_1c_3 \right]I(m_\pi,\Delta,\mu) \nonumber \\
&  & \mbox{} + \left[-\frac{1}{9}a_1^2-\frac{1}{9} a_1 c_3 \right]I(m_K,\Delta,\mu), \label{eq:msig}
\end{eqnarray}
\end{widetext}
which in the limit $\Delta\to 0$ reduces to the value already found \cite{rfm09}. Theoretical expressions like Eq.~(\ref{eq:msig}) are quite useful when comparing our results with the ones obtained in the framework of chiral perturbation theory \cite{jen92,geng,geng2,tib}. It has been already shown that there is a one-to-one correspondence between the parameters of the $1/N_c$ baryon chiral Lagrangian at $N_c=3$ \cite{jen96} and the octet and decuplet chiral Lagrangian \cite{jm255,jm259}. The baryon-meson couplings are related to the coefficients of the $1/N_c$ expansion of $A^{ia}$, Eq.~(\ref{eq:akc}), at $N_c=3$ by
\begin{subequations}
\label{eq:rel1}
\begin{eqnarray}
&  & D = \frac12 a_1 + \frac16 b_3, \\
&  & F = \frac13 a_1 + \frac16 b_2 + \frac19 b_3, \\
&  & \mathcal{C} = - a_1 - \frac12 c_3, \\
&  & \mathcal{H} = - \frac32 a_1 - \frac32 b_2 - \frac52 b_3.
\end{eqnarray}
\end{subequations}

For octet baryons, the magnetic moments computed in Ref.~\cite{jen92} can be rewritten as
\begin{eqnarray}
\mu_i & = & \alpha_i + \sum_{X=\pi, K}\beta_i^{(X)}I(m_X,0,\mu) \nonumber \\
&  & \mbox{} + \sum_{X=\pi, K}\beta_i^{\prime (X)}I(m_X,\Delta,\mu) \nonumber \\
&  & \mbox{} + \sum_{X=\pi, K, \eta} \frac{1}{32\pi^2 f^2} (\overline{\gamma}_i^{(X)}-2\overline{\lambda}_i^{(X)}\alpha_i)m_X^2\ln\frac{m_X^2}{\mu^2}, \nonumber \\
\label{eq:chiral}
\end{eqnarray}
where $\alpha_i$ corresponds to the tree-level value of baryon $i$, $\beta_i^{(X)}$ and $\beta_i^{\prime (X)}$ are the contributions arising from loop graphs of Fig.~\ref{fig:mmloop1}, and the remaining coefficients come from loop graphs of Fig.~\ref{fig:mmloop2}. For $\mu_{\Sigma^-}$ the corresponding chiral coefficients listed in Ref.~\cite{jen92} read
\begin{eqnarray}
&  & \beta_{\Sigma^-}^{(\pi)} = \frac23 D^2 + 2 F^2, \qquad \beta_{\Sigma^-}^{(K)} = (D-F)^2, \nonumber \\
&  & \beta_{\Sigma^-}^{\prime (\pi)} = -\frac{1}{18} \mathcal{C}^2, \qquad \beta_{\Sigma^-}^{\prime (K)} = -\frac19 \mathcal{C}^2.
\end{eqnarray}
Under identifications (\ref{eq:rel1}), the above chiral coefficients coincide with their corresponding analogs in Eq.~(\ref{eq:msig}). The same agreement is found in all expressions for octet baryons. As for decuplet baryons and decuplet-octet transitions, the comparison is not as simple as in the previous case, so we prefer to perform a numerical comparison instead. This will be discussed in the next section.

On the other hand, corrections of order $\mathcal{O}(m_q^{1/2}N_c)$ with a nonvanishing $\Delta$ have some important effects on the Coleman--Glashow relations referred to in the introductory section. First, the term that comes along with $A_0$, $M_{\mathbf{8},\textrm{loop 1}}^{kQ}$ in Eq.~(\ref{eq:loop1}), yields baryon magnetic moments that satisfy relations (\ref{eq:cg}), whereas violations to them are due to the terms that accompany to $A_1$ and $A_2$, which are $M_{\mathbf{8},\textrm{loop 1}}^{k\bar{Q}}$ and $M_{\mathbf{10}+\overline{\mathbf{10}},\textrm{loop 1}}^{kQ}$, respectively. For instance, for the first relation, one has
\begin{widetext}
\begin{eqnarray}
\mu_{\Sigma^+}^{\textrm{(loop 1)}} - \mu_p^{\textrm{(loop 1)}} & = & \left[ -\frac{11}{36}[I(m_K,0,\mu)-I(m_\pi,0,\mu)]-\frac{5}{18}[I(m_K,\Delta,\mu)-I(m_\pi,\Delta,\mu)] \right]a_1^2 \nonumber \\
&  & \mbox{} - \frac{1}{18}[I(m_K,0,\mu)-I(m_\pi,0,\mu)]a_1b_2 + \frac{1}{36}[I(m_K,0,\mu)-I(m_\pi,0,\mu)]b_2^2 \nonumber \\
&  & \mbox{}  - \frac{11}{54}[I(m_K,0,\mu)-I(m_\pi,0,\mu)]a_1b_3 - \frac{1}{54}[I(m_K,0,\mu)-I(m_\pi,0,\mu)]b_2b_3 \nonumber \\
&  & \mbox{} - \frac{5}{18}[I(m_K,\Delta,\mu)-I(m_\pi,\Delta,\mu)]a_1c_3.
\end{eqnarray}
\end{widetext}
Analogous results are obtained for the remaining relations and will not be listed here.

In addition, we can verify that the sum rules derived by Caldi and Pagels \cite{caldi} are also satisfied for $\Delta\neq 0$ in our approach, namely,
\begin{equation}
\mu_{\Sigma^+}^{\textrm{(loop 1)}} + 2\mu_\Lambda^{\textrm{(loop 1)}} + \mu_{\Sigma^-}^{\textrm{(loop 1)}} = 0, \label{eq:cp1}
\end{equation}
\begin{equation}
\mu_{\Xi^0}^{\textrm{(loop 1)}} + \mu_{\Xi^-}^{\textrm{(loop 1)}} + \mu_n^{\textrm{(loop 1)}} - 2\mu_\Lambda^{\textrm{(loop 1)}} + \mu_p^{\textrm{(loop 1)}} = 0, \label{eq:cp2}
\end{equation}
and
\begin{equation}
\mu_\Lambda^{\textrm{(loop 1)}} - \sqrt{3} \mu_{\Lambda\Sigma^0}^{\textrm{(loop 1)}} - \mu_{\Xi^0}^{\textrm{(loop 1)}} -
\mu_n^{\textrm{(loop 1)}} = 0. \label{eq:cp3}
\end{equation}

In turn, the isospin relation
\begin{equation}
\mu_{\Sigma^+}^{\textrm{(loop 1)}} - 2\mu_{\Sigma^0}^{\textrm{(loop 1)}} + \mu_{\Sigma^-}^{\textrm{(loop 1)}} = 0 \label{eq:is1}
\end{equation}
also holds to this order, as it should.

Similarly, for decuplet baryons we find that the $I=2$ sum rules introduced in Ref.~\cite{lb} are also satisfied,
\begin{equation}
\mu_{\Delta^{++}}^{\textrm{(loop 1)}} - \mu_{\Delta^+}^{\textrm{(loop 1)}} - \mu_{\Delta^0}^{\textrm{(loop 1)}} + \mu_{\Delta^-}^{\textrm{(loop 1)}} = 0,
\end{equation}
\begin{equation}
\mu_{{\Sigma^*}^+}^{\textrm{(loop 1)}} - 2 \mu_{{\Sigma^*}^0}^{\textrm{(loop 1)}} + \mu_{{\Sigma^*}^-}^{\textrm{(loop 1)}} = 0,
\end{equation}
whereas for $I=3$
\begin{equation}
\mu_{\Delta^{++}}^{\textrm{(loop 1)}} - 3 \mu_{\Delta^+}^{\textrm{(loop 1)}} + 3 \mu_{\Delta^0}^{\textrm{(loop 1)}} - \mu_{\Delta^-}^{\textrm{(loop 1)}} = 0.
\end{equation}
For transition magnetic moments, the isotensor combinations for $I=2$ read \cite{lb}
\begin{equation}
\mu_{\Delta^{+}p}^{\textrm{(loop 1)}} - \mu_{\Delta^{0}n}^{\textrm{(loop 1)}} = 0,
\end{equation}
and
\begin{equation}
\mu_{{\Sigma^{*}}^+\Sigma^+}^{\textrm{(loop 1)}} - 2 \mu_{{\Sigma^{*}}^0\Sigma^0}^{\textrm{(loop 1)}} + \mu_{{\Sigma^{*}}^-\Sigma^-}^{\textrm{(loop 1)}} = 0. \label{eq:is6}
\end{equation}

In summary, the introduction of a nonvanishing $\Delta$ does not modify the sum rules between magnetic moments derived in previous works.

\subsection{\label{sec:mqlnmq}Diagrams of order $\mathcal{O}(m_q\ln m_q)$}

The loop diagrams displayed in Fig.~\ref{fig:mmloop2} contribute to order $\mathcal{O}(m_q\ln m_q)$ to the baryon magnetic moments. To incorporate the effects of a nonvanishing $\Delta$, the same approach as in the previous case could be followed. This task, however, is rather involved. We will follow a more pragmatic approach instead by using a simple argument: due to the fact that the baryon axial vector current operator and the baryon magnetic moment operator share the same kinematical properties in the large-$N_c$ limit, then the analysis of the former presented in Ref.~\cite{rfm12} will help us save a substantial amount of effort in the present analysis.

Thus, in a close analogy with Eq.~(14) of Ref.~\cite{rfm12}, the operator that yields the one-loop correction to the baryon magnetic moment from diagrams in Fig.~\ref{fig:mmloop2}(a)--(d) can be cast into the single expression\footnote{Equation (53) of Ref.~\cite{rfm09} is the analog of the first summand in Eq.~(\ref{eq:dmkc}). However, in that reference's Eq.~(53), the 1/2 factor was absorbed into the loop integral, where a minus sign is missing. This will be pointed out in a forthcoming erratum.}
\begin{eqnarray}
\delta M^{kc}_{\textrm{loop 2(a-d)}} & = & \frac12 \left[A^{ja},\left[A^{jb},M^{kc}\right]\right] \Pi_{(1)}^{ab} \nonumber \\
&  & \mbox{} - \frac12 \left\{ A^{ja}, \left[M^{kc},\left[\mathcal{M},A^{jb}\right] \right] \right\} \Pi_{(2)}^{ab} \nonumber \\
&  & \mbox{} + \frac16 \big(\left[A^{ja}, \left[\left[\mathcal{M}, \left[ \mathcal{M},A^{jb}\right]\right],M^{kc}\right] \right] \nonumber \\
&  & \mbox{} - \frac12 \left[\left[\mathcal{M},A^{ja}\right], \left[\left[\mathcal{M},A^{jb}\right],M^{kc}\right]\right]\big) \Pi_{(3)}^{ab} \nonumber \\
&  & \mbox{} + \ldots , \label{eq:dmkc}
\end{eqnarray}
so, the actual correction $\delta M^k_{\textrm{loop 2(a-d)}}$ can be obtained as
\begin{eqnarray}
\delta M^k_{\textrm{loop 2(a-d)}} = \delta M^{kQ}_{\textrm{loop 2(a-d)}}. \label{eq:magl2}
\end{eqnarray}

Let us notice that in Eq.~(\ref{eq:dmkc}), $A^{ja}$ and $A^{jb}$ represent the meson-baryon vertices, and $M^{kc}$ denotes an insertion of the baryon magnetic moment operator. Similarly, $\mathcal{M}$ is the baryon mass operator, and $\Pi_{(n)}^{ab}$ represents a symmetric tensor which decomposes into flavor singlet, flavor $\mathbf{8}$, and flavor $\mathbf{27}$ representations as \cite{jen96}
\begin{eqnarray}
\Pi_{(n)}^{ab} & = & F_\mathbf{1}^{(n)} \delta^{ab} + F_\mathbf{8}^{(n)} d^{ab8} \nonumber \\
&  & \mbox{} + F_\mathbf{27}^{(n)} \left[ \delta^{a8} \delta^{b8} - \frac18 \delta^{ab} - \frac35 d^{ab8} d^{888}\right], \label{eq:pisym}
\end{eqnarray}
where
\begin{subequations}
\begin{eqnarray}
F_\mathbf{1}^{(n)} & = & \frac18 [3F^{(n)}(m_\pi,0,\mu) + 4F^{(n)}(m_K,0,\mu) \nonumber \\
&  & \mbox{} + F^{(n)}(m_\eta,0,\mu)], \label{eq:F1} \\
F_\mathbf{8}^{(n)} & = & \frac{2\sqrt 3}{5} \bigg[\frac32 F^{(n)}(m_\pi,0,\mu) - F^{(n)}(m_K,0,\mu) \nonumber \\
&  & \mbox{} - \frac12 F^{(n)}(m_\eta,0,\mu) \bigg], \label{eq:F8} \\
F_\mathbf{27}^{(n)} & = & \frac13 F^{(n)}(m_\pi,0,\mu) - \frac43 F^{(n)}(m_K,0,\mu) \nonumber \\
&  & \mbox{} + F^{(n)}(m_\eta,0,\mu). \label{eq:F27}
\end{eqnarray}
\label{eq:loopffs}
\end{subequations}
Here $F^{(n)}(m_\Pi,0,\mu)$ represents the degeneracy limit $\Delta/m_\Pi \to 0$ of the general function $F^{(n)}(m_\Pi,\Delta,\mu)$, defined as
\begin{equation}
F^{(n)}(m_\Pi,\Delta,\mu) \equiv \frac{\partial^n F(m_\Pi,\Delta,\mu)}{\partial \Delta^n}, \label{eq:fn}
\end{equation}
where $\mu$ is the scale parameter of dimensional regularization. The function $F(m,\Delta,\mu)$ along with its derivatives is given explicitly in Appendix A of Ref.~\cite{rfm12}. In the degeneracy limit, one finds
\begin{subequations}
\label{eq:fprimes}
\begin{eqnarray}
F^{(1)} (m, 0, \mu) & = & - \frac{m^2}{16\pi^2f^2} \ln{\frac{m^2}{\mu^2}}, \label{eq:fprime} \\
F^{(2)}(m, 0, \mu) & = & - \frac{1}{8 \pi f^2}m, \label{eq:fprime2} \\
F^{(3)} (m, 0, \mu) & = & \frac{1}{4\pi^2 f^2}\ln{\frac{m^2}{\mu^2}}. \label{eq:fprime3}
\end{eqnarray}
\label{eq:loopf}
\end{subequations}
Notice that in Eq.~(\ref{eq:fprimes}) we have kept nonanalytic terms in the quark mass explicitly. Analytic terms are scheme dependent and have the same form as higher-dimension terms in the chiral Lagrangian, so they have been omitted.

The computation of the group theoretic structure involved in the loop graphs of Fig.~\ref{fig:mmloop2} can be performed following the lines of Ref.~\cite{rfm12}. Our interest here is computing corrections of relative order $\mathcal{O}(1/N_c^2)$ to $M^{kc}$, which is order $\mathcal{O}(N_c)$. In other words, we need to retain terms up to order $\mathcal{O}(1/N_c^3)$ in $\delta M^{kc}$ in Eq.~(\ref{eq:dmkc}). For vanishing $\Delta$, we will borrow the expressions listed in Appendix B of Ref.~\cite{rfm09}.

For a nonvanishing $\Delta$, however, the insertion of the operator $M^{kc}$, which introduces different coefficients in the expansion compared to $A^{kc}$, does not allow us to straightforwardly borrow the expressions listed in Appendix B of Ref.~\cite{rfm12}. We thus have to take a few steps backward and recalculate some operator reductions. We should stress the fact that in Refs.~\cite{rfm09} and \cite{rfm12}, we inadvertently kept the operator
\begin{eqnarray}
&  & \{J^2,[G^{kc},\{J^r,G^{r8}\}]\} - \{J^2,[G^{k8},\{J^r,G^{rc}\}]\} \nonumber \\
&  & \mbox{} + \{[J^2,G^{kc}],\{J^r,G^{r8}\}\}  - \{[J^2,G^{k8}],\{J^r,G^{rc}\}\} \nonumber \\
&  & \mbox{} - \{J^k,[\{J^m,G^{mc}\},\{J^r,G^{r8}\}]\},
\end{eqnarray}
which vanishes identically. So its presence does not affect any of the expressions where it appears.

After a long, tedious, but otherwise standard, calculation, the one-loop correction to the baryon magnetic moment operator arising from graphs in Fig.~\ref{fig:mmloop2}(a)-(d) can be organized as
\begin{equation}
\delta M_{\textrm{loop 2(a-d)}}^{kc} = \delta M_{\mathbf{1}}^{kc} + \delta M_{\mathbf{8}}^{kc} + \delta M_{\mathbf{27}}^{kc}, \label{eq:dasplit}
\end{equation}
where
\begin{equation}
\delta M_{\mathbf{1}}^{kc} = \sum_{i=1}^{7} x_i X_i^{kc}, \label{eq:das}
\end{equation}
\begin{equation}
\delta M_{\mathbf{8}}^{kc} = \sum_{i=1}^{30} y_i Y_i^{kc}, \label{eq:dao}
\end{equation}
and
\begin{equation}
\delta M_{\mathbf{27}}^{kc} = \sum_{i=1}^{47} z_i Z_i^{kc}. \label{eq:dat}
\end{equation}
The subscript in each summand in Eq.~(\ref{eq:dasplit}) denotes the SU(3) flavor representation it comes from. The operator bases $X_i$, $Y_i$, and $Z_i$ along with the coefficients that accompany them, $x_i$, $y_i$, and $z_i$ are listed in Appendix \ref{app:OpBasis} for the sake of completeness.

As for the one-loop contribution arising from Fig.~\ref{fig:mmloop2}(e), following Refs.~\cite{rfm09,rfm12} and fixing signs and factors, the correction can be written as\footnote{Equation (63) of Ref.~\cite{rfm09} is the analog of Eq.~(\ref{eq:l2e}). However, in that reference's Eq.~(63) the 1/2 factor was absorbed into the loop integral, where a minus sign is missing. This will be pointed out in a forthcoming erratum.}
\begin{equation}
\delta M^k_{\textrm{loop 2(e)}} = -\frac12 [T^a,[T^b,M^k]]\Pi^{ab}, \label{eq:l2e}
\end{equation}
where $\Pi^{ab}$ is a symmetric tensor similar to the one introduced in Eq.~(\ref{eq:pisym}), except that now the integral over the loop is \cite{rfm06}
\begin{equation}
G(m,\mu) = \frac{i}{f^2} \int\frac{d^4k}{(2\pi)^4}\frac{1}{k^2-m^2} = \frac{m^2}{16\pi^2f^2}\left[\ln\frac{m^2}{\mu^2}-1\right]. \label{eq:loopg}
\end{equation}
Following Ref.~\cite{rfm09}, $\delta M^k_{\textrm{loop 2(e)}}$ can be decomposed as
\begin{equation}
\delta M_{\textrm{loop 2(e)}}^k = G_\mathbf{1} M_{\mathbf{1},\textrm{loop 2(e)}}^{kQ} + G_\mathbf{8} M_{\mathbf{8},\textrm{loop 2(e)}}^{kQ} + G_\mathbf{27} M_{\mathbf{27},\textrm{loop 2(e)}}^{kQ}, \label{eq:loop2sum2}
\end{equation}
where the group structures of the double commutator read as follows:

(1) flavor singlet contribution
\begin{eqnarray}
M_{\mathbf{1},\textrm{loop 2(e)}}^{kc} & = & -\frac12 [T^a,[T^a,M^{kc}]] \nonumber \\
& = & -\frac32 M^{kc}; \label{eq:sind}
\end{eqnarray}

(2) flavor octet contribution
\begin{eqnarray}
M_{\mathbf{8},\textrm{loop 2(e)}}^{kc} & = & -\frac12 d^{ab8} [T^a,[T^b,M^{kc}]] \nonumber \\
& = & -\frac34 d^{c8e} M^{ke}; \label{eq:octd}
\end{eqnarray}

(3) flavor $\mathbf{27}$ contribution
\begin{eqnarray}
M_{\mathbf{27},\textrm{loop 2(e)}}^{kc} & = & -\frac12 [T^8,[T^8,M^{kc}]] \nonumber \\
& = & -\frac12 f^{c8e} f^{8eg} M^{kg}. \label{eq:27d}
\end{eqnarray}

Let us notice that in order for $M_{\textbf{27},\textrm{loop 2(e)}}^{kc}$ to be a truly $\mathbf{27}$ contribution singlet and octet pieces must be subtracted off.

Similarly, the functions $G_{\mathbf{1}}$, $G_{\mathbf{8}}$, and $G_{\mathbf{27}}$ have the same structure as their counterparts given by Eqs.~(\ref{eq:F1}), (\ref{eq:F8}), and (\ref{eq:F27}), respectively, written in terms of $G(m,\mu)$. Let us notice that by retaining only the nonanalytic terms in $m_q$ in the loop integrals $F^{(1)}(m_\Pi,0,\mu) = -G(m_\Pi,\mu)$.

\section{\label{sec:su3}The baryon magnetic moment with perturbative SU(3) symmetry breaking}

In the conventional chiral momentum counting scheme, tree diagrams involving higher-order vertices will also contribute to the magnetic moments \cite{krause,ms97} along with the one-loop contributions already discussed. Some of them are needed as counterterms for the divergent parts of the integrals over the loops and are accompanied by low-energy constants, which introduce more unknowns to the low-energy expansion. The leading SU(3) breaking effects of the magnetic moments thus will also have contributions from the effective Lagrangian of order $p^4$ \cite{krause,ms97}, which yield contributions linear in the quark mass. The dependence of the loop integrals on the renormalization scale $\mu$ are of the forms $\ln \mu^2$ for $F^{(3)}$, $m^2\ln \mu^2$ for $F^{(1)}$ and $G$, $\Delta \ln \mu^2$ for $F^{(2)}$ and $I$, and $\Delta^2\ln \mu^2$ also for $F^{(1)}$, where the functions $I(m,\Delta,\mu)$, $F^{(n)}(m,\Delta,\mu)$, and $G(m,\mu)$ are given in Eqs.~(\ref{eq:loopi}), (\ref{eq:fn}), and (\ref{eq:loopg}), respectively. In most of the cases, the $\mu$ dependence of the loop integrals can be compensated by the lowest-order coupling constants, except for the term $m^2\ln \mu^2$, which is formally canceled by the counterterms of order $\mathcal{O}(m_q)$.

In the combined formalism we work with, a convenient way of accounting for terms of order $\mathcal{O}(m_q)$ springs from the fact that flavor SU(3) symmetry breaking transforms as a flavor octet. Thus,  we need to incorporate SB to the baryon magnetic moment operator to linear order in $\epsilon \propto m_s/\Lambda_\chi$.\footnote{$\epsilon$ is a dimensionless measure of SU(3) symmetry breaking; we consider $\epsilon \sim 30\%$ for definiteness.}

Before proceeding any further, we would like to comment on the comparison between the heavy baryon Lagrangian with a $1/N_c$ expansion and the heavy baryon Lagrangian without a $1/N_c$ expansion. More precisely, we want to point out how the different diagrams occurring in heavy baryon chiral perturbation theory (i.e., without $1/N_c$ expansion) are related to our combined formalism. In fact, we have already pointed out that there is a one-to-one correspondence between the parameters of the octet and decuplet chiral Lagrangian and the coefficients of the $1/N_c$ baryon chiral Lagrangian at the physical value $N_c=3$. The relation between the flavor octet baryon-pion couplings $D$, $F$, ${\mathcal C}$, ${\mathcal H}$ and the coefficients of the $1/N_c$ baryon chiral Lagrangian has been provided by Eq.~(\ref{eq:rel1}). If one further includes the SU(3) invariant couplings $\mu_D$, $\mu_F$, $\mu_C$, $\mu_T$ of heavy baryon chiral perturbation theory [but still neglects SU(3) breaking effects], then the correspondence is given by Eq.~(\ref{eq:rel2}). Finally, if one includes SU(3) symmetry breaking effects at linear order in the quark mass matrix, seven new independent terms arise in the heavy baryon Lagrangian at order $p^4$ \cite{jen92}. The seven new effective constants accompanying these terms are related to the various coefficients of the $1/N_c$ expansion that account for SB and that we present below. While the exact correspondence is not needed here, we emphasize that these additional coefficients---and the additional tree-level diagrams occurring at order $p^4$ in heavy baryon chiral perturbation theory---are encoded in the SB coefficients in our combined framework and therefore are accounted for in our numerical analysis.

The issue of SB for a spin-1 object that transforms as a flavor octet under SU(3) has been analyzed in detail in Ref.~\cite{jen12}. This study was then used in the construction of the corrections to the baryon axial vector operator of Ref.~\cite{rfm12}. Thus, the analysis of SB for the baryon magnetic moment operator is then straightforward if we follow the lines of the previous analyses.

If we neglect isospin breaking and include first-order SU(3) symmetry breaking, then $M^{kc}$ has pieces transforming according to all SU(3) representations contained in the tensor product $(1,\mathbf{8}\otimes \mathbf{8})=(1,\mathbf{1}) \oplus (1,\mathbf{8}_S) \oplus (1,\mathbf{8}_A) \oplus (1,\mathbf{10}+\overline{\mathbf{10}}) \oplus (1,\mathbf{27})$, namely,
\begin{equation}
\delta M_{\mathrm{SB}}^{kc} = \delta M_{\mathrm{SB},\mathbf{\mathbf{1}}}^{kc} + \delta M_{\mathrm{SB},\mathbf{\mathbf{8}}}^{kc} + \delta M_{\mathrm{SB},\mathbf{\mathbf{10}+\overline{\mathbf{10}}}}^{kc} + \delta M_{\mathrm{SB},\mathbf{\mathbf{27}}}^{kc}. \label{eq:akcsb}
\end{equation}

The operators in the different representations are given as follows:

\subsection{$(1,\mathbf{1})$}

The $1/N_c$ expansion for the $(1,\mathbf{1})$ operator, to relative order $1/N_c^2$, reads
\begin{equation}
\delta M_{\mathrm{SB},\mathbf{1}}^{kc} = m_1^{1,\mathbf{1}} \delta^{c8}J^k + m_3^{1,\mathbf{1}} \frac{1}{N_c^2} \delta^{c8} \{J^2,J^k\}, \label{eq:ex1}
\end{equation}
where the superscripts attached to the coefficients $m_i^{1,\mathbf{1}}$ indicate the spin-flavor representation. Higher-order terms can be obtained by anticommuting the operators retained with $J^2/N_c^2$.

\subsection{$(1,\mathbf{8})$}

The $1/N_c$ expansion for the $(1,\mathbf{8})$ operator is written as
\begin{eqnarray}
\delta M_{\mathrm{SB},\mathbf{8}}^{kc} & = & n_1^{1,\mathbf{8}} d^{ce8} G^{ke} + n_2^{1,\mathbf{8}} \frac{1}{N_c} d^{ce8} \mathcal{D}_2^{ke} \nonumber \\
&  & \mbox{} + n_3^{1,\mathbf{8}} \frac{1}{N_c^2} d^{ce8} \mathcal{D}_3^{ke} + n_4^{1,\mathbf{8}} \frac{1}{N_c^2} d^{ce8} \mathcal{O}_3^{ke}. \label{eq:ex8}
\end{eqnarray}
Time reversal rules out a similar series with the $d$ symbol replaced by the $f$ symbol. There is another series for the $(1,\mathbf{8})$ operator, which begins with
\begin{equation}
\bar{n}_2^{1,\mathbf{8}} \frac{1}{N_c} f^{ce8}\epsilon^{ijk}\{J^i,G^{je}\}, \label{eq:oth8}
\end{equation}
and higher-order terms can be constructed by anticommuting the leading operator with $J^2/N_c^2$. Let us notice that
\begin{equation}
f^{ce8}\epsilon^{ijk}\{J^i,G^{je}\} = [J^2,[T^8,G^{kc}]]. \label{eq:iden8}
\end{equation}
The right-hand side of Eq.~(\ref{eq:iden8}) shows that the operator only contributes to processes where both spin and strangeness are changed. These processes have not been observed, so the series (\ref{eq:oth8}) will be excluded.

\subsection{$(1,\mathbf{10}+\overline{\mathbf{10}})$}

To relative order $1/N_c^2$, the series for the $(1,\mathbf{10}+\overline{\mathbf{10}})$ symmetry breaking term can be written as
\begin{eqnarray}
\delta M_{\mathrm{SB},\mathbf{\mathbf{10}+\overline{\mathbf{10}}}}^{kc} & = & m_2^{1,\mathbf{10}+\overline{\mathbf{10}}} \frac{1}{N_c} \big(\{G^{kc},T^8\}-\{G^{k8},T^c\} \big) \nonumber \\
&  & \mbox{} + m_3^{1,\mathbf{10}+\overline{\mathbf{10}}} \frac{1}{N_c^2} \big(\{G^{kc},\{J^r,G^{r8}\}\} \nonumber \\
&  & \mbox{} - \{G^{k8},\{J^r,G^{rc}\}\} \big), \label{eq:sb8}
\end{eqnarray}
where the subtractions of the flavor-octet operators off Eq.~(\ref{eq:sb8}) are found to be proportional to the operator $[J^2,[T^8,G^{kc}]]$ and will be ignored \cite{rfm12}.

\subsection{$(1,\mathbf{27})$}

To relative order $1/N_c^2$, the series for the $(1,\mathbf{27})$ operator is written as
\begin{eqnarray}
\delta M_{\mathrm{SB},\mathbf{27}}^{kc} & = & m_2^{1,\mathbf{27}} \frac{1}{N_c} \left(\{G^{kc},T^8\}+\{G^{k8},T^c\} \right) \nonumber \\
&  & \mbox{} + m_3^{1,\mathbf{27}} \frac{1}{N_c^2} \{J^k,\{T^c,T^8\}\} \nonumber \\
&  & \mbox{} + \bar{m}_3^{1,\mathbf{27}} \frac{1}{N_c^2} \big( \{G^{kc},\{J^r,G^{r8}\}\} \nonumber \\
&  & \mbox{} + \{G^{k8},\{J^r,G^{rc}\}\} \big). \label{eq:ex27}
\end{eqnarray}
The subtractions of the flavor-singlet and flavor-octet pieces off Eq.~(\ref{eq:ex27}) are found to be already contained in Eqs.~(\ref{eq:ex1}) and (\ref{eq:ex8}), so Eq.~(\ref{eq:ex27}) can be considered as final \cite{rfm12}.
\section{\label{sec:checks}Total correction to the baryon magnetic moment and consistency checks}

The total corrections to the baryon magnetic moment $M^k$ arise from both one-loop and SB corrections. The one-loop correction, $\delta M_{\mathrm{1L}}^k$, which comes from Figs.~\ref{fig:mmloop1} and \ref{fig:mmloop2}, is obtained by adding up $\delta M_{\textrm{loop 1}}$, given by Eq.~(\ref{eq:corrloop1}), and $\delta M_{\textrm{loop 2}}$, which is the resultant of adding up $M_{\textrm{loop 2(a-d)}}$ and $M_{\textrm{loop 2(e)}}$, given by Eqs.~(\ref{eq:dasplit}) and (\ref{eq:loop2sum2}), respectively. In turn, SB corrections come from Eq.~(\ref{eq:akcsb}). The overall correction to the baryon magnetic moment thus amounts to
\begin{equation}
M^k + \delta M^k = M^{kQ} + \delta M_{\mathrm{1L}}^{kQ} + \delta M_{\mathrm{SB}}^{kQ}. \label{eq:totalmkc}
\end{equation}

The matrix elements of operator (\ref{eq:totalmkc}) between SU(6) symmetric baryon states give the actual values of the baryon magnetic moments. The rather long expressions obtained can indeed shed light on the role SU(3) symmetry breaking plays compared to the SU(3) symmetric case. In this regard, we can perform a series of consistency checks of our expressions using the Coleman--Glashow relations (\ref{eq:cg}), the Caldi--Pagels sum rules (\ref{eq:cp1})--(\ref{eq:cp3}), and the isotensor combinations among baryon magnetic moments (\ref{eq:is1})--(\ref{eq:is6}).

The Coleman--Glashow relations, valid in the limit of exact SU(3) symmetry, thus get corrections from both one-loop and SB. The former contributes with the $\mathbf{8}$ and $\mathbf{27}$ components, whereas the singlet component respects these relations. On the other hand, all the components of SB are present in these relations.

The Caldi--Pagels sum rules are valid up to one-loop corrections of order $\mathcal{O}(m_q^{1/2})$, so corrections to them must arise from one-loop corrections of order $\mathcal{O}(m_q\ln m_q)$ and SB. Explicitly, we find that only the $\mathbf{8}$ and $\mathbf{27}$ components of Fig.~\ref{fig:mmloop2}(a--d) correct these sum rules, whereas Fig.~\ref{fig:mmloop2}(e) does not play any role here. Similarly, SB corrects these sum rules with the $\mathbf{1}$ and $\mathbf{27}$ components, whereas the $\mathbf{8}$ and $\mathbf{10}+\overline{\mathbf{10}}$ respect them. This is in agreement with the $1/N_c$ power counting presented in Table VIII of Ref.~\cite{jen12}, where it is pointed out that the $\mathbf{8}$ and $\mathbf{10}+\overline{\mathbf{10}}$ components of SB contribute at order $\mathcal{O}(m_q^{1/2})$, whereas the $\mathbf{1}$ and $\mathbf{27}$ contribute at order $\mathcal{O}(m_q\ln m_q)$.

Finally, the isotensor combinations are respected both by one-loop and SB corrections, as expected.

We are now in a position of performing a detailed comparison of our theoretical expressions with the experimental data \cite{part} through various fits. This is now discussed in the next section.

\section{\label{sec:num}Fits to the experimental data}

We now proceed to perform a numerical comparison of the theoretical expressions obtained here with the available experimental data through a least-squares fit. Nowadays, only 10 out of 27 possible magnetic moments are reported in the Review of Particle Physics \cite{part}. They correspond to the magnetic moments of the octet baryons (excluding $\mu_{\Sigma^0}$, which has not been measured) and the transition magnetic moment $\mu_{\Lambda\Sigma^0}$, along with $\mu_{\Omega^-}$ and $\mu_{\Delta^+p}$. To diversify the data, we use $\mu_{\Delta^{++}}$ reported in Ref.~\cite{lopez}, which was obtained from radiative $\pi^+p$ scattering with a dynamical model. We also use two more data, $\mu_{\Sigma^{*0}\Lambda}$ and $\mu_{\Sigma^{*+}\Sigma^+}$, measured recently by the CLAS Collaboration \cite{clas1,clas2}. We thus have 13 data points about magnetic moments at our disposal. All this information is displayed in the third column (from left to right) of Table \ref{t:fits}.

We can perform a number of fits to compare theory and experiment. However, we consider pertinent it to perform those fits which somehow display information on the departure from exact SU(3) symmetry. In doing this, we find some limitations about the number of magnetic moments measured and the number of unknown parameters we need to determine: at tree level, there are four parameters, namely, $m_1,\ldots, m_4$. One-loop corrections introduce four more parameters, the ones which come along the axial vector current operator, namely, $a_1$, $b_2$, $b_3$, and $c_3$. SB introduces 11 more parameters, $m_1^{1,\mathbf{1}}$, $m_3^{1,\mathbf{1}}$, $n_1^{1,\mathbf{8}},\ldots,n_4^{1,\mathbf{8}}$, $m_2^{1,\mathbf{10}+\overline{\mathbf{10}}}$, $m_3^{1,\mathbf{10}+\overline{\mathbf{10}}}$, $m_2^{1,\mathbf{27}}$, $m_3^{1,\mathbf{27}}$, and $\bar{m}_3^{1,\mathbf{27}}$. We thus need to implement some criteria which allows us to reduce the number of parameters compared to the number of measured quantities. Let us discuss briefly what can be done.

At tree level, the operators that accompany the coefficients $m_1$, $m_2$, $m_3$, and $m_4$ are of orders $\mathcal{O}(N_c)$, $\mathcal{O}(1/N_c)$, $\mathcal{O}(1/N_c)$, and $\mathcal{O}(1/N_c)$, respectively, and so are the operators that come along the coefficients $n_1^{1,\mathbf{8}}$, $n_2^{1,\mathbf{8}}$, $n_3^{1,\mathbf{8}}$, and $n_4^{1,\mathbf{8}}$ \cite{jen12}. Similarly, $m_1^{1,\mathbf{1}}$ and $m_3^{1,\mathbf{1}}$ are accompanied by operators which are of orders $\mathcal{O}(1)$ and $\mathcal{O}(1/N_c^2)$, respectively. In turn, $m_2^{1,\mathbf{10}+\overline{\mathbf{10}}}$ and $m_3^{1,\mathbf{10}+\overline{\mathbf{10}}}$ come along with operators of orders $\mathcal{O}(1)$ and $\mathcal{O}(1/N_c)$, respectively. Finally, $m_2^{1,\mathbf{27}}$, $m_3^{1,\mathbf{27}}$ and $\bar{m}_3^{1,\mathbf{27}}$ go with operators of orders $\mathcal{O}(1)$, $\mathcal{O}(1/N_c^2)$ and $\mathcal{O}(1/N_c)$, respectively \cite{jen12}. This apparent complexity suggests some patterns about the terms one needs to retain for a consistent numerical analysis.

\subsection{SU(3) symmetric fit}

The simplest fit we can perform is an SU(3) symmetric fit. For this task we keep only the terms that come along with $M^k$ at tree level, namely, $m_1$, $m_2$, $m_3$, and $m_4$. This is identical to a fit using the SU(3) invariant couplings $\mu_D$, $\mu_F$, $\mu_C$, $\mu_T$ of heavy baryon chiral perturbation theory \cite{jen92}, neglecting all SU(3) breaking effects.

Without further ado, the fit yields
\begin{eqnarray}
\begin{array}{ll}
m_1=5.03\pm 0.51,   & m_2=0.72\pm 1.54, \\[2mm]
m_3=-0.30\pm 0.98, & m_4=4.06 \pm 1.49,
\label{eq:fita}
\end{array}
\end{eqnarray}
or equivalently, $\mu_D=2.47\pm 1.17$, $\mu_F=1.76\pm 1.11$, $\mu_C=2.63\pm 0.81$, and $\mu_T=-14.12\pm 5.04$. Here a theoretical error of $\delta\mu_{\textrm{th}}=0.362\, \mu_N$ has been added in quadrature in order to achieve $\chi^2=1/\textrm{degrees of freedom}$. The best-fit parameters listed in Eq.~(\ref{eq:fita}) depart noticeably from the expected order $\mathcal{O}(N_c^0)$ values; needless to say, the numerical values of the SU(3) invariant couplings do not match the ones found in the original paper \cite{jen92}. This is not a withdrawal of our approach. Actually, we could have scaled all the theoretical expressions by dividing them by a factor, let us say, $\alpha_0=2\mu_p^{\textrm{exp}}$, in the same way we did in Ref.~\cite{rfm09}. We prefer not to do so in order to compare our outputs with the ones of Ref.~\cite{jen12}. Indeed, fit A in the present case is equivalent to fit A of this reference, and our best-fit parameters (\ref{eq:fita}) are comparable to those obtained there. 

The predicted magnetic moments are listed in the column labeled fit A in Table \ref{t:fits}. A quick glance at these results shows that the magnetic moments are poorly determined in the limit of exact SU(3) symmetry.

\subsection{Perturbative SU(3) symmetry breaking}

The next fit consists of taking into account only the SB effects. Strictly speaking, there are 15 free parameters, which exceed the available data. We can perform a kind of a restricted fit if we ignore factors of order $1/N_c^2$ in the $1/N_c$ expansion, which is equivalent to rule out the terms that come along with $m_3^{1,\mathbf{1}}$ and $m_3^{1,\mathbf{27}}$. We can reduce by one more parameter if we neglect the $1/N_c$ contribution of the $\mathbf{27}$ and leave only the order $\mathcal{O}(1)$ term, namely, $m_2^{1,\mathbf{27}}$. We are thus left with 12 parameters. The fit yields
\begin{eqnarray}
\begin{array}{ll}
m_1=4.48\pm 0.14, & m_2=0.83\pm 0.33, \\[2mm]
m_3=0.08\pm 0.32, & m_4=5.86\pm 2.14, \\[2mm]
m_1^{1,\mathbf{1}}=0.12\pm 0.12, & \\[2mm]
n_1^{1,\mathbf{8}}=1.18\pm 0.27, & n_2^{1,\mathbf{8}}=-0.26\pm 0.55, \\[2mm]
n_3^{1,\mathbf{8}}=0.54\pm 0.71, & n_4^{1,\mathbf{8}}=-4.89\pm 5.37, \\[2mm]
m_2^{1,\mathbf{10}+\overline{\mathbf{10}}}=0.42\pm 0.14, & m_3^{1,\mathbf{10}+\overline{\mathbf{10}}}= 1.66\pm 2.40, \\[2mm]
m_2^{1,\mathbf{27}}=0.06\pm 0.26. &
\end{array}
\label{eq:fitb}
\end{eqnarray}
The theoretical error added in quadrature to get $\chi^2=1/\textrm{degrees of freedom}$ this time is $\delta\mu_{\textrm{th}}=0.062\, \mu_N$, which is considerably smaller than the one added in the previous case. This output is equivalent to fit F of Ref.~\cite{jen12} and our best-fit parameters are fairly comparable to the ones obtained there. We notice some rearrangements in the leading-order parameters compared to the symmetric case, except for $m_4$, which remains ill determined (even its value worsens in this case). The parameters arising from SB are roughly speaking according to the expected $\mathcal{O}(\epsilon)\sim 30\%$ measure of SB, except for $n_4^{1,\mathbf{8}}$, which turns larger than expected. With these best-fit parameters, the predicted magnetic moments are listed in Table \ref{t:fits}, labeled as fit B. In this case, the agreement between theory and experiment is good.

\subsection{Total correction}

The next relevant fit we can perform consists of adding one-loop corrections to the previous cases. We split this analysis into two parts. In a first stage we consider the degenerate case, namely, $\Delta=0$. In a second stage, we consider a nonvanishing $\Delta$, which we set to $\Delta=0.231 \, \textrm{GeV}/c^2$ for definiteness. This will allow us to quantify the effects of $\Delta$. Let us recall that one-loop corrections depend also on the quantities that parametrize the baryon axial vector current. In other words, we need the values of $a_1$, $b_2$, $b_3$, and $c_3$. The impossibility of extracting them from the current data forces us to use them from other sources. For this purpose we use the best-fit values reported in Ref.~\cite{rfm12}, where the renormalization of the baryon axial vector current was computed at the very same order of approximation in $1/N_c$ as we have done for the baryon magnetic moment in the present analysis. The values obtained there are $a_1=0.64$, $b_2=0.21$, $b_3=1.35$, and $c_3=1.90$. For definiteness, we use the physical masses of the pseudoscalar mesons listed in Ref.~\cite{part}.

Thus, for $\Delta=0$ we find
\begin{eqnarray}
\begin{array}{ll}
m_1=7.27\pm 0.11, & m_2=-1.92\pm 0.19, \\[2mm]
m_3=0.76\pm 0.24, & m_4=9.44\pm 1.32, \\[2mm]
m_1^{1,\mathbf{1}}=0.31\pm 0.15, & \\[2mm]
n_1^{1,\mathbf{8}}=-0.53\pm 0.31, & n_2^{1,\mathbf{8}}=1.50\pm 0.64, \\[2mm]
n_3^{1,\mathbf{8}}=1.62\pm 0.84, & n_4^{1,\mathbf{8}}=-12.39\pm 5.39, \\[2mm]
m_2^{1,\mathbf{10}+\overline{\mathbf{10}}}=-1.39\pm 0.17, & m_3^{1,\mathbf{10}+\overline{\mathbf{10}}}= 1.51\pm 2.71, \\[2mm]
m_2^{1,\mathbf{27}}=-0.44\pm 0.32. &
\end{array}
\label{eq:fitc}
\end{eqnarray}
The theoretical error added in quadrature to get $\chi^2=1/\textrm{degrees of freedom}$ is $\delta\mu_{\textrm{th}}=0.075\, \mu_N$. The predicted magnetic moments are listed in Table \ref{t:fits} labeled as fit C and the corresponding tree level and SU(3) breaking components are listed in Table \ref{t:fitc} for the sake of completeness.

On the other hand, for $\Delta=0.231 \, \textrm{GeV}/c^2$, we find
\begin{eqnarray}
\begin{array}{ll}
m_1=6.61\pm 0.13, & m_2=-6.36\pm 0.19, \\[2mm]
m_3=4.96\pm 0.27, & m_4=9.41\pm 1.63, \\[2mm]
m_1^{1,\mathbf{1}}=0.99\pm 0.12, & \\[2mm]
n_1^{1,\mathbf{8}}=1.77\pm 0.27, & n_2^{1,\mathbf{8}}=-4.78\pm 0.47, \\[2mm]
n_3^{1,\mathbf{8}}=-0.14\pm 0.72, & n_4^{1,\mathbf{8}}=-7.22\pm 5.19, \\[2mm]
m_2^{1,\mathbf{10}+\overline{\mathbf{10}}}=-0.21\pm 0.14, & m_3^{1,\mathbf{10}+\overline{\mathbf{10}}}=-2.32\pm 2.39, \\[2mm]
m_2^{1,\mathbf{27}}=-0.14\pm 0.24. &
\end{array}
\label{eq:fitd}
\end{eqnarray}
The theoretical error added in quadrature to get $\chi^2=1/\textrm{degrees of freedom}$ is $\delta\mu_{\textrm{th}}=0.058\, \mu_N$. The predicted magnetic moments are listed in Table \ref{t:fits} labeled as fit D and the corresponding tree level and SU(3) breaking components are listed in Table \ref{t:fitd}, also for the sake of completeness.

The numerical values of the baryon magnetic moments obtained with the inclusion of one-loop corrections (fits C and D) are in good agreement with the experimental ones. However, the predicted values for the unmeasured ones differ between them in some cases rather remarkably. For instance, the most important differences are observed in the magnetic moments $\mu_{\Delta^0}$, $\mu_{\Delta^-}$, $\mu_{\Sigma^{*-}}$, $\mu_{\Xi^{*0}}$, and in the transition magnetic moments $\mu_{\Sigma^{*-}\Sigma^-}$ and $\mu_{\Xi^{*-}\Xi^-}$, for which the values are radically different with the inclusion of $\Delta$.

\begin{table*}
\caption{\label{t:fits}Numerical values of baryon magnetic moments found in this work. Comparisons with other determinations are also included. The entries are given in nuclear magnetons.}
\begin{ruledtabular}
\begin{tabular}{llrrrrrrrrrrr}
& \textrm{Baryon} & \textrm{Experimental data} & \textrm{fit A} & \textrm{fit B} & \textrm{fit C} & \textrm{fit D} & Ref.~\cite{ms97} &Ref.~\cite{geng2} & Ref.~\cite{lb} & Ref.~\cite{jen12}\footnote{fit F of this reference} & Ref.~\cite{keller} \\
\hline
$ 1$ & $n$                   & $-1.913 \pm 0.000$ & $-1.644$ & $-1.931$ & $-1.936$ & $-1.929$ & $-1.91$ & & & $-1.93$ & \\
$ 2$ & $p$                   & $ 2.793 \pm 0.000$ & $ 2.587$ & $ 2.793$ & $ 2.793$ & $ 2.793$ & $2.79$ & & & $2.70$ & \\
$ 3$ & $\Sigma^-$            & $-1.160 \pm 0.025$ & $-0.943$ & $-1.155$ & $-1.154$ & $-1.155$ & $-1.16$ & & & $-1.15$ & \\
$ 4$ & $\Sigma^0$            &                    & $ 0.822$ & $ 0.654$ & $ 0.655$ & $ 0.653$ & $0.65$ & & $ 0.77(10)$ & $0.65$ & \\
$ 5$ & $\Sigma^+$            & $ 2.458 \pm 0.010$ & $ 2.587$ & $ 2.463$ & $ 2.464$ & $ 2.462$ & $2.46$ & & & $2.46$ & \\
$ 6$ & $\Xi^-$               & $-0.651 \pm 0.003$ & $-0.943$ & $-0.651$ & $-0.651$ & $-0.651$ & $-0.65$ & & & $-0.65$ & \\
$ 7$ & $\Xi^0$               & $-1.250 \pm 0.014$ & $-1.644$ & $-1.269$ & $-1.273$ & $-1.267$ & $-1.25$ & & & $-1.27$ & \\
$ 8$ & $\Lambda$             & $-0.613 \pm 0.004$ & $-0.822$ & $-0.586$ & $-0.579$ & $-0.589$ & $-0.61$ & & & $-0.59$ & \\
$ 9$ & $\Lambda\Sigma^0$     & $ 1.61  \pm 0.08 $ & $ 1.424$ & $ 1.529$ & $ 1.526$ & $ 1.530$ & $1.40$ & & & $-1.53$ & \\
$10$ & $\Delta^{++}$         & $ 6.14  \pm 0.51 $\footnote{Value reported in Ref.~\cite{lopez}} & $ 5.252$ & $ 6.140$ & $6.140$ & $ 6.140$ & & $ 6.04(13)$ & & $6.14$ & \\
$11$ & $\Delta^+$            &                    & $ 2.626$ & $ 2.857$ & $ 2.252$ & $ 3.058$ & & $2.84(2)$ & $ 3.04(13)$ & $2.79$ & \\
$12$ & $\Delta^0$            &                    & $ 0.000$ & $-0.427$ & $-1.636$ & $-0.023$ & & $-0.36(9)$ & $ 0.00(10)$ & $-0.56$ & \\
$13$ & $\Delta^-$            &                    & $-2.626$ & $-3.710$ & $-5.523$ & $-3.105$ & & $-3.56(20)$ & $-3.04(13)$ & $-3.91$ & \\
$14$ & $\Sigma^{*+}$         &                    & $ 2.626$ & $ 3.350$ & $ 3.896$ & $ 2.520$ & & $3.07(12)$ & $ 3.35(13)$ & $3.49$ & \\
$15$ & $\Sigma^{*0}$         &                    & $ 0.000$ & $ 0.102$ & $-0.268$ & $-0.159$ & & $0$ & $ 0.32(11)$ & $0.10$ & \\
$16$ & $\Sigma^{*-}$         &                    & $-2.626$ & $-3.147$ & $-4.433$ & $-2.838$ & & $-3.07(12)$ & $-2.70(13)$ & $-3.28$ & \\
$17$ & $\Xi^{*0}$            &                    & $ 0.000$ & $ 0.630$ & $ 1.195$ & $-0.117$ & & $0.36(9)$ & $ 0.64(11)$ & $0.77$ & \\
$18$ & $\Xi^{*-}$            &                    & $-2.626$ & $-2.583$ & $-3.265$ & $-2.476$ & & $-2.56(6)$ & $-2.36(14)$ & $-2.65$ & \\
$19$ & $\Omega^-$            & $-2.02  \pm 0.05 $ & $-2.626$ & $-2.020$ & $-2.020$ & $-2.020$ & & $-2.02$ & & $-2.02$ & \\
$20$ & $\Delta^+p$           & $ 3.51  \pm 0.09 $ & $ 3.329$ & $ 3.510$ & $ 3.510$ & $ 3.510$ & & & & $3.51$ & \\
$21$ & $\Delta^0n$           &                    & $ 3.329$ & $ 3.510$ & $ 3.510$ & $ 3.510$ & & & $3.51(11)$ & $3.51$ & \\
$22$ & $\Sigma^{*0}\Lambda$  & $ 2.73  \pm 0.25 $\footnote{Value extracted from Ref.~\cite{clas1}} & $ 2.883$ & $ 2.730$ & $ 2.732$ & $ 2.731$ & & & $ 2.93(11)$ & $2.74$ & 2.68(04) \\
$23$ & $\Sigma^{*0}\Sigma^0$ &                    & $ 1.665$ & $ 1.919$ & $ 2.389$ & $ 1.592$ & & &  $ 1.39(11)$ & $2.01$ & 1.61(07) \\
$24$ & $\Sigma^{*+}\Sigma^+$ & $ 3.17  \pm 0.36 $\footnote{Value extracted from Ref.~\cite{clas2}} & $ 3.329$ & $ 3.170$ & $ 3.166$ & $ 3.168$ & & & $ 2.97(11)$ & $3.22$ & 3.22(05) \\
$25$ & $\Sigma^{*-}\Sigma^-$ &                    & $ 0.000$ & $ 0.667$ & $ 1.611$ & $ 0.016$ & & & $-0.19(11)$ & $0.79$ & 0.0(20) \\
$26$ & $\Xi^{*0}\Xi^0$       &                    & $ 3.329$ & $ 3.137$ & $ 3.533$ & $ 2.787$ & & & $ 2.96(12)$ & $3.25$ & 3.21(15) \\
$27$ & $\Xi^{*-}\Xi^-$       &                    & $ 0.000$ & $ 0.667$ & $ 1.568$ & $ 0.033$ & & & $-0.19(11)$ & $0.79$ &
\end{tabular}
\end{ruledtabular}
\end{table*}

\begin{table*}
\caption{\label{t:fitc}SU(3) flavor contributions to the baryon magnetic moments obtained for fit C.}
\begin{ruledtabular}
\begin{tabular}{lrrrrrrrrrr}
& & & & & \multicolumn{3}{c}{Fig.~\ref{fig:mmloop2}(a-d)} & \multicolumn{3}{c}{Fig.~\ref{fig:mmloop2}(e)} \\
 & Total & Tree & SB & Fig.~\ref{fig:mmloop1} & $\mathbf{1}$ & $\mathbf{8}$ & $\mathbf{27}$ & $\mathbf{1}$ & $\mathbf{8}$ & $\mathbf{27}$ \\ \hline
$n$                   & $-1.936$ & $-2.508$ & $ 0.294$ & $ 1.020$ & $-0.359$ & $-0.018$ & $ 0.001$ & $-0.678$ & $ 0.310$ & $ 0.002$ \\
$p$                   & $ 2.793$ & $ 3.443$ & $-0.345$ & $-1.407$ & $ 0.347$ & $ 0.042$ & $-0.006$ & $ 0.930$ & $-0.226$ & $ 0.014$ \\
$\Sigma^-$            & $-1.154$ & $-0.934$ & $-0.241$ & $-0.142$ & $ 0.012$ & $ 0.082$ & $-0.005$ & $-0.252$ & $ 0.310$ & $ 0.018$ \\
$\Sigma^0$            & $ 0.655$ & $ 1.254$ & $ 0.089$ & $-1.359$ & $ 0.180$ & $ 0.016$ & $ 0.001$ & $ 0.339$ & $ 0.113$ & $ 0.022$ \\
$\Sigma^+$            & $ 2.464$ & $ 3.443$ & $ 0.418$ & $-2.575$ & $ 0.347$ & $-0.050$ & $ 0.008$ & $ 0.930$ & $-0.084$ & $ 0.027$ \\
$\Xi^-$               & $-0.651$ & $-0.934$ & $ 0.190$ & $ 0.639$ & $ 0.012$ & $-0.058$ & $ 0.008$ & $-0.252$ & $-0.226$ & $-0.029$ \\
$\Xi^0$               & $-1.273$ & $-2.508$ & $-0.117$ & $ 2.466$ & $-0.359$ & $ 0.050$ & $-0.011$ & $-0.678$ & $-0.084$ & $-0.032$ \\
$\Lambda$             & $-0.579$ & $-1.254$ & $ 0.090$ & $ 1.359$ & $-0.180$ & $-0.124$ & $ 0.004$ & $-0.339$ & $-0.113$ & $-0.022$ \\
$\Lambda\Sigma^0$     & $ 1.526$ & $ 2.172$ & $ 0.001$ & $-1.228$ & $ 0.311$ & $-0.121$ & $-0.005$ & $ 0.587$ & $-0.195$ & $ 0.004$ \\
$\Delta^{++}$         & $ 6.140$ & $ 6.615$ & $ 0.823$ & $-5.329$ & $ 3.408$ & $-0.838$ & $-0.068$ & $ 1.787$ & $-0.298$ & $ 0.039$ \\
$\Delta^+$            & $ 2.252$ & $ 3.308$ & $ 0.016$ & $-3.161$ & $ 1.704$ & $-0.499$ & $-0.041$ & $ 0.893$ & $ 0.000$ & $ 0.033$ \\
$\Delta^0$            & $-1.636$ & $ 0.000$ & $-0.791$ & $-0.993$ & $ 0.000$ & $-0.161$ & $-0.014$ & $ 0.000$ & $ 0.298$ & $ 0.026$ \\
$\Delta^-$            & $-5.523$ & $-3.308$ & $-1.598$ & $ 1.175$ & $-1.704$ & $ 0.177$ & $ 0.013$ & $-0.893$ & $ 0.595$ & $ 0.020$ \\
$\Sigma^{*+}$         & $ 3.896$ & $ 3.308$ & $ 1.327$ & $-2.168$ & $ 1.704$ & $-0.873$ & $-0.003$ & $ 0.893$ & $-0.298$ & $ 0.007$ \\
$\Sigma^{*0}$         & $-0.268$ & $ 0.000$ & $ 0.268$ & $ 0.000$ & $ 0.000$ & $-0.517$ & $-0.019$ & $ 0.000$ & $ 0.000$ & $ 0.000$ \\
$\Sigma^{*-}$         & $-4.433$ & $-3.308$ & $-0.791$ & $ 2.168$ & $-1.704$ & $-0.161$ & $-0.035$ & $-0.893$ & $ 0.298$ & $-0.007$ \\
$\Xi^{*0}$            & $ 1.195$ & $ 0.000$ & $ 1.327$ & $ 0.993$ & $ 0.000$ & $-0.873$ & $ 0.072$ & $ 0.000$ & $-0.298$ & $-0.026$ \\
$\Xi^{*-}$            & $-3.265$ & $-3.308$ & $ 0.016$ & $ 3.161$ & $-1.704$ & $-0.499$ & $-0.005$ & $-0.893$ & $ 0.000$ & $-0.033$ \\
$\Omega^-$            & $-2.020$ & $-3.308$ & $ 0.823$ & $ 4.155$ & $-1.704$ & $-0.838$ & $ 0.101$ & $-0.893$ & $-0.298$ & $-0.059$ \\
$\Delta^+p$           & $ 3.510$ & $ 5.654$ & $-2.121$ & $-3.064$ & $ 2.530$ & $-0.478$ & $-0.040$ & $ 1.527$ & $-0.509$ & $ 0.011$ \\
$\Delta^0n$           & $ 3.510$ & $ 5.654$ & $-2.121$ & $-3.064$ & $ 2.530$ & $-0.478$ & $-0.040$ & $ 1.527$ & $ 0.011$ & $-0.509$ \\
$\Sigma^{*0}\Lambda$  & $ 2.732$ & $ 4.896$ & $-1.585$ & $-3.464$ & $ 2.191$ & $-0.230$ & $ 0.033$ & $ 1.323$ & $-0.441$ & $ 0.010$ \\
$\Sigma^{*0}\Sigma^0$ & $ 2.389$ & $ 2.827$ & $ 0.915$ & $-3.833$ & $ 1.265$ & $ 0.133$ & $ 0.014$ & $ 0.764$ & $ 0.254$ & $ 0.050$ \\
$\Sigma^{*+}\Sigma^+$ & $ 3.166$ & $ 5.654$ & $ 0.053$ & $-6.770$ & $ 2.530$ & $ 0.085$ & $ 0.031$ & $ 1.527$ & $ 0.000$ & $ 0.056$ \\
$\Sigma^{*-}\Sigma^-$ & $ 1.611$ & $ 0.000$ & $ 1.777$ & $-0.896$ & $ 0.000$ & $ 0.180$ & $-0.003$ & $ 0.000$ & $ 0.509$ & $ 0.045$ \\
$\Xi^{*0}\Xi^0$       & $ 3.533$ & $ 5.654$ & $ 0.291$ & $-6.770$ & $ 2.530$ & $ 0.213$ & $ 0.033$ & $ 1.527$ & $ 0.000$ & $ 0.056$ \\
$\Xi^{*-}\Xi^-$       & $ 1.568$ & $ 0.000$ & $ 1.777$ & $-0.896$ & $ 0.000$ & $ 0.180$ & $-0.047$ & $ 0.000$ & $ 0.509$ & $ 0.045$
\end{tabular}
\end{ruledtabular}
\end{table*}

\begin{turnpage}

\begin{table*}
\caption{\label{t:fitd}SU(3) flavor contributions to the baryon magnetic moments obtained for fit D.}
\begin{ruledtabular}
\begin{tabular}{lrrrrrrrrrrrrrrrrr}
& & & & \multicolumn{2}{c}{Fig.~\ref{fig:mmloop1}} & \multicolumn{3}{c}{Fig.~\ref{fig:mmloop2}(a-d), $\mathcal{O}(\Delta^0)$} & \multicolumn{3}{c}{Fig.~\ref{fig:mmloop2}(a-d), $\mathcal{O}(\Delta)$} & \multicolumn{3}{c}{Fig.~\ref{fig:mmloop2}(a-d), $\mathcal{O}(\Delta^2)$} & \multicolumn{3}{c}{Fig.~\ref{fig:mmloop2}(e)} \\
 & Total & Tree & SB & $\mathcal{O}(\Delta^0)$ & $\mathcal{O}(\Delta)$ & $\mathbf{1}$ & $\mathbf{8}$ & $\mathbf{27}$ & $\mathbf{1}$ & $\mathbf{8}$ & $\mathbf{27}$ & $\mathbf{1}$ & $\mathbf{8}$ & $\mathbf{27}$ & $\mathbf{1}$ & $\mathbf{8}$ & $\mathbf{27}$ \\
\hline
$n$                   & $-1.929$ & $-2.755$ & $ 0.313$ & $ 0.418$ & $ 0.568$ & $-0.743$ & $ 0.140$ & $ 0.001$ & $ 0.597$ & $-0.355$ & $-0.004$ & $ 0.248$ & $ 0.113$ & $ 0.001$ & $-0.744$ & $ 0.276$ & $-0.003$ \\
$p$                   & $ 2.793$ & $ 3.073$ & $ 0.527$ & $-1.382$ & $-0.095$ & $ 0.311$ & $-0.047$ & $-0.011$ & $ 0.680$ & $-0.491$ & $-0.015$ & $-0.233$ & $-0.114$ & $-0.001$ & $ 0.830$ & $-0.248$ & $ 0.009$ \\
$\Sigma^-$            & $-1.155$ & $-0.318$ & $ 0.236$ & $ 0.297$ & $-0.379$ & $ 0.431$ & $ 0.051$ & $ 0.000$ & $-1.277$ & $-0.411$ & $ 0.015$ & $-0.015$ & $ 0.001$ & $ 0.001$ & $-0.086$ & $ 0.276$ & $ 0.021$ \\
$\Sigma^0$            & $ 0.653$ & $ 1.378$ & $ 0.120$ & $-0.781$ & $-0.473$ & $ 0.371$ & $ 0.064$ & $ 0.010$ & $-0.298$ & $-0.198$ & $ 0.010$ & $-0.124$ & $ 0.050$ & $ 0.006$ & $ 0.372$ & $ 0.124$ & $ 0.024$ \\
$\Sigma^+$            & $ 2.462$ & $ 3.073$ & $ 0.005$ & $-1.860$ & $-0.568$ & $ 0.311$ & $ 0.076$ & $ 0.020$ & $ 0.680$ & $ 0.015$ & $ 0.005$ & $-0.233$ & $ 0.099$ & $ 0.010$ & $ 0.830$ & $-0.029$ & $ 0.028$ \\
$\Xi^-$               & $-0.651$ & $-0.318$ & $ 0.604$ & $ 0.940$ & $-0.284$ & $ 0.431$ & $ 0.042$ & $ 0.006$ & $-1.277$ & $-0.435$ & $ 0.013$ & $-0.015$ & $-0.002$ & $ 0.001$ & $-0.086$ & $-0.248$ & $-0.025$ \\
$\Xi^0$               & $-1.267$ & $-2.755$ & $-0.073$ & $ 1.587$ & $ 0.757$ & $-0.743$ & $-0.013$ & $-0.020$ & $ 0.597$ & $-0.042$ & $ 0.007$ & $ 0.248$ & $-0.013$ & $-0.002$ & $-0.744$ & $-0.029$ & $-0.030$ \\
$\Lambda$             & $-0.589$ & $-1.378$ & $ 0.453$ & $ 0.781$ & $ 0.473$ & $-0.371$ & $-0.159$ & $-0.007$ & $ 0.298$ & $-0.300$ & $-0.009$ & $ 0.124$ & $ 0.027$ & $-0.002$ & $-0.372$ & $-0.124$ & $-0.024$ \\
$\Lambda\Sigma^0 $    & $ 1.530$ & $ 2.386$ & $ 0.288$ & $-0.706$ & $-0.492$ & $ 0.643$ & $-0.193$ & $-0.002$ & $-0.517$ & $-0.088$ & $ 0.006$ & $-0.215$ & $-0.020$ & $ 0.004$ & $ 0.645$ & $-0.215$ & $ 0.005$ \\
$\Delta^{++}$         & $ 6.140$ & $ 8.521$ & $-0.236$ & $-3.107$ & $-1.073$ & $ 0.029$ & $-0.132$ & $-0.067$ & $-0.041$ & $ 0.006$ & $-0.066$ & $ 0.295$ & $ 0.049$ & $-0.006$ & $ 2.302$ & $-0.383$ & $ 0.051$ \\
$\Delta^+$            & $ 3.058$ & $ 4.260$ & $ 0.779$ & $-1.843$ & $-0.905$ & $ 0.014$ & $-0.243$ & $-0.049$ & $-0.020$ & $-0.255$ & $-0.015$ & $ 0.147$ & $-0.001$ & $-0.005$ & $ 1.151$ & $ 0.000$ & $ 0.042$ \\
$\Delta^0$            & $-0.023$ & $ 0.000$ & $ 1.795$ & $-0.579$ & $-0.736$ & $ 0.000$ & $-0.354$ & $-0.031$ & $ 0.000$ & $-0.516$ & $ 0.036$ & $ 0.000$ & $-0.051$ & $-0.004$ & $ 0.000$ & $ 0.383$ & $ 0.034$ \\
$\Delta^-$            & $-3.105$ & $-4.260$ & $ 2.810$ & $ 0.685$ & $-0.568$ & $-0.014$ & $-0.464$ & $-0.013$ & $ 0.020$ & $-0.777$ & $ 0.087$ & $-0.147$ & $-0.101$ & $-0.003$ & $-1.151$ & $ 0.767$ & $ 0.025$ \\
$\Sigma^{*+}$         & $ 2.520$ & $ 4.260$ & $-0.076$ & $-1.264$ & $-0.168$ & $ 0.014$ & $-0.601$ & $-0.025$ & $-0.020$ & $-0.464$ & $-0.035$ & $ 0.147$ & $-0.025$ & $ 0.001$ & $ 1.151$ & $-0.383$ & $ 0.008$ \\
$\Sigma^{*0}$         & $-0.159$ & $ 0.000$ & $ 0.859$ & $ 0.000$ & $ 0.000$ & $ 0.000$ & $-0.477$ & $-0.017$ & $ 0.000$ & $-0.490$ & $ 0.003$ & $ 0.000$ & $-0.038$ & $ 0.001$ & $ 0.000$ & $ 0.000$ & $ 0.000$ \\
$\Sigma^{*-}$         & $-2.838$ & $-4.260$ & $ 1.795$ & $ 1.264$ & $ 0.168$ & $-0.014$ & $-0.354$ & $-0.009$ & $ 0.020$ & $-0.516$ & $ 0.042$ & $-0.147$ & $-0.051$ & $ 0.000$ & $-1.151$ & $ 0.383$ & $-0.008$ \\
$\Xi^{*0}$            & $-0.117$ & $ 0.000$ & $-0.076$ & $ 0.579$ & $ 0.736$ & $ 0.000$ & $-0.601$ & $ 0.082$ & $ 0.000$ & $-0.464$ & $ 0.082$ & $ 0.000$ & $-0.025$ & $-0.013$ & $ 0.000$ & $-0.383$ & $-0.034$ \\
$\Xi^{*-}$            & $-2.476$ & $-4.260$ & $ 0.779$ & $ 1.843$ & $ 0.905$ & $-0.014$ & $-0.243$ & $ 0.029$ & $ 0.020$ & $-0.255$ & $ 0.057$ & $-0.147$ & $-0.001$ & $ 0.004$ & $-1.151$ & $ 0.000$ & $-0.042$ \\
$\Omega^-$            & $-2.020$ & $-4.260$ & $-0.236$ & $ 2.423$ & $ 1.641$ & $-0.014$ & $-0.132$ & $ 0.100$ & $ 0.020$ & $ 0.006$ & $ 0.132$ & $-0.147$ & $ 0.049$ & $ 0.009$ & $-1.151$ & $-0.383$ & $-0.076$ \\
$\Delta^+p$           & $ 3.510$ & $ 5.335$ & $-0.908$ & $-0.244$ & $-2.874$ & $ 2.952$ & $-0.761$ & $-0.043$ & $-1.272$ & $ 0.735$ & $ 0.043$ & $-0.308$ & $-0.117$ & $-0.001$ & $ 1.441$ & $-0.480$ & $ 0.011$ \\
$\Delta^0n$           & $ 3.510$ & $ 5.335$ & $-0.908$ & $-0.244$ & $-2.874$ & $ 2.952$ & $-0.761$ & $-0.043$ & $-1.272$ & $ 0.735$ & $ 0.043$ & $-0.308$ & $-0.117$ & $-0.001$ & $ 1.441$ & $ 0.011$ & $-0.480$ \\
$\Sigma^{*0}\Lambda$  & $ 2.731$ & $ 4.620$ & $-0.433$ & $-0.296$ & $-2.987$ & $ 2.556$ & $-0.405$ & $ 0.034$ & $-1.102$ & $ 0.255$ & $-0.024$ & $-0.267$ & $-0.065$ & $ 0.002$ & $ 1.248$ & $-0.416$ & $ 0.009$ \\
$\Sigma^{*0}\Sigma^0$ & $ 1.592$ & $ 2.667$ & $ 0.250$ & $-0.328$ & $-2.875$ & $ 1.476$ & $ 0.234$ & $ 0.035$ & $-0.636$ & $-0.147$ & $ 0.020$ & $-0.154$ & $ 0.038$ & $ 0.004$ & $ 0.721$ & $ 0.240$ & $ 0.047$ \\
$\Sigma^{*+}\Sigma^+$ & $ 3.168$ & $ 5.335$ & $ 0.333$ & $-0.597$ & $-5.174$ & $ 2.952$ & $ 0.384$ & $ 0.069$ & $-1.272$ & $-0.120$ & $-0.001$ & $-0.308$ & $ 0.065$ & $ 0.009$ & $ 1.441$ & $ 0.000$ & $ 0.053$ \\
$\Sigma^{*-}\Sigma^-$ & $ 0.016$ & $ 0.000$ & $ 0.167$ & $-0.058$ & $-0.575$ & $ 0.000$ & $ 0.084$ & $ 0.001$ & $ 0.000$ & $-0.175$ & $ 0.041$ & $ 0.000$ & $ 0.010$ & $-0.002$ & $ 0.000$ & $ 0.480$ & $ 0.042$ \\
$\Xi^{*0}\Xi^0$       & $ 2.787$ & $ 5.335$ & $ 0.408$ & $-0.597$ & $-5.174$ & $ 2.952$ & $ 0.293$ & $ 0.065$ & $-1.272$ & $-0.440$ & $-0.015$ & $-0.308$ & $ 0.042$ & $ 0.006$ & $ 1.441$ & $ 0.000$ & $ 0.053$ \\
$\Xi^{*-}\Xi^-$       & $ 0.033$ & $ 0.000$ & $ 0.167$ & $-0.058$ & $-0.575$ & $ 0.000$ & $ 0.084$ & $-0.026$ & $ 0.000$ & $-0.175$ & $ 0.080$ & $ 0.000$ & $ 0.010$ & $ 0.004$ & $ 0.000$ & $ 0.480$ & $ 0.042$ \\
\end{tabular}
\end{ruledtabular}
\end{table*}

\end{turnpage}

\begin{table*}
\caption{\label{t:fitAll}SU(3) flavor contributions to the baryon magnetic moments.}
\begin{ruledtabular}
\begin{tabular}{lrrrrrrrrrrrrr}
& \multicolumn{3}{c}{fit B} & \multicolumn{5}{c}{fit C} & \multicolumn{5}{c}{fit D} \\ \hline
 & Tree & SB & Total & Tree & SB & Loop 1 & Loop 2 & Total & Tree & SB & Loop 1 & Loop 2 & Total \\
\hline
$n$                    & $-1.503$ & $-0.428$ & $-1.931$ & $-2.508$ & $ 0.294$ & $ 1.020$ & $-0.741$ & $-1.936$ & $-2.755$ & $ 0.313$ & $ 0.986$ & $-0.473$ & $-1.929$ \\
$p$                    & $ 2.392$ & $ 0.401$ & $ 2.793$ & $ 3.443$ & $-0.345$ & $-1.407$ & $ 1.102$ & $ 2.793$ & $ 3.073$ & $ 0.527$ & $-1.476$ & $ 0.670$ & $ 2.793$ \\
$\Sigma^-$             & $-0.889$ & $-0.266$ & $-1.155$ & $-0.934$ & $-0.241$ & $-0.142$ & $ 0.164$ & $-1.154$ & $-0.318$ & $ 0.236$ & $-0.082$ & $-0.991$ & $-1.155$ \\
$\Sigma^0$             & $ 0.751$ & $-0.097$ & $ 0.654$ & $ 1.254$ & $ 0.089$ & $-1.359$ & $ 0.671$ & $ 0.655$ & $ 1.378$ & $ 0.120$ & $-1.255$ & $ 0.410$ & $ 0.653$ \\
$\Sigma^+$             & $ 2.392$ & $ 0.071$ & $ 2.463$ & $ 3.443$ & $ 0.418$ & $-2.575$ & $ 1.178$ & $ 2.464$ & $ 3.073$ & $ 0.005$ & $-2.428$ & $ 1.812$ & $ 2.462$ \\
$\Xi^-$                & $-0.889$ & $ 0.238$ & $-0.651$ & $-0.934$ & $ 0.190$ & $ 0.639$ & $-0.545$ & $-0.651$ & $-0.318$ & $ 0.604$ & $ 0.656$ & $-1.593$ & $-0.651$ \\
$\Xi^0$                & $-1.503$ & $ 0.234$ & $-1.269$ & $-2.508$ & $-0.117$ & $ 2.466$ & $-1.114$ & $-1.273$ & $-2.755$ & $-0.073$ & $ 2.345$ & $-0.783$ & $-1.267$ \\
$\Lambda$              & $-0.751$ & $ 0.165$ & $-0.586$ & $-1.254$ & $ 0.090$ & $ 1.359$ & $-0.774$ & $-0.579$ & $-1.378$ & $ 0.453$ & $ 1.255$ & $-0.919$ & $-0.589$ \\
$\Lambda\Sigma^0$      & $ 1.301$ & $ 0.227$ & $ 1.529$ & $ 2.172$ & $ 0.001$ & $-1.228$ & $ 0.580$ & $ 1.526$ & $ 2.386$ & $ 0.288$ & $-1.198$ & $ 0.054$ & $ 1.530$ \\
$\Delta^{++}$          & $ 5.440$ & $ 0.700$ & $ 6.140$ & $ 6.615$ & $ 0.823$ & $-5.329$ & $ 4.031$ & $ 6.140$ & $ 8.521$ & $-0.236$ & $-4.181$ & $ 2.036$ & $ 6.140$ \\
$\Delta^+$             & $ 2.720$ & $ 0.137$ & $ 2.857$ & $ 3.308$ & $ 0.016$ & $-3.161$ & $ 2.090$ & $ 2.252$ & $ 4.260$ & $ 0.779$ & $-2.748$ & $ 0.766$ & $ 3.058$ \\
$\Delta^0$             & $ 0.000$ & $-0.427$ & $-0.427$ & $ 0.000$ & $-0.791$ & $-0.993$ & $ 0.148$ & $-1.636$ & $ 0.000$ & $ 1.795$ & $-1.316$ & $-0.503$ & $-0.023$ \\
$\Delta^-$             & $-2.720$ & $-0.990$ & $-3.710$ & $-3.308$ & $-1.598$ & $ 1.175$ & $-1.793$ & $-5.523$ & $-4.260$ & $ 2.810$ & $ 0.117$ & $-1.772$ & $-3.105$ \\
$\Sigma^{*+}$          & $ 2.720$ & $ 0.630$ & $ 3.350$ & $ 3.308$ & $ 1.327$ & $-2.168$ & $ 1.430$ & $ 3.896$ & $ 4.260$ & $-0.076$ & $-1.433$ & $-0.232$ & $ 2.520$ \\
$\Sigma^{*0}$          & $ 0.000$ & $ 0.102$ & $ 0.102$ & $ 0.000$ & $ 0.268$ & $ 0.000$ & $-0.536$ & $-0.268$ & $ 0.000$ & $ 0.859$ & $ 0.000$ & $-1.018$ & $-0.159$ \\
$\Sigma^{*-}$          & $-2.720$ & $-0.427$ & $-3.147$ & $-3.308$ & $-0.791$ & $ 2.168$ & $-2.502$ & $-4.433$ & $-4.260$ & $ 1.795$ & $ 1.433$ & $-1.805$ & $-2.838$ \\
$\Xi^{*0}$             & $ 0.000$ & $ 0.630$ & $ 0.630$ & $ 0.000$ & $ 1.327$ & $ 0.993$ & $-1.125$ & $ 1.195$ & $ 0.000$ & $-0.076$ & $ 1.316$ & $-1.357$ & $-0.117$ \\
$\Xi^{*-}$             & $-2.720$ & $ 0.137$ & $-2.583$ & $-3.308$ & $ 0.016$ & $ 3.161$ & $-3.135$ & $-3.265$ & $-4.260$ & $ 0.779$ & $ 2.748$ & $-1.744$ & $-2.476$ \\
$\Omega^-$             & $-2.720$ & $ 0.700$ & $-2.020$ & $-3.308$ & $ 0.823$ & $ 4.155$ & $-3.690$ & $-2.020$ & $-4.260$ & $-0.236$ & $ 4.064$ & $-1.587$ & $-2.020$ \\
$\Delta^+p$            & $ 3.495$ & $ 0.015$ & $ 3.510$ & $ 5.654$ & $-2.121$ & $-3.064$ & $ 3.041$ & $ 3.510$ & $ 5.335$ & $-0.908$ & $-3.117$ & $ 2.200$ & $ 3.510$ \\
$\Delta^0n$            & $ 3.495$ & $ 0.015$ & $ 3.510$ & $ 5.654$ & $-2.121$ & $-3.064$ & $ 3.041$ & $ 3.510$ & $ 5.335$ & $-0.908$ & $-3.117$ & $ 2.200$ & $ 3.510$ \\
$\Sigma^{*0}\Lambda$   & $ 3.027$ & $-0.297$ & $ 2.730$ & $ 4.896$ & $-1.585$ & $-3.464$ & $ 2.886$ & $ 2.732$ & $ 4.620$ & $-0.433$ & $-3.283$ & $ 1.827$ & $ 2.731$ \\
$\Sigma^{*0}\Sigma^0$  & $ 1.747$ & $ 0.171$ & $ 1.919$ & $ 2.827$ & $ 0.915$ & $-3.833$ & $ 2.480$ & $ 2.389$ & $ 2.667$ & $ 0.250$ & $-3.202$ & $ 1.877$ & $ 1.592$ \\
$\Sigma^{*+}\Sigma^+$  & $ 3.495$ & $-0.325$ & $ 3.170$ & $ 5.654$ & $ 0.053$ & $-6.770$ & $ 4.229$ & $ 3.166$ & $ 5.335$ & $ 0.333$ & $-5.771$ & $ 3.271$ & $ 3.168$ \\
$\Sigma^{*-}\Sigma^-$  & $ 0.000$ & $ 0.667$ & $ 0.667$ & $ 0.000$ & $ 1.777$ & $-0.896$ & $ 0.731$ & $ 1.611$ & $ 0.000$ & $ 0.167$ & $-0.633$ & $ 0.482$ & $ 0.016$ \\
$\Xi^{*0}\Xi^0$        & $ 3.495$ & $-0.358$ & $ 3.137$ & $ 5.654$ & $ 0.291$ & $-6.770$ & $ 4.358$ & $ 3.533$ & $ 5.335$ & $ 0.408$ & $-5.771$ & $ 2.816$ & $ 2.787$ \\
$\Xi^{*-}\Xi^-$        & $ 0.000$ & $ 0.667$ & $ 0.667$ & $ 0.000$ & $ 1.777$ & $-0.896$ & $ 0.687$ & $ 1.568$ & $ 0.000$ & $ 0.167$ & $-0.633$ & $ 0.499$ & $ 0.033$ \\
\end{tabular}
\end{ruledtabular}
\end{table*}

We can summarize our findings displayed in Tables \ref{t:fitc} and \ref{t:fitd}, into the combined Table \ref{t:fitAll} by adding up the different flavor contributions from the loops. In all these cases, SB represents an important contribution to the total value. Besides, although individual contributions from Figs.~\ref{fig:mmloop1} and \ref{fig:mmloop2} might be large compared to the tree-level values, in general there are numerical cancellations between these two contributions, so the one-loop net result is consistent with being a quantum correction. This observation is not apparent in the previous Tables \ref{t:fitc} and \ref{t:fitd}.

\begin{figure}[ht]
\scalebox{0.8}{\includegraphics{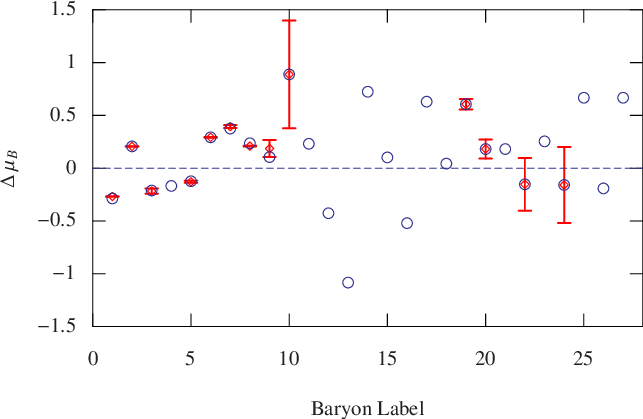}}
\caption{\label{fig:plotB}Deviation of baryon magnetic moments (in units of $\mu_N$) relative to the SU(3) symmetric fit. The open circles are from fit B. The open diamonds are from the experimental values. The baryon labels are indicated in the horizontal axis, cf.~Table \ref{t:fits}.}
\end{figure}

\begin{figure}[ht]
\scalebox{0.8}{\includegraphics{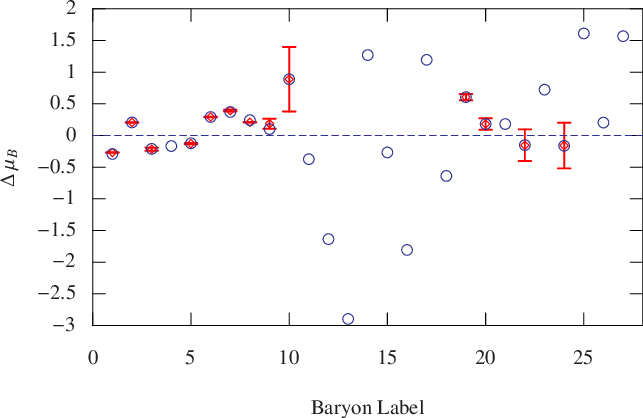}}
\caption{\label{fig:plotC}Deviation of baryon magnetic moments (in units of $\mu_N$) relative to the SU(3) symmetric fit. The open circles are from fit C. The open diamonds are from the experimental values. The baryon labels are indicated in the horizontal axis, cf.~Table \ref{t:fits}.}
\end{figure}

\begin{figure}[ht]
\scalebox{0.8}{\includegraphics{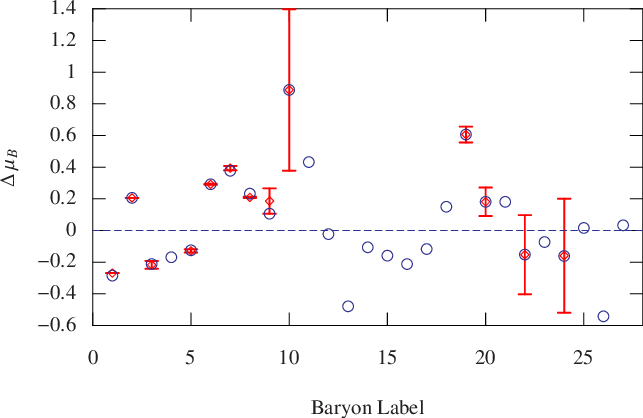}}
\caption{\label{fig:plotD}Deviation of baryon magnetic moments (in units of $\mu_N$) relative to the SU(3) symmetric fit. The open circles are from fit D. The open diamonds are from the experimental values. The baryon labels are indicated in the horizontal axis, cf.~Table \ref{t:fits}.}
\end{figure}

Some other interesting features extracted from the fits can be better seen by plotting the deviations $\Delta \mu_B=\mu_B^{\textrm{fit X}}-\mu_B^{\textrm{SU(3)}}$, where $\mu_B^{\textrm{fit X}}$ is the magnetic moment of baryon $B$ predicted by fit X (X = B, C, D) and $\mu_B^{\textrm{SU(3)}}$ is the magnetic moment given by the SU(3) symmetric fit, namely fit A. We plot $\Delta \mu_B$ in Figs.~\ref{fig:plotB}, \ref{fig:plotC}, and \ref{fig:plotD} for X = B, C, and D, respectively; in these graphs we also plot $\Delta\mu_B^{\textrm{exp}}$, defined on the same footing as its theoretical counterpart. On general grounds, large deviations from the SU(3) estimates are found in the decuplet baryons and in particular $\mu_{\Delta^{++}}$ and $\mu_{\Delta^-}$ exhibit the largest ones in Fig.~\ref{fig:plotB}. One-loop corrections in the degenerate limit plus SB, Fig.~\ref{fig:plotC}, do not improve the situation, but even worsen it for $\mu_{\Delta^-}$, $\mu_{\Sigma^{*0}}$, $\mu_{\Sigma^{*-}\Sigma^-}$, and $\mu_{\Xi^{*-}\Xi^-}$. The inclusion of $\Delta$ seems to correct the situation for the whole sector, as can be seen in Fig.~\ref{fig:plotD}. If we plotted the relative deviation\footnote{Although plotting the relative deviation could be more enlightening, important pieces of information would be lost for those magnetic moments which are zero at the SU(3) symmetry limit. We prefer to plot the absolute values of the deviations instead.} $\Delta \mu_B/\mu_B^{\textrm{SU(3)}}$, we would realize that corrections to the SU(3) symmetric case fall in the $\pm 40\%$, $\pm 60\%$ (except for $\mu_{\Delta^0}$ which acquires a sizable correction greater than 100\%), and $\pm30\%$ ranges for fits B, C, and D, respectively. Definitely, one-loop corrections of orders $\mathcal{O}(m_q^{1/2})$ and $\mathcal{O}(m_q\ln m_q)$ with a nonvanishing $\Delta$ taken into account simultaneously with SB yield SU(3) breaking corrections consistent with expectations: we naively assume that order $\mathcal{O}(m_s)$ and order $\mathcal{O}(1/N_c)$ corrections are both order $\mathcal{O}(\epsilon)\sim 30\%$.

In Ref.~\cite{rfm09} one-loop corrections in the degenerate case were analyzed without the inclusion of SB corrections. It was found that the fit was quite unstable in the sense that slight departures from the initial values of the parameters would yield rather different results. In the present analysis, a stable fit is obtained by adding SB corrections. Even better, taking also into account a nonvanishing $\Delta$ provides the fit with stability and robustness hardly attainable otherwise.

To close this section, we can numerically compare our results with others in the literature. This comparison is displayed in the last five columns (from left to right) in Table \ref{t:fits}. For instance, Refs.~\cite{ms97,geng2} compute one-loop corrections in baryon chiral perturbation theory to orders $p^4$ and $p^3$, respectively. Except for the lowest-order terms, the analytical comparison is not possible term by term, so we content ourselves with performing a numerical comparison instead. In this respect, the numerical findings of Ref.~\cite{ms97} are in remarkable agreement to our fit D (which includes terms of order $p^4$ and contributions from a nonvanishing $\Delta$) for all the octet baryons. Similarly, Refs.~\cite{lb} and \cite{jen12} perform their analyses in the context of the $1/N_c$ expansion. Their results are comparable to fit B in our case. Finally, Ref.~\cite{keller} provides calculations of some decuplet-octet transition magnetic moments from the CLAS experimental results. If we compare our predictions from fit D with these ones, the agreement is very good. Unfortunately, we cannot compare the complete set of predictions of our fit D because there are no other analyses available in the literature computed under the same order of approximation.

\section{\label{sec:final}Concluding remarks}

In this paper we evaluated the magnetic moments of baryons within large-$N_c$ chiral perturbation theory, including one-loop corrections of orders $\mathcal{O}(m_q^{1/2})$ and $\mathcal{O}(m_q\ln m_q)$ by following the lines of Ref.~\cite{rfm09}. The present analysis complements the previous one in the sense that here we considered the effects of a nonvanishing baryon decuplet-octet mass difference $\Delta$ and also the effects of SB corrections. In the large-$N_c$ limit, $\Delta \propto 1/N_c$ so the degeneracy case constitutes a very good first approximation. However, a more realistic situation should consider $\Delta \neq 0$.

In a complete parallelism to Ref.~\cite{rfm09}, we constructed the baryon operator that describes the order $\mathcal{O}(m_q^{1/2})$ correction to baryon magnetic moments. This correction arises from the Feynman diagrams depicted in Fig.~\ref{fig:mmloop1}. The explicit dependence on $\Delta$ is contained in the definition of the baryon propagator (\ref{eq:barprop}). After a long, tedious, but otherwise standard calculation, we obtained the spin-dependent terms, Eqs.~(\ref{eq:loop1Ared}) and (\ref{eq:loop1Bred}), which have to be combined with the spin-independent ones already computed in Ref.~\cite{rfm09}. Expressions like Eq.~(\ref{eq:msig}) are thus obtained for $\mu_B^{\textrm{(loop 1)}}$ for all 27 possible magnetic moments.

On the other hand, corrections of order $\mathcal{O}(m_q\ln m_q)$ were computed following the lines of Ref.~\cite{rfm12}, where corrections to the baryon axial vector current within large-$N_c$ chiral perturbation theory were presented. We took advantage of the fact that the baryon axial vector current and the baryon magnetic moment operators share the same kinematical properties in the large-$N_c$ limit; one might think that the only change is to replace the $A^{kc}$ operator by the $M^{k}$ operator in the corresponding expressions for the one-loop corrections. However, the matter was not quite that simple. We had to take a few steps back in order to recalculate some operator structures, taking as a starting point the operator structures already presented in Ref.~\cite{rfm12}.

The final analytical expressions were compared with the experimental data \cite{part} through a least-squares fit and also cross-checked with other calculations within the $1/N_c$ expansion \cite{lb,jen12} and chiral perturbation theory \cite{jen92,geng2}. Although the fit is good and seemingly stable, somehow we still consider it rather unsatisfactory from a theoretical point of view. In particular, we cannot explain why the parameters $m_4$ and $n_4^{1,\mathbf{8}}$ are rather large. On the other hand, we should stress that for octet baryons the comparison between analytical expressions was possible, whereas for the other cases, it was performed through numerical estimates. The overall comparison has been a successful one.

A very clear result emerges from the present analysis: in order to have a complete understanding of SU(3) flavor symmetry breaking in the magnetic moments of baryons in the context of baryon chiral perturbation theory in the large-$N_c$, one-loop corrections of orders $\mathcal{O}(m_q^{1/2})$ and $\mathcal{O}(m_q\ln m_q)$, together with SB, must be taken into account.

\acknowledgments

The authors wish to express their gratitude to A.~V.~Manohar and J.~M.~Camalich for enlightening correspondence and to J.~Goity for useful discussions. This work has been partially supported by Consejo Nacional de Ciencia y Tecnolog{\'\i}a and Fondo de Apoyo a la Investigaci\'on (Universidad Aut\'onoma de San Luis Potos{\'\i}), Mexico.

\appendix

\section{\label{app:reduc1}Reduction of baryon operators---Structures from Fig.~\ref{fig:mmloop1}}

In this section we present the reduction of the spin-dependent operator structures contained in Eqs.~(\ref{eq:m8l1}) and (\ref{eq:m10l1}). The spin-independent contributions can be found in Appendix A of Ref.~\cite{rfm09}. The analysis, although long and tedious, is otherwise straightforward. We find:

(1) flavor $\mathbf{8}$ representation

\begin{widetext}

\begin{equation}
\epsilon^{ijk} f^{abc} G^{ia}J^2G^{jb} = - \frac12 (N_c+N_f) G^{kc} + \frac12 (N_f+1) \mathcal{D}_2^{kc} - \frac18 (N_c+N_f) \mathcal{D}_3^{kc} - \frac14 (N_c+N_f) \mathcal{O}_3^{kc} + \frac14 \mathcal{D}_4^{kc},
\end{equation}

\begin{equation}
\epsilon^{ijk} f^{abc} ( G^{ia}J^2\mathcal{D}_2^{jb} + \mathcal{D}_2^{ia}J^2G^{jb} ) = -\frac14 N_f \mathcal{D}_3^{kc} - (N_f+1) \mathcal{O}_3^{kc} - \frac12 \mathcal{O}_5^{kc},
\end{equation}

\begin{equation}
\epsilon^{ijk} f^{abc} \mathcal{D}_2^{ia}J^2\mathcal{D}_2^{jb} = -\frac14 N_f \mathcal{D}_4^{kc},
\end{equation}

\begin{equation}
\epsilon^{ijk} f^{abc} ( G^{ia}J^2\mathcal{D}_3^{jb} + \mathcal{D}_3^{ia}J^2G^{jb} ) = -\frac12 (N_c+N_f) \mathcal{D}_3^{kc} - 3 (N_c+N_f) \mathcal{O}_3^{kc} - \frac12 (N_f-2) \mathcal{D}_4^{kc} - \frac12 (N_c+N_f) \mathcal{O}_5^{kc},
\end{equation}

\begin{eqnarray}
\epsilon^{ijk} f^{abc} ( G^{ia}J^2\mathcal{O}_3^{jb} + \mathcal{O}_3^{ia}J^2G^{jb} ) & = & 3 N_f \mathcal{D}_2^{kc} - \frac32 (N_c+N_f) \mathcal{D}_3^{kc} - \frac12 (N_c+N_f) \mathcal{O}_3^{kc} + \frac14 (7 N_f+12) \mathcal{D}_4^{kc} \nonumber \\
&  & \mbox{} - \frac14 (N_c+N_f) \mathcal{D}_5^{kc} - \frac14 (N_c+N_f) \mathcal{O}_5^{kc} + \frac12 \mathcal{D}_6^{kc},
\end{eqnarray}

\begin{equation}
\epsilon^{ijk} f^{abc} (\mathcal{D}_2^{ia}J^2\mathcal{D}_3^{jb} + \mathcal{D}_3^{ia}J^2\mathcal{D}_2^{jb} ) = -\frac12 N_f \mathcal{D}_5^{kc},
\end{equation}

\begin{equation}
\epsilon^{ijk} f^{abc} (\mathcal{D}_2^{ia}J^2\mathcal{O}_3^{jb} + \mathcal{O}_3^{ia}J^2\mathcal{D}_2^{jb} ) = -(N_f+1) \mathcal{O}_5^{kc} - \frac12 \mathcal{O}_7^{kc}.
\end{equation}

(2) flavor $\mathbf{10}+\overline{\mathbf{10}}$ representation

\begin{eqnarray}
&  & \epsilon^{ijk} ( f^{aec}d^{be8} - f^{bec}d^{ae8} - f^{abe}d^{ec8} ) G^{ia} J^2 G^{jb} \nonumber \\
&  & \mbox{\hglue0.4truecm} = - \frac12 \{G^{kc},T^8\} + \frac12 \{G^{k8},T^c\} + \frac{1}{N_f} [J^2,[T^8,G^{kc}]] \nonumber \\
&  & \mbox{\hglue0.8truecm} - \frac{1}{4}\{J^2,\{G^{kc},T^8\}\} + \frac{1}{4}\{J^2,\{G^{k8},T^c\}\} + \frac{1}{2N_f} \{J^2,[J^2,[T^8,G^{kc}]]\},
\end{eqnarray}

\begin{eqnarray}
&  & \epsilon^{ijk} ( f^{aec}d^{be8} - f^{bec}d^{ae8} - f^{abe}d^{ec8} )( G^{ia} J^2 \mathcal{D}_2^{jb} + \mathcal{D}_2^{ia} J^2 G^{jb} ) \nonumber \\
&  & \mbox{\hglue0.4truecm} = \frac{N_c+N_f}{N_f} [J^2,[T^8,G^{kc}]] - \{G^{kc},\{J^r,G^{r8}\}\} + \{G^{k8},\{J^r,G^{rc}\}\} + \frac{N_c+N_f}{2N_f} \{J^2,[J^2,[T^8,G^{kc}]]\} \nonumber \\
&  & \mbox{\hglue0.4truecm} - \frac12 \{J^2,\{G^{kc},\{J^r,G^{r8}\}\}\} + \frac12 \{J^2,\{G^{k8},\{J^r,G^{rc}\}\}\},
\end{eqnarray}

\begin{equation}
\epsilon^{ijk} ( f^{aec}d^{be8} - f^{bec}d^{ae8} - f^{abe}d^{ec8} ) \mathcal{D}_2^{ia} J^2 \mathcal{D}_2^{jb} = 0,
\end{equation}

\begin{eqnarray}
&  & \epsilon^{ijk} ( f^{aec}d^{be8} - f^{bec}d^{ae8} - f^{abe}d^{ec8} ) ( G^{ia} J^2 \mathcal{D}_3^{jb} + \mathcal{D}_3^{ia} J^2 G^{jb} ) \nonumber \\
&  & \mbox{\hglue0.4truecm} = - 3 \{J^2,\{G^{kc},T^8\}\} + 3 \{J^2,\{G^{k8},T^c\}\} - 2 \{\mathcal{D}_2^{kc},\{J^r,G^{r8}\}\} + 2 \{\mathcal{D}_2^{k8},\{J^r,G^{rc}\}\} + \frac{6}{N_f} \{J^2,[J^2,[T^8,G^{kc}]]\} \nonumber \\
&  & \mbox{\hglue0.8truecm} - \frac12 \{J^2,\{J^2,\{G^{kc},T^8\}\}\} + \frac12 \{J^2,\{J^2,\{G^{k8},T^c\}\}\} - \frac12 \{J^2,\{\mathcal{D}_2^{kc},\{J^r,G^{r8}\}\}\} + \frac12 \{J^2,\{\mathcal{D}_2^{k8},\{J^r,G^{rc}\}\}\} \nonumber \\
&  & \mbox{\hglue0.8truecm} + \frac{1}{N_f} \{J^2,\{J^2,[J^2,[T^8,G^{kc}]]\}\}, \nonumber \\
\end{eqnarray}

\begin{eqnarray}
&  & \epsilon^{ijk} ( f^{aec}d^{be8} - f^{bec}d^{ae8} - f^{abe}d^{ec8} ) ( G^{ia} J^2 \mathcal{O}_3^{jb} + \mathcal{O}_3^{ia} J^2 G^{jb} ) \nonumber \\
&  & \mbox{\hglue0.4truecm}  = - \frac12 \{J^2,\{G^{kc},T^8\}\} + \frac12 \{J^2,\{G^{k8},T^c\}\} + \frac52 \{\mathcal{D}_2^{kc},\{J^r,G^{r8}\}\} - \frac52 \{\mathcal{D}_2^{k8},\{J^r,G^{rc}\}\} + \frac{1}{N_f} \{J^2,[J^2,[T^8,G^{kc}]]\} \nonumber \\
&  & \mbox{\hglue0.8truecm} - \frac14 \{J^2,\{J^2,\{G^{kc},T^8\}\}\} + \frac14 \{J^2,\{J^2,\{G^{k8},T^c\}\}\} + \frac14 \{J^2,\{\mathcal{D}_2^{kc},\{J^r,G^{r8}\}\}\} - \frac14 \{J^2,\{\mathcal{D}_2^{k8},\{J^r,G^{rc}\}\}\} \nonumber \\
&  & \mbox{\hglue0.8truecm} + \frac{1}{2N_f} \{J^2,\{J^2,[J^2,[T^8,G^{kc}]]\}\},
\end{eqnarray}

\begin{equation}
\epsilon^{ijk} ( f^{aec}d^{be8} - f^{bec}d^{ae8} - f^{abe}d^{ec8} ) (\mathcal{D}_2^{ia} J^2 \mathcal{D}_3^{jb} + \mathcal{D}_3^{ia} J^2 \mathcal{D}_2^{jb} ) = 0,
\end{equation}

\begin{eqnarray}
&  & \epsilon^{ijk} ( f^{aec}d^{be8} - f^{bec}d^{ae8} - f^{abe}d^{ec8} ) (\mathcal{D}_2^{ia} J^2 \mathcal{O}_3^{jb} + \mathcal{O}_3^{ia} J^2 \mathcal{D}_2^{jb} ) \nonumber \\
&  & \mbox{\hglue0.4truecm} = \frac{N_c+N_f}{N_f} \{J^2,[J^2,[T^8,G^{kc}]]\} + \frac{N_c+N_f}{2N_f} \{J^2,\{J^2,[J^2,[T^8,G^{kc}]]\}\} - \{J^2,\{G^{kc},\{J^r,G^{r8}\}\}\} \nonumber \\
&  & \mbox{\hglue0.8truecm} + \{J^2,\{G^{k8},\{J^r,G^{rc}\}\}\} - \frac12 \{J^2,\{J^2,\{G^{kc},\{J^r,G^{r8}\}\}\}\} + \frac12 \{J^2,\{J^2,\{G^{k8},\{J^r,G^{rc}\}\}\}\}.
\end{eqnarray}

\section{\label{app:reduc2}Reduction of baryon operators---Structures from Fig.~\ref{fig:mmloop2}}

The evaluation of the commutator-anticommutator structure
\begin{eqnarray}
\left\{ A^{ja}, \left[A^{kc}, \left[ \mathcal{M}, A^{jb} \right] \right] \right\}, \nonumber
\end{eqnarray}
which represents the leading contribution to the renormalized baryon axial vector current for finite decuplet-octet mass difference, has been computed in Ref.~\cite{rfm12}. To use those results for baryon magnetic moments, we need to replace the $A^{kc}$ operator with $M^{kc}$. This indeed introduces some changes in the original expressions, so that only partial results can be used in the present case. Those which require computation are the following:

(1) Flavor $\mathbf{1}$ contribution

\begin{equation}
\{G^{ia},[\mathcal{D}_2^{kc},[J^2,G^{ia}]]\} = (N_c+N_f) G^{kc} + \frac12 [N_c(N_c+2N_f)-7 N_f] \mathcal{D}_2^{kc} + \frac12 (N_c+N_f) \mathcal{D}_3^{kc} - 2 \mathcal{D}_4^{kc},
\end{equation}

\begin{equation}
\{\mathcal{D}_2^{ia},[G^{kc},[J^2,G^{ia}]]\} = (N_c+N_f) G^{kc} - (N_f+1) \mathcal{D}_2^{kc},
\end{equation}

\begin{eqnarray}
\{G^{ia},[\mathcal{D}_3^{kc},[J^2,G^{ia}]]\} & = & [N_c(N_c+2N_f)+2N_f] G^{kc} + 11(N_c+N_f) \mathcal{D}_2^{kc} + \frac12[2N_c(N_c+2N_f)-15N_f-2] \mathcal{D}_3^{kc} \nonumber \\
&  & \mbox{} - (N_f-2) \mathcal{O}_3^{kc} + (N_c+N_f) \mathcal{D}_4^{kc} - 3 \mathcal{D}_5^{kc},
\end{eqnarray}

\begin{equation}
\{\mathcal{D}_3^{ia},[G^{kc},[J^2,G^{ia}]]\} = [N_c(N_c+2N_f)+2N_f] G^{kc} - (N_c+N_f) \mathcal{D}_2^{kc} - N_f \mathcal{D}_3^{kc} - (N_f-2) \mathcal{O}_3^{kc},
\end{equation}

\begin{equation}
\{G^{ia},[\mathcal{O}_3^{kc},[J^2,G^{ia}]]\} = \frac32 N_c(N_c+2 N_f) G^{kc} - \frac32 (N_c+N_f) \mathcal{D}_2^{kc} - \frac12 N_f \mathcal{D}_3^{kc}
 + [N_c(N_c+2N_f)-8N_f-3] \mathcal{O}_3^{kc} - 3 \mathcal{O}_5^{kc},
\end{equation}

\begin{equation}
\{G^{ia},[G^{kc},[J^2,\mathcal{O}_3^{ia}]]\} = \frac32 (N_c+N_f) \mathcal{D}_2^{kc} - \frac34 N_f \mathcal{D}_3^{kc} - \frac12(N_f-2) \mathcal{O}_3^{kc} + \frac12 (N_c+N_f) \mathcal{D}_4^{kc} - \frac12 \mathcal{D}_5^{kc} - \mathcal{O}_5^{kc},
\end{equation}

\begin{eqnarray}
\{\mathcal{O}_3^{ia},[G^{kc},[J^2,G^{ia}]]\} & = & -\frac32 N_c (N_c+2 N_f) G^{kc} + 3 (N_c+N_f) \mathcal{D}_2^{kc} - \frac14 N_f \mathcal{D}_3^{kc} + (2N_f+3) \mathcal{O}_3^{kc} \nonumber \\
&  & \mbox{} + \frac12 (N_c+N_f) \mathcal{D}_4^{kc} - \frac12 \mathcal{D}_5^{kc} - \mathcal{O}_5^{kc},
\end{eqnarray}

(2) Flavor $\mathbf{8}$ contribution

\begin{eqnarray}
d^{ab8} \{G^{ia},[\mathcal{D}_2^{kc},[J^2,G^{ib}]]\} & = & \frac12 (N_c+N_f) d^{c8e} G^{ke} - \frac14 (5N_f+2) d^{c8e} \mathcal{D}_2^{ke} - \frac{1}{N_f} [J^2,[T^8,G^{kc}]] + \frac12 \{G^{kc},T^8\} \nonumber \\
&  & \mbox{} - \frac12 (N_f-1) \{G^{k8},T^c\} + \frac14 (N_c+N_f) d^{c8e} \mathcal{D}_3^{ke} + \frac14 (N_c+N_f) \{J^k,\{T^c,T^8\}\} \nonumber \\
&  & \mbox{} - \frac12 d^{c8e} \mathcal{D}_4^{ke} - \frac{N_f+1}{N_f} \{\mathcal{D}_2^{kc},\{J^r,G^{r8}\}\} + \frac12 \{\mathcal{D}_2^{k8},\{J^r,G^{rc}\}\} \nonumber \\
&  & \mbox{} - \frac{N_f-2}{2N_f} \{J^2,\{G^{k8},T^c\}\},
\end{eqnarray}

\begin{eqnarray}
d^{ab8} \{\mathcal{D}_2^{ia},[G^{kc},[J^2,G^{ib}]]\} & = & \frac12 (N_c+N_f) d^{c8e} G^{ke} - \frac12 (N_f+1) d^{c8e} \mathcal{D}_2^{ke} + \frac12 \{G^{kc},T^8\} - \frac12 \{G^{k8},T^c\} \nonumber \\
&  & \mbox{} - \frac14 N_f [J^2,[T^8,G^{kc}]] - \frac{(N_c+N_f)(N_f-2)}{2N_f} \{G^{k8},\{J^r,G^{rc}\}\} \nonumber \\
&  & \mbox{} + \frac{(N_c+N_f)(N_f-2)}{2N_f} \{J^k,\{G^{rc},G^{r8}\}\},
\end{eqnarray}

\begin{eqnarray}
&  & d^{ab8} \{G^{ia},[\mathcal{D}_3^{kc},[J^2,G^{ib}]]\} \nonumber \\
&  & \mbox{\hglue0.2truecm} = N_f d^{c8e} G^{ke} + \frac{11N_c(N_c+2N_f)}{2N_f} \delta^{c8} J^k + \frac{11}{2} (N_c+N_f) d^{c8e} \mathcal{D}_2^{ke} \nonumber \\
&  & \mbox{\hglue0.6truecm} + (N_c+N_f) \{G^{kc},T^8\} - \frac12 (N_c+N_f) [J^2,[T^8,G^{kc}]] - \frac14 (5N_f-2) d^{c8e} \mathcal{D}_3^{ke} + \frac{N_f+2}{N_f} d^{c8e} \mathcal{O}_3^{ke} \nonumber \\
&  & \mbox{\hglue0.6truecm} - \frac{N_f^2-2N_f+4}{2N_f} \{G^{kc},\{J^r,G^{r8}\}\} - \frac{3N_f^2-2N_f-4}{2N_f} \{G^{k8},\{J^r,G^{rc}\}\} + \frac{11}{4} \{J^k,\{T^c,T^8\}\} \nonumber \\
&  & \mbox{\hglue0.6truecm} - (3N_f+5) \{J^k,\{G^{rc},G^{r8}\}\} + \frac{N_c(N_c+2N_f)-12N_f+2}{2N_f} \delta^{c8} \{J^2,J^k\} + \frac12 (N_c+N_f) d^{c8e} \mathcal{D}_4^{ke} \nonumber \\
&  & \mbox{\hglue0.6truecm} + 2 (N_c+N_f) \{\mathcal{D}_2^{k8},\{J^r,G^{rc}\}\} - \frac12 d^{c8e} \mathcal{D}_5^{ke} - \frac{N_f-2}{N_f} \{J^2,\{G^{k8},\{J^r,G^{rc}\}\}\} + \frac14 \{J^2,\{J^k,\{T^c,T^8\}\}\} \nonumber \\
&  & \mbox{\hglue0.6truecm} - \{J^2,\{J^k,\{G^{rc},G^{r8}\}\}\} - \frac{3N_f+2}{2N_f} \{J^k,\{\{J^r,G^{rc}\},\{J^m,G^{m8}\}\}\} - \frac{1}{N_f} \delta^{c8} \{J^2,\{J^2,J^k\}\},
\end{eqnarray}

\begin{eqnarray}
&  & d^{ab8} \{\mathcal{D}_3^{ia},[G^{kc},[J^2,G^{ib}]]\} \nonumber \\
&  & \mbox{\hglue0.2truecm} = N_f d^{c8e} G^{ke} - \frac{N_c(N_c+2N_f)}{2N_f} \delta^{c8} J^k - \frac12 (N_c+N_f) d^{c8e} \mathcal{D}_2^{ke} + (N_c+N_f) \{G^{kc},T^8\} \nonumber \\
&  & \mbox{\hglue0.6truecm} - \frac12 (N_c+N_f) [J^2,[T^8,G^{kc}]] - \frac{N_f^2-2N_f+4}{2N_f} \{G^{kc},\{J^r,G^{r8}\}\} - \frac{3N_f^2-2N_f-4}{2N_f} \{G^{k8},\{J^r,G^{rc}\}\} \nonumber \\
&  & \mbox{\hglue0.6truecm} - \frac12 d^{c8e} \mathcal{D}_3^{ke} + \frac{N_f+2}{N_f} d^{c8e} \mathcal{O}_3^{ke} - \frac14 \{J^k,\{T^c,T^8\}\} + (N_f-1) \{J^k,\{G^{rc},G^{r8}\}\} + \frac{N_f-1}{N_f} \delta^{c8} \{J^2,J^k\} \nonumber \\
&  & \mbox{\hglue0.6truecm} - \frac{N_f-2}{N_f} \{J^2,\{G^{k8},\{J^r,G^{rc}\}\}\} + \frac{N_f-2}{N_f} \{J^2,\{J^k,\{G^{rc},G^{r8}\}\}\},
\end{eqnarray}

\begin{eqnarray}
&  & d^{ab8} \{G^{ia},[\mathcal{O}_3^{kc},[J^2,G^{ib}]]\} \nonumber \\
&  & \mbox{\hglue0.2truecm} = - \frac{3 N_c (N_c+2 N_f)}{4N_f} \delta^{c8} J^k - \frac34 (N_c+N_f) d^{c8e} \mathcal{D}_2^{ke} + \frac32 (N_c+N_f) \{G^{kc},T^8\} \nonumber \\
&  & \mbox{\hglue0.6truecm} + 2 (N_c+N_f) [J^2,[T^8,G^{kc}]] - \frac14 (11N_f+6) \{G^{kc},\{J^r,G^{r8}\}\} + \frac14 (5N_f+6) \{G^{k8},\{J^r,G^{rc}\}\} \nonumber \\
&  & \mbox{\hglue0.6truecm} - \frac38 \{J^k,\{T^c,T^8\}\} + \frac{2N_f^2+N_f-4}{2N_f} \{J^k,\{G^{rc},G^{r8}\}\} - \frac{N_f-4}{4N_f} d^{c8e} \mathcal{D}_3^{ke} - \frac14 (5N_f+6) d^{c8e} \mathcal{O}_3^{ke} \nonumber \\
&  & \mbox{\hglue0.6truecm} - (N_c+N_f) \{\mathcal{D}_2^{k8},\{J^r,G^{rc}\}\} + (N_c+N_f) \{J^2,\{G^{kc},T^8\}\} + \frac14 (N_c+N_f) \{J^2,[J^2,[T^8,G^{kc}]]\} \nonumber \\
&  & \mbox{\hglue0.6truecm} - \frac12 d^{c8e} \mathcal{O}_5^{ke} - \frac52 \{J^2,\{G^{kc},\{J^r,G^{r8}\}\}\} + \frac12 \{J^2,\{G^{k8},\{J^r,G^{rc}\}\}\} + \frac{N_f-2}{2N_f} \{J^2,\{J^k,\{G^{rc},G^{r8}\}\}\} \nonumber \\
&  & \mbox{\hglue0.6truecm} + \frac{N_f^2-N_f+4}{2N_f^2} \delta^{c8} \{J^2,J^k\} + \frac{3N_f+2}{4N_f} \{J^k,\{\{J^r,G^{rc}\},\{J^m,G^{m8}\}\}\},
\end{eqnarray}

\begin{eqnarray}
&  & d^{ab8} \{G^{ia},[G^{kc},[J^2,\mathcal{O}_3^{ib}]]\} \nonumber \\
&  & \mbox{\hglue0.2truecm} = \frac{3N_c(N_c+2 N_f)}{4N_f} \delta^{c8} J^k + \frac34 (N_c+N_f) d^{c8e} \mathcal{D}_2^{ke} - \frac{3N_f^2-2N_f-8}{8N_f} d^{c8e} \mathcal{D}_3^{ke} - \frac{N_f-4}{4} d^{c8e} \mathcal{O}_3^{ke} \nonumber \\
&  & \mbox{\hglue0.6truecm} + \frac38 \{J^k,\{T^c,T^8\}\} - \frac{N_f+4}{2N_f} \{J^k,\{G^{rc},G^{r8}\}\} + \frac{N_cN_f(N_c+2N_f)-6N_f^2+2N_f+8}{4N_f^2} \delta^{c8} \{J^2,J^k\} \nonumber \\
&  & \mbox{\hglue0.6truecm} + \frac14 (N_c+N_f) d^{c8e} \mathcal{D}_4^{ke} + \frac14 (N_c+N_f) \{J^2,[J^2,[T^8,G^{kc}]]\} - \frac14 d^{c8e} \mathcal{D}_5^{ke} - \frac12 d^{c8e} \mathcal{O}_5^{ke} - \frac12 \{J^2,\{G^{kc},\{J^r,G^{r8}\}\}\} \nonumber \\
&  & \mbox{\hglue0.6truecm} + \frac{1}{N_f} \{J^2,\{G^{k8},\{J^r,G^{rc}\}\}\} + \frac18 \{J^2,\{J^k,\{T^c,T^8\}\}\} - \frac{1}{N_f} \{J^2,\{J^k,\{G^{rc},G^{r8}\}\}\} - \frac{1}{2N_f} \delta^{c8} \{J^2,\{J^2,J^k\}\}, \nonumber \\
\end{eqnarray}

\begin{eqnarray}
&  & d^{ab8} \{\mathcal{O}_3^{ia},[G^{kc},[J^2,G^{ib}]]\} \nonumber \\
&  & \mbox{\hglue0.2truecm} = \frac{3N_c(N_c+2N_f)}{2N_f} \delta^{c8} J^k + \frac32 (N_c+N_f) d^{c8e} \mathcal{D}_2^{ke} - \frac32 (N_c+N_f) \{G^{kc},T^8\} \nonumber \\
&  & \mbox{\hglue0.6truecm} - \frac12 (N_c+N_f) [J^2,[T^8,G^{kc}]] - \frac18 (3N_f-4) d^{c8e} \mathcal{D}_3^{ke} - \frac14 (N_f-6) d^{c8e} \mathcal{O}_3^{ke} + \frac14 (5N_f+6) \{G^{kc},\{J^r,G^{r8}\}\} \nonumber \\
&  & \mbox{\hglue0.6truecm} + \frac14 (N_f-6) \{G^{k8},\{J^r,G^{rc}\}\} + \frac34 \{J^k,\{T^c,T^8\}\} - (N_f+1) \{J^k,\{G^{rc},G^{r8}\}\} \nonumber \\
&  & \mbox{\hglue0.6truecm} + \frac{N_c(N_c+2N_f)-8N_f+4}{4N_f} \delta^{c8} \{J^2,J^k\} + \frac14 (N_c+N_f) d^{c8e} \mathcal{D}_4^{ke} + \frac14 (N_c+N_f) \{J^2,[J^2,[T^8,G^{kc}]]\} \nonumber \\
&  & \mbox{\hglue0.6truecm} - \frac14 d^{c8e} \mathcal{D}_5^{ke} - \frac12 d^{c8e} \mathcal{O}_5^{ke} - \frac12 \{J^2,\{G^{kc},\{J^r,G^{r8}\}\}\} + \frac12 \{J^2,\{G^{k8},\{J^r,G^{rc}\}\}\} + \frac18 \{J^2,\{J^k,\{T^c,T^8\}\}\} \nonumber \\
&  & \mbox{\hglue0.6truecm} - \frac12 \{J^2,\{J^k,\{G^{rc},G^{r8}\}\}\} - \frac{1}{2N_f} \delta^{c8} \{J^2,\{J^2,J^k\}\};
\end{eqnarray}

(3) flavor $\mathbf{27}$ contribution

\begin{eqnarray}
\{G^{i8},[\mathcal{D}_2^{kc},[J^2,G^{i8}]]\} & = & - 3 f^{c8e} f^{8eg} \mathcal{D}_2^{kg} + \frac12 i f^{c8e} [G^{ke},\{J^r,G^{r8}\}] - \frac12 f^{c8e} f^{8eg} \mathcal{D}_4^{kg} + \{\mathcal{D}_2^{kc},\{G^{r8},G^{r8}\}\} \nonumber \\
&  & \mbox{} - \frac12 \{\{J^r,G^{r8}\},\{G^{k8},T^c\}\} + \frac12 i f^{c8e} \{J^k,[\{J^i,G^{ie}\},\{J^r,G^{r8}\}]\},
\end{eqnarray}

\begin{equation}
\{\mathcal{D}_2^{i8},[G^{kc},[J^2,G^{i8}]]\} = - \frac34 f^{c8e} f^{8eg} \mathcal{D}_2^{kg} - i f^{c8e} [G^{k8},\{J^r,G^{re}\}] + \{\mathcal{D}_2^{k8},\{G^{rc},G^{r8}\}\} - \frac12 \{\{J^r,G^{rc}\},\{G^{k8},T^8\}\},
\end{equation}

\begin{eqnarray}
&  & \{G^{i8},[\mathcal{D}_3^{kc},[J^2,G^{i8}]]\} \nonumber \\
&  & \mbox{\hglue0.2truecm} = \frac32 f^{c8e} f^{8eg} G^{kg} - \frac12 d^{c8e} d^{8eg} G^{kg} - \frac{1}{2N_f} d^{c88} J^k + \frac14 f^{c8e} f^{8eg} \mathcal{D}_3^{kg} - 2 d^{c8e} d^{8eg} \mathcal{D}_3^{kg} + \frac{4}{N_f} \delta^{c8} \mathcal{D}_3^{k8} \nonumber \\
&  & \mbox{\hglue0.6truecm} - \frac{4}{N_f} \delta^{88} \mathcal{D}_3^{kc} + \frac12 f^{c8e} f^{8eg} \mathcal{O}_3^{kg} - \frac12 d^{c8e} d^{8eg} \mathcal{O}_3^{kg} + \frac{1}{N_f} \delta^{c8} \mathcal{O}_3^{k8} - \frac{1}{N_f} \delta^{88} \mathcal{O}_3^{kc} + 2 \{G^{kc},\{G^{r8},G^{r8}\}\} \nonumber \\
&  & \mbox{\hglue0.6truecm} - 2 \{G^{k8},\{G^{rc},G^{r8}\}\} + 7 d^{c8e} \{J^k,\{G^{re},G^{r8}\}\} - 3 d^{88e} \{J^k,\{G^{rc},G^{re}\}\} + \frac12 d^{c8e} \{G^{ke},\{J^r,G^{r8}\}\} \nonumber \\
&  & \mbox{\hglue0.6truecm} + d^{c8e} \{G^{k8},\{J^r,G^{re}\}\} - \frac12 d^{88e} \{G^{kc},\{J^r,G^{re}\}\} - \frac12 d^{88e} \{G^{ke},\{J^r,G^{rc}\}\} - \frac{4}{N_f} d^{c88} \{J^2,J^k\} \nonumber \\
&  & \mbox{\hglue0.6truecm} + \frac12 \epsilon^{kim} f^{c8e} \{T^e,\{J^i,G^{m8}\}\} - \{\{J^r,G^{rc}\},\{G^{k8},\{J^i,G^{i8}\}\}\} + 2 \{J^k,\{\{J^i,G^{ic}\},\{G^{r8},G^{r8}\}\}\} \nonumber \\
&  & \mbox{\hglue0.6truecm} - \{J^k,\{\{J^i,G^{i8}\},\{G^{rc},G^{r8}\}\}\} - \frac12 d^{c8e} \{\mathcal{D}_3^{k8},\{J^r,G^{re}\}\} + d^{c8e} \{J^2,\{J^k,\{G^{re},G^{r8}\}\}\},
\end{eqnarray}

\begin{eqnarray}
&  & \{\mathcal{D}_3^{i8},[G^{kc},[J^2,G^{i8}]]\} \nonumber \\
&  & \mbox{\hglue0.2truecm} = \frac32 f^{c8e} f^{8eg} G^{kg} - \frac14 f^{c8e} f^{8eg} \mathcal{D}_3^{kg} + \frac12 f^{c8e} f^{8eg} \mathcal{O}_3^{kg} - \frac12 d^{c8e} d^{8eg} \mathcal{O}_3^{kg} + \frac{1}{N_f} \delta^{c8} \mathcal{O}_3^{k8} \nonumber \\
&  & \mbox{\hglue0.6truecm} - \frac{1}{N_f} \delta^{88} \mathcal{O}_3^{kc} + 2 \{G^{kc},\{G^{r8},G^{r8}\}\} - 2 \{G^{k8},\{G^{rc},G^{r8}\}\} - d^{c8e} \{J^k,\{G^{re},G^{r8}\}\} + d^{88e} \{J^k,\{G^{rc},G^{re}\}\} \nonumber \\
&  & \mbox{\hglue0.6truecm} + d^{c8e} \{G^{k8},\{J^r,G^{re}\}\} - \frac12 d^{88e} \{G^{kc},\{J^r,G^{re}\}\} - \frac12 d^{88e} \{G^{ke},\{J^r,G^{rc}\}\} + \frac12 \epsilon^{kim} f^{c8e} \{T^e,\{J^i,G^{m8}\}\} \nonumber \\
&  & \mbox{\hglue0.6truecm} - \{\{J^r,G^{rc}\},\{G^{k8},\{J^i,G^{i8}\}\}\} + \{J^k,\{\{J^i,G^{i8}\},\{G^{rc},G^{r8}\}\}\},
\end{eqnarray}

\begin{eqnarray}
&  & \{G^{i8},[\mathcal{O}_3^{kc},[J^2,G^{i8}]]\} \nonumber \\
&  & \mbox{\hglue0.2truecm} = - \frac18 f^{c8e} f^{8eg} \mathcal{D}_3^{kg} - \frac{1}{N_f} \delta^{c8} \mathcal{D}_3^{k8} - \frac54 f^{c8e} f^{8eg} \mathcal{O}_3^{kg} - \frac34 d^{c8e} d^{8eg} \mathcal{O}_3^{kg} - \frac{17}{2N_f} \delta^{c8} \mathcal{O}_3^{k8} \nonumber \\
&  & \mbox{\hglue0.6truecm} - \frac{7}{2N_f} \delta^{88} \mathcal{O}_3^{kc} + 2 \{G^{kc},\{G^{r8},G^{r8}\}\} - d^{c8e} \{J^k,\{G^{re},G^{r8}\}\} + d^{88e} \{J^k,\{G^{rc},G^{re}\}\} + \frac72  d^{c8e} \{G^{ke},\{J^r,G^{r8}\}\} \nonumber \\
&  & \mbox{\hglue0.6truecm} - \frac72 d^{c8e} \{G^{k8},\{J^r,G^{re}\}\} - \frac74 d^{88e} \{G^{kc},\{J^r,G^{re}\}\} + \frac34 d^{88e} \{G^{ke},\{J^r,G^{rc}\}\} - \frac14 \epsilon^{kim} f^{c8e} \{J^2,\{T^e,\{J^i,G^{m8}\}\}\} \nonumber \\
&  & \mbox{\hglue0.6truecm} - \frac32 \epsilon^{kim} f^{c8e} \{T^e,\{J^i,G^{m8}\}\}  - \frac{1}{N_f} \delta^{c8} \mathcal{O}_5^{k8}
- \{G^{kc},\{\{J^i,G^{i8}\},\{J^r,G^{r8}\}\}\} + \frac12 \{\{J^r,G^{rc}\},\{G^{k8},\{J^i,G^{i8}\}\}\} \nonumber \\
&  & \mbox{\hglue0.6truecm} - \{J^k,\{\{J^i,G^{ic}\},\{G^{r8},G^{r8}\}\}\} + \frac12 \{J^k,\{\{J^i,G^{i8}\},\{G^{rc},G^{r8}\}\}\} + 2 \{J^2,\{G^{kc},\{G^{r8},G^{r8}\}\}\} \nonumber \\
&  & \mbox{\hglue0.6truecm} + \frac14 d^{c8e} \{\mathcal{D}_3^{k8},\{J^r,G^{re}\}\} - \frac14 d^{c8e} \{J^2,\{G^{k8},\{J^r,G^{re}\}\}\},
\end{eqnarray}

\begin{eqnarray}
\{G^{i8},[G^{kc},[J^2,\mathcal{O}_3^{i8}]]\} & = & - \frac14 d^{c8e} d^{8eg} G^{kg} - \frac{1}{4N_f} d^{c88} J^k - \frac12 d^{c8e} d^{8eg} \mathcal{D}_3^{kg} +  d^{c8e} \{J^k,\{G^{re},G^{r8}\}\} \nonumber \\
&  & \mbox{\hglue0.6truecm} + \frac14 d^{c8e} \{G^{ke},\{J^r,G^{r8}\}\} - \frac{1}{N_f} d^{c88} \{J^2,J^k\} - \frac14 \epsilon^{kim} f^{c8e} \{J^2,\{T^e,\{J^i,G^{m8}\}\}\} \nonumber \\
&  & \mbox{\hglue0.6truecm} - \frac{1}{N_f} \delta^{c8} \mathcal{O}_5^{k8} + \frac12 d^{c8e} \{J^2,\{J^k,\{G^{re},G^{r8}\}\}\} - \frac14 d^{c8e} \{J^2,\{G^{k8},\{J^r,G^{re}\}\}\},
\end{eqnarray}

\begin{eqnarray}
&  & \{\mathcal{O}_3^{i8},[G^{kc},[J^2,G^{i8}]]\} \nonumber \\
&  & \mbox{\hglue0.2truecm} = - \frac14 d^{c8e} d^{8eg} G^{kg} - \frac{1}{4N_f} d^{c88} J^k + \frac18 f^{c8e} f^{8eg} \mathcal{D}_3^{kg} - \frac12 d^{c8e} d^{8eg} \mathcal{D}_3^{kg} + \frac{1}{N_f} \delta^{c8} \mathcal{D}_3^{k8} - \frac14 f^{c8e} f^{8eg} \mathcal{O}_3^{kg} \nonumber \\
&  & \mbox{\hglue0.6truecm} + \frac14 d^{c8e} d^{8eg} \mathcal{O}_3^{kg} + \frac{3}{2N_f} \delta^{c8} \mathcal{O}_3^{k8} + \frac{5}{2N_f} \delta^{88} \mathcal{O}_3^{kc} - 2 \{G^{kc},\{G^{r8},G^{r8}\}\} + 2 d^{c8e} \{J^k,\{G^{re},G^{r8}\}\} \nonumber \\
&  & \mbox{\hglue0.6truecm} - d^{88e} \{J^k,\{G^{rc},G^{re}\}\} - \frac14 d^{c8e} \{G^{ke},\{J^r,G^{r8}\}\} + \frac12 d^{c8e} \{G^{k8},\{J^r,G^{re}\}\} + \frac54 d^{88e} \{G^{kc},\{J^r,G^{re}\}\} \nonumber \\
&  & \mbox{\hglue0.6truecm} - \frac14 d^{88e} \{G^{ke},\{J^r,G^{rc}\}\} - \frac{1}{N_f} d^{c88} \{J^2,J^k\} + \frac12 \epsilon^{kim} f^{c8e} \{T^e,\{J^i,G^{m8}\}\} - \frac{1}{N_f} \delta^{c8} \mathcal{O}_5^{k8} \nonumber \\
&  & \mbox{\hglue0.6truecm} + \frac12 \{\{J^r,G^{rc}\},\{G^{k8},\{J^i,G^{i8}\}\}\} + \frac12 d^{c8e} \{J^2,\{J^k,\{G^{re},G^{r8}\}\}\} - \frac14 d^{c8e} \{J^2,\{G^{k8},\{J^r,G^{re}\}\}\} \nonumber \\
&  & \mbox{\hglue0.6truecm} - \frac14 \epsilon^{kim} f^{c8e} \{J^2,\{T^e,\{J^i,G^{m8}\}\}\} - \frac12 \{J^k,\{\{J^i,G^{i8}\},\{G^{rc},G^{r8}\}\}\}.
\end{eqnarray}

The next-to-leading order contribution to the baryon magnetic moment for finite decuplet-octet mass difference involves the two operator structures,
\begin{equation*}
\left[A^{ja}, \left[\left[\mathcal{M}, \left[ \mathcal{M},A^{jb}\right]\right],M^{kc}\right] \right],
\quad \mbox{and} \quad \left[\left[\mathcal{M},A^{ja}\right], \left[\left[\mathcal{M},A^{jb}\right],M^{kc}\right]\right],
\end{equation*}
with two mass insertions. For the latter the results listed in Appendix B of Ref.~\cite{rfm12} can be directly used. For the former the expressions not listed in this reference read as follows:

(1) flavor $\mathbf{1}$ contribution

\begin{equation}
[G^{ia},[[J^2,[J^2,G^{ia}]],\mathcal{D}_2^{kc}]] = - \frac32 [N_c(N_c+2N_f)-4N_f] \mathcal{D}_2^{kc} - \frac52 (N_c+N_f) \mathcal{D}_3^{kc} - (N_c+N_f) \mathcal{O}_3^{kc} + 3(N_f+2) \mathcal{D}_4^{kc},
\end{equation}

\begin{equation}
[\mathcal{D}_2^{ia},[[J^2,[J^2,G^{ia}]],G^{kc}]] = 3 N_f \mathcal{D}_2^{kc} - (N_c+N_f) \mathcal{D}_3^{kc} - (N_c+N_f) \mathcal{O}_3^{kc} + 2 \mathcal{D}_4^{kc}.
\end{equation}

(2) flavor $\mathbf{8}$ contribution

\begin{eqnarray}
d^{ab8} [G^{ia},[[J^2,[J^2,G^{ib}]],\mathcal{D}_2^{kc}]] & = & 3 N_f d^{c8e} \mathcal{D}_2^{ke} - \frac54 (N_c+N_f) d^{c8e} \mathcal{D}_3^{ke} - \frac12 (N_c+N_f) d^{c8e} \mathcal{O}_3^{ke} \nonumber \\
&  & \mbox{} - \frac34 (N_c+N_f) \{J^k,\{T^c,T^8\}\} + \frac12 (N_f+5) d^{c8e} \mathcal{D}_4^{ke} \nonumber \\
&  & \mbox{} + \frac{N_f^2+6N_f+4}{2N_f} \{\mathcal{D}_2^{kc},\{J^r,G^{r8}\}\} - 2 \{\mathcal{D}_2^{k8},\{J^r,G^{rc}\}\} - \frac12 \{J^2,\{G^{kc},T^8\}\} \nonumber \\
&  & \mbox{} + \frac{N_f^2+N_f-4}{2N_f} \{J^2,\{G^{k8},T^c\}\} + \frac{1}{N_f} \{J^2,[J^2,[T^8,G^{kc}]]\},
\end{eqnarray}

\begin{eqnarray}
d^{ab8} [\mathcal{D}_2^{ia},[[J^2,[J^2,G^{ib}]],G^{kc}]] & = & \frac32 N_f d^{c8e} \mathcal{D}_2^{ke} + \frac12 (N_f-2) [J^2,[T^8,G^{kc}]] - \frac{N_c+N_f}{N_f} d^{c8e} \mathcal{D}_3^{ke} - \frac{N_c+N_f}{N_f} d^{c8e} \mathcal{O}_3^{ke} \nonumber \\
&  & \mbox{} - \frac{(N_c+N_f)(N_f-2)}{2N_f} \{G^{kc},\{J^r,G^{r8}\}\} + \frac{(N_c+N_f)(N_f-2)}{2N_f} \{G^{k8},\{J^r,G^{rc}\}\} \nonumber \\
&  & \mbox{} - \frac{(N_c+N_f)(N_f-2)}{N_f} \{J^k,\{G^{rc},G^{r8}\}\} + \frac{(N_c+N_f)(N_f-2)}{N_f^2} \delta^{c8} \{J^2,J^k\} \nonumber \\
&  & \mbox{} + d^{c8e} \mathcal{D}_4^{ke} + \frac12 \{\mathcal{D}_2^{kc},\{J^r,G^{r8}\}\} - \frac12 \{\mathcal{D}_2^{k8},\{J^r,G^{rc}\}\} - \frac12 \{J^2,\{G^{kc},T^8\}\} \nonumber \\
&  & \mbox{} + \frac12 \{J^2,\{G^{k8},T^c\}\} + \frac12 \{J^2,[J^2,[T^8,G^{kc}]]\};
\end{eqnarray}

(3) flavor $\mathbf{27}$ contribution

\begin{eqnarray}
[G^{i8},[[J^2,[J^2,G^{i8}]],\mathcal{D}_2^{kc}]] & = & 6 f^{c8e} f^{8eg} \mathcal{D}_2^{kg} + \frac72 f^{c8e} f^{8eg} \mathcal{D}_4^{kg} + \frac{2}{N_f} \delta^{88} \mathcal{D}_4^{kc} - 2 \{\mathcal{D}_2^{kc},\{G^{r8},G^{r8}\}\} \nonumber \\
&  & \mbox{} + \frac12 d^{88e} \{J^2,\{G^{ke},T^c\}\} + \frac12 d^{88e} \{\mathcal{D}_2^{kc},\{J^r,G^{re}\}\},
\end{eqnarray}

\begin{eqnarray}
[\mathcal{D}_2^{i8},[[J^2,[J^2,G^{i8}]],G^{kc}]] & = & \frac{3}{2} f^{c8e} f^{8eg} \mathcal{D}_2^{kg} + \frac12 i f^{c8e} [G^{k8},\{J^r,G^{re}\}] + \frac12 f^{c8e} f^{8eg} \mathcal{D}_4^{kg} + \frac{2}{N_f} \delta^{c8} \mathcal{D}_4^{k8} \nonumber \\
&  & \mbox{} - 2 \{\mathcal{D}_2^{k8},\{G^{rc},G^{r8}\}\} + \frac12 d^{c8e} \{J^2,\{G^{ke},T^8\}\} + \frac12 d^{c8e} \{\mathcal{D}_2^{k8},\{J^r,G^{re}\}\} \nonumber \\
&  & \mbox{} + \frac12 \{\{J^r,G^{rc}\},\{G^{k8},T^8\}\} - \frac12 \{\{J^r,G^{r8}\},\{G^{kc},T^8\}\} \nonumber \\
&  & \mbox{} - 3 i f^{c8e} \{J^k,[\{J^i,G^{ie}\},\{J^r,G^{r8}\}]\} - i f^{c8e} \{\{J^r,G^{re}\},[J^2,G^{k8}]\} \nonumber \\
&  & \mbox{} + \frac12 i f^{c8e}\{\{J^r,G^{r8}\},[J^2,G^{ke}]\}.
\end{eqnarray}

\section{\label{app:OpBasis}Flavor contributions from Fig.~\ref{fig:mmloop2}}

Here we discuss the different flavor contributions that make up the one-loop correction to the baryon magnetic moment operator from Fig.~\ref{fig:mmloop2}(a)--(d), Eq.~(\ref{eq:dasplit}).

For the flavor $\mathbf{1}$ representation, the operators that occur at this order are
\begin{eqnarray}
X_{1}^{kc} = G^{kc}, &
X_{2}^{kc} = \mathcal{D}_2^{kc}, \quad
X_{3}^{kc} = \mathcal{D}_3^{kc}, \quad
X_{4}^{kc} = \mathcal{O}_3^{kc}, \quad
X_{5}^{kc} = \mathcal{D}_4^{kc}, \quad
X_{6}^{kc} = \mathcal{D}_5^{kc}, \quad
X_{7}^{kc} = \mathcal{O}_5^{kc}. \nonumber
\end{eqnarray}
The matrix elements are listed in Tables I--III of Ref.~\cite{rfm09}. The corresponding coefficients are
\begin{eqnarray}
x_{1} & = & \Big[ \frac{23}{24}a_1^2 m_1 + \frac{N_c+3}{6N_c}a_1 b_2 m_1 - \frac{N_c+3}{2N_c}a_1^2 m_2 - \frac{3}{N_c^2}a_1 b_2 m_2 + \frac{N_c^2+6N_c-18}{12N_c^2} b_2^2 m_1 + \frac{1}{N_c^2}a_1 b_3 m_1 \nonumber \\
&  & \mbox{} - \frac{N_c^2+6N_c+4}{2N_c^2}a_1^2 m_3 - \frac{N_c^2+6N_c-3}{2N_c^2}a_1^2 m_4 - \frac{2(N_c+3)}{N_c^3}b_2 b_3 m_1 - \frac{2(N_c+3)}{N_c^3}a_1 b_3 m_2 - \frac{2(N_c+3)}{N_c^3}a_1 b_2 m_3 \Big] F_{\mathbf{1}}^{(1)} \nonumber \\
&  & \mbox{} + \Big[ \frac{1}{4}a_1^2 m_1 - \frac{N_c+3}{2N_c}a_1 b_2 m_1 - \frac{N_c+3}{2N_c}a_1^2 m_2 -\frac{N_c^2+6N_c+6}{2N_c^2}a_1 b_3 m_1 - \frac{N_c^2+6N_c+6}{2N_c^2}a_1^2 m_3 +\frac{3(N_c+6)}{4N_c}a_1 c_3 m_1 \nonumber \\
&  & \mbox{} - \frac{3(N_c+6)}{4N_c}a_1^2 m_4 \Big] \frac{\Delta}{N_c} F_{\mathbf{1}}^{(2)} + \Big[ \frac{1}{12} \left(N_c^2+6N_c-3\right)a_1^2 m_1 \Big] \frac{\Delta^2}{N_c^2} F_{\mathbf{1}}^{(3)}, \nonumber \\
\end{eqnarray}

\begin{eqnarray}
x_{2} & = & \Big[ \frac{5}{4N_c}a_1 b_2 m_1 + \frac{71}{24N_c}a_1^2 m_2 + \frac{2 (N_c+3)}{3N_c^2}a_1 b_2 m_2 + \frac{N_c+3}{2N_c^2}a_1 b_3 m_1 - \frac{2 (N_c+3)}{N_c^2}a_1^2 m_3 - \frac{3 (N_c+3)}{4N_c^2}a_1 c_3 m_1 \nonumber \\
&  & \mbox{} + \frac{N_c+3}{2N_c^2}a_1^2 m_4 + \frac{N_c^2+6N_c-18}{12N_c^3}b_2^2 m_2 - \frac{1}{N_c^3}b_2 b_3 m_1 + \frac{N_c^2+6N_c+6}{2N_c^3}a_1 b_3 m_2 - \frac{1}{N_c^3}a_1 b_2 m_3 + \frac{9}{2N_c^3}b_2 c_3 m_1 \nonumber \\
&  & \mbox{} - \frac{3(N_c^2+6N_c-12)}{4N_c^3}a_1 c_3 m_2 + \frac{9}{2N_c^3}a_1 b_2 m_4 \Big] F_{\mathbf{1}}^{(1)} \nonumber \\
&  & \mbox{} + \Big[ - \frac{1}{4} (N_c+3)a_1^2 m_1 + \frac{2}{N_c}a_1 b_2 m_1 - \frac{N_c^2+6N_c-21}{4N_c}a_1^2 m_2 + \frac{N_c+3}{2N_c^2}a_1 b_3 m_1 - \frac{11 (N_c+3)}{2N_c^2}a_1^2 m_3 \nonumber \\
&  & \mbox{} - \frac{9(N_c+3)}{4N_c^2}a_1 c_3 m_1 + \frac{3 (N_c+3)}{4N_c^2}a_1^2 m_4 \Big] \frac{\Delta}{N_c} F_{\mathbf{1}}^{(2)} \nonumber \\
&  & \mbox{} + \Big[ - \frac{11}{24} (N_c+3)a_1^2 m_1 +\frac{3}{2N_c}a_1 b_2 m_1 - \frac{3(N_c^2+6N_c-12)}{8N_c}a_1^2 m_2 \Big] \frac{\Delta^2}{N_c^2} F_{\mathbf{1}}^{(3)},
\end{eqnarray}

\begin{eqnarray}
x_{3} & = & \Big[ \frac{3}{4N_c^2}a_1 b_2 m_2 + \frac{5}{8N_c^2}b_2^2 m_1 + \frac{11}{12N_c^2}a_1 b_3 m_1 + \frac{131}{24N_c^2}a_1^2 m_3 + \frac{1}{N_c^2}a_1 c_3 m_1 + \frac{7(N_c+3)}{6N_c^3}b_2 b_3 m_1 \nonumber \\
&  & \mbox{} + \frac{N_c+3}{2N_c^3}a_1 b_3 m_2 + \frac{7(N_c+3)}{6N_c^3}a_1 b_2 m_3 - \frac{N_c+3}{2N_c^3}b_2 c_3 m_1 - \frac{5(N_c+3)}{4N_c^3}a_1 c_3 m_2 - \frac{N_c+3}{2N_c^3}a_1 b_2 m_4 \Big] F_{\mathbf{1}}^{(1)} \nonumber \\
&  & \mbox{} + \Big[ \frac{1}{4}a_1^2 m_1 - \frac{N_c+3}{4N_c}a_1^2 m_2 + \frac{3}{2N_c^2}a_1 b_3 m_1 - \frac{2N_c^2+12N_c-47}{4N_c^2}a_1^2 m_3 + \frac{3}{2N_c^2}a_1 c_3 m_1 + \frac{3}{4N_c^2}a_1^2 m_4 \Big] \frac{\Delta}{N_c} F_{\mathbf{1}}^{(2)} \nonumber \\
&  & \mbox{} + \Big[ \frac{1}{2}a_1^2 m_1 - \frac{N_c+3}{6N_c}a_1 b_2 m_1 - \frac{5 (N_c+3)}{8N_c}a_1^2 m_2 \Big] \frac{\Delta^2}{N_c^2} F_{\mathbf{1}}^{(3)},
\end{eqnarray}

\begin{eqnarray}
x_{4} & = & \Big[ - \frac{1}{N_c^2}a_1 b_2 m_2 + \frac{7}{4N_c^2}b_2^2 m_1 + \frac{7}{6N_c^2}a_1 b_3 m_1 + \frac{3}{2N_c^2}a_1 c_3 m_1 + \frac{131}{24N_c^2}a_1^2 m_4 + \frac{11(N_c+3)}{3N_c^3}b_2 b_3 m_1 \nonumber \\
&  & \mbox{} - \frac{N_c+3}{N_c^3}a_1 b_3 m_2 - \frac{N_c+3}{N_c^3}a_1 b_2 m_3- \frac{N_c+3}{2N_c^3}b_2 c_3 m_1 - \frac{N_c+3}{2N_c^3}a_1 c_3 m_2 + \frac{2 (N_c+3)}{3N_c^3}a_1 b_2 m_4 \Big] F_{\mathbf{1}}^{(1)} \nonumber \\
&  & \mbox{} + \Big[ \frac{1}{2}a_1^2 m_1 - \frac{5}{2N_c^2}a_1 b_2 m_2 + \frac{1}{2N_c^2}a_1 b_3 m_1 + \frac{1}{2N_c^2}a_1^2 m_3 - \frac{17}{4N_c^2}a_1 c_3 m_1 - \frac{(N_c-3)(N_c+9)}{2N_c^2}a_1^2 m_4
\Big] \frac{\Delta}{N_c} F_{\mathbf{1}}^{(2)} \nonumber \\
&  & \mbox{} + \Big[ \frac{2}{3}a_1^2 m_1 - \frac{N_c+3}{6N_c}a_1 b_2 m_1 - \frac{N_c+3}{6N_c}a_1^2 m_2 \Big] \frac{\Delta^2}{N_c^2} F_{\mathbf{1}}^{(3)},
\end{eqnarray}

\begin{eqnarray}
x_{5} & = & \Big[ \frac{5}{4N_c^3}b_2^2 m_2 + \frac{1}{2N_c^3}b_2 b_3 m_1 + \frac{5}{6N_c^3}a_1 b_3 m_2 + \frac{1}{2N_c^3}a_1 b_2 m_3 + \frac{1}{N_c^3}b_2 c_3 m_1 + \frac{15}{2N_c^3}a_1 c_3 m_2 + \frac{1}{N_c^3}a_1 b_2 m_4 \Big] F_{\mathbf{1}}^{(1)} \nonumber \\
&  & \mbox{} + \Big[ \frac{1}{N_c}a_1^2 m_2 - \frac{N_c+3}{2N_c^2}a_1^2 m_3 - \frac{N_c+3}{2N_c^2}a_1 c_3 m_1 \Big] \frac{\Delta}{N_c} F_{\mathbf{1}}^{(2)} + \Big[ \frac{1}{3N_c}a_1 b_2 m_1 + \frac{15}{4N_c}a_1^2 m_2 \Big] \frac{\Delta^2}{N_c^2} F_{\mathbf{1}}^{(3)},
\end{eqnarray}

\begin{equation}
x_{6} = \Big[ \frac{3}{2N_c^2}a_1^2 m_3 + \frac{1}{2N_c^2}a_1 c_3 m_1 \Big] \frac{\Delta}{N_c} F_{\mathbf{1}}^{(2)},
\end{equation}

\begin{equation}
x_{7} = \Big[ \frac{1}{N_c^2}a_1 c_3 m_1 + \frac{3}{2N_c^2}a_1^2 m_4 \Big] \frac{\Delta}{N_c} F_{\mathbf{1}}^{(2)}.
\end{equation}

For the flavor $\mathbf{8}$ representation, the relevant operators are listed in Sec.~IV.A of Ref.~\cite{rfm09}. A nonvanishing $\Delta$ requires that the list be complemented with the following operators
\begin{eqnarray}
&  & Y_{25}^{kc} = d^{c8e} \mathcal{D}_5^{ke}, \quad 
Y_{26}^{kc} = d^{c8e} \mathcal{O}_5^{ke}, \quad
Y_{27}^{kc} = \{J^2,\{J^k,\{T^c,T^8\}\}\}, \quad
Y_{28}^{kc} = \{J^2,\{J^k,\{G^{rc},G^{r8}\}\}\}, \nonumber \\
&  & Y_{29}^{kc} = \{J^k,\{\{J^r,G^{rc}\},\{J^m,G^{m8}\}\}\}, \quad 
Y_{30}^{kc} = \delta^{c8} \{J^2,\{J^2,J^k\}\}. \nonumber
\end{eqnarray}
The matrix elements of these operators are listed in Tables \ref{t:mm833O}, \ref{t:mm833T}, and \ref{t:mm833TO} for the magnetic moments of octet and decuplet baryons and the transition magnetic moment of decuplet-octet baryons, respectively. These tables are to be considered as continuations of Tables IV, V, and VI, respectively, of Ref.~\cite{rfm09}.

\begin{table*}
\caption{\label{t:mm833O}nontrivial matrix elements of the operators involved in the magnetic moments of octet baryons: Flavor $\mathbf{8}$ and $\mathbf{10}+\overline{\mathbf{10}}$ representations. The entries correspond to $48\sqrt{3}\langle Y_{m}^{33}\rangle$ and $48\langle Y_{m}^{38}\rangle$.}
\begin{ruledtabular}
\begin{tabular}{lccccccccc}
& $\displaystyle n$ & $\displaystyle p$ & $\displaystyle \Sigma^-$ & $\displaystyle \Sigma^0$ & $\displaystyle \Sigma^+$ & $\displaystyle \Xi^-$ & $\displaystyle \Xi^0$ & $\displaystyle \Lambda$ & $\displaystyle \Lambda\Sigma^0$  \\[2mm]
\hline
$\langle Y_{25}^{33}\rangle$ & $-90$ & $90$ & $-72$ & $0$ & $72$ & $18$ & $-18$ & $0$ & $36 \sqrt{3}$ \\
$\langle Y_{26}^{33}\rangle$ & $0$ & $0$ & $0$ & $0$ & $0$ & $0$ & $0$ & $0$ & $0$ \\
$\langle Y_{27}^{33}\rangle$ & $-108$ & $108$ & $0$ & $0$ & $0$ & $108$ & $-108$ & $0$ & $0$ \\
$\langle Y_{28}^{33}\rangle$ & $-45$ & $45$ & $-144$ & $0$ & $144$ & $-99$ & $99$ & $0$ & $-36 \sqrt{3}$ \\
$\langle Y_{29}^{33}\rangle$ & $-90$ & $90$ & $-144$ & $0$ & $144$ & $-54$ & $54$ & $0$ & $0$ \\
$\langle Y_{30}^{33}\rangle$ & $0$ & $0$ & $0$ & $0$ & $0$ & $0$ & $0$ & $0$ & $0$  \\
$\langle Y_{25}^{38}\rangle$ & $-18$ & $-18$ & $-36$ & $-36$ & $-36$ & $54$ & $54$ & $36$ & $0$ \\
$\langle Y_{26}^{38}\rangle$ & $0$ & $0$ & $0$ & $0$ & $0$ & $0$ & $0$ & $0$ & $0$ \\
$\langle Y_{27}^{38}\rangle$ & $108$ & $108$ & $0$ & $0$ & $0$ & $108$ & $108$ & $0$ & $0$ \\
$\langle Y_{28}^{38}\rangle$ & $9$ & $9$ & $108$ & $108$ & $108$ & $153$ & $153$ & $36$ & $0$ \\
$\langle Y_{29}^{38}\rangle$ & $18$ & $18$ & $72$ & $72$ & $72$ & $162$ & $162$ & $72$ & $0$ \\
$\langle Y_{30}^{38}\rangle$ & $54$ & $54$ & $54$ & $54$ & $54$ & $54$ & $54$ & $54$ & $0$
\end{tabular}
\end{ruledtabular}
\end{table*}

\begin{table*}
\caption{\label{t:mm833T}nontrivial matrix elements of the operators involved in the magnetic moments of decuplet baryons: Flavor $\mathbf{8}$ and $\mathbf{10}+\overline{\mathbf{10}}$ representations. The entries correspond to $16\sqrt{3}\langle Y_{m}^{33}\rangle$ and $16\langle Y_{m}^{38}\rangle$.}
\begin{ruledtabular}
\begin{tabular}{lcccccccccc}
& $\displaystyle \Delta^{++}$ & $\displaystyle \Delta^+$ & $\displaystyle \Delta^0$ & $\displaystyle \Delta^-$ & $\displaystyle {\Sigma^*}^+$ & $\displaystyle {\Sigma^*}^0$ & $\displaystyle {\Sigma^*}^-$ & $\displaystyle {\Xi^*}^0$ & $\displaystyle {\Xi^*}^-$ & $\displaystyle \Omega^-$  \\[2mm]
\hline
$\langle Y_{25}^{33}\rangle$ & $1350$ & $450$ & $-450$ & $-1350$ & $900$ & $0$ & $-900$ & $450$ & $-450$ & $0$ \\
$\langle Y_{26}^{33}\rangle$ & $0$ & $0$ & $0$ & $0$ & $0$ & $0$ & $0$ & $0$ & $0$ & $0$ \\
$\langle Y_{27}^{33}\rangle$ & $1620$ & $540$ & $-540$ & $-1620$ & $0$ & $0$ & $0$ & $-540$ & $540$ & $0$ \\
$\langle Y_{28}^{33}\rangle$ & $675$ & $225$ & $-225$ & $-675$ & $180$ & $0$ & $-180$ & $-45$ & $45$ & $0$ \\
$\langle Y_{29}^{33}\rangle$ & $1350$ & $450$ & $-450$ & $-1350$ & $0$ & $0$ & $0$ & $-450$ & $450$ & $0$ \\
$\langle Y_{30}^{33}\rangle$ & $0$ & $0$ & $0$ & $0$ & $0$ & $0$ & $0$ & $0$ & $0$ & $0$  \\
$\langle Y_{25}^{38}\rangle$ & $-450$ & $-450$ & $-450$ & $-450$ & $0$ & $0$ & $0$ & $450$ & $450$ & $900$ \\
$\langle Y_{26}^{38}\rangle$ & $0$ & $0$ & $0$ & $0$ & $0$ & $0$ & $0$ & $0$ & $0$ & $0$ \\
$\langle Y_{27}^{38}\rangle$ & $540$ & $540$ & $540$ & $540$ & $0$ & $0$ & $0$ & $540$ & $540$ & $2160$ \\
$\langle Y_{28}^{38}\rangle$ & $225$ & $225$ & $225$ & $225$ & $180$ & $180$ & $180$ & $405$ & $405$ & $900$ \\
$\langle Y_{29}^{38}\rangle$ & $450$ & $450$ & $450$ & $450$ & $0$ & $0$ & $0$ & $450$ & $450$ & $1800$ \\
$\langle Y_{30}^{38}\rangle$ & $1350$ & $1350$ & $1350$ & $1350$ & $1350$ & $1350$ & $1350$ & $1350$ & $1350$ & $1350$  \\
\end{tabular}
\end{ruledtabular}
\end{table*}

\begin{table*}
\caption{\label{t:mm833TO}nontrivial matrix elements of the operators involved in the decuplet to octet transition magnetic moments: Flavor $\mathbf{8}$ and $\mathbf{10}+\overline{\mathbf{10}}$ representations. The entries correspond to $12\sqrt{6}\langle Y_{m}^{33}\rangle$ and $12\sqrt{2}\langle Y_{m}^{38}\rangle$.}
\begin{ruledtabular}
\begin{tabular}{lcccccccc}
& $\displaystyle \Delta^+p$ & $\displaystyle \Delta^0n$ & $\displaystyle {\Sigma^*}^0\Lambda$ & $\displaystyle {\Sigma^*}^0\Sigma^0$ & $\displaystyle {\Sigma^*}^+\Sigma^+$ & $\displaystyle {\Sigma^*}^-\Sigma^-$ & $\displaystyle {\Xi^*}^0\Xi^0$ & $\displaystyle {\Xi^*}^-\Xi^-$  \\[2mm]
\hline
$\langle Y_{25}^{33}\rangle$ & $0$ & $0$ & $0$ & $0$ & $0$ & $0$ & $0$ & $0$ \\
$\langle Y_{26}^{33}\rangle$ & $162$ & $162$ & $81 \sqrt{3}$ & $0$ & $81$ & $-81$ & $81$ & $-81$ \\
$\langle Y_{27}^{33}\rangle$ & $0$ & $0$ & $0$ & $0$ & $0$ & $0$ & $0$ & $0$ \\
$\langle Y_{28}^{33}\rangle$ & $0$ & $0$ & $0$ & $0$ & $0$ & $0$ & $0$ & $0$ \\
$\langle Y_{29}^{33}\rangle$ & $0$ & $0$ & $0$ & $0$ & $0$ & $0$ & $0$ & $0$ \\
$\langle Y_{30}^{33}\rangle$ & $0$ & $0$ & $0$ & $0$ & $0$ & $0$ & $0$ & $0$  \\
$\langle Y_{25}^{38}\rangle$ & $0$ & $0$ & $0$ & $0$ & $0$ & $0$ & $0$ & $0$ \\
$\langle Y_{26}^{38}\rangle$ & $0$ & $0$ & $0$ & $-81$ & $-81$ & $-81$ & $-81$ & $-81$ \\
$\langle Y_{27}^{38}\rangle$ & $0$ & $0$ & $0$ & $0$ & $0$ & $0$ & $0$ & $0$ \\
$\langle Y_{28}^{38}\rangle$ & $0$ & $0$ & $0$ & $0$ & $0$ & $0$ & $0$ & $0$ \\
$\langle Y_{29}^{38}\rangle$ & $0$ & $0$ & $0$ & $0$ & $0$ & $0$ & $0$ & $0$ \\
$\langle Y_{30}^{38}\rangle$ & $0$ & $0$ & $0$ & $0$ & $0$ & $0$ & $0$ & $0$
\end{tabular}
\end{ruledtabular}
\end{table*}

The coefficients that accompany the operator basis read
\begin{eqnarray}
y_{1} & = & \Big[ \frac{11}{48}a_1^2 m_1 - \frac{N_c+3}{12N_c}a_1 b_2 m_1 - \frac{N_c+3}{4N_c}a_1^2 m_2 - \frac{3}{2N_c^2}a_1 b_2 m_2 - \frac{3}{4N_c^2}b_2^2 m_1 - \frac{1}{2N_c^2}a_1 b_3 m_1 - \frac{2}{N_c^2}a_1^2 m_3 + \frac{3}{4N_c^2}a_1^2 m_4 \nonumber \\
&  & \mbox{} - \frac{N_c+3}{N_c^3}b_2 b_3 m_1 - \frac{N_c+3}{N_c^3}a_1 b_3 m_2 - \frac{N_c+3}{N_c^3}a_1 b_2 m_3 \Big] F_{\mathbf{8}}^{(1)} \nonumber \\
&  & \mbox{} + \Big[ - \frac{1}{8}a_1^2 m_1 - \frac{N_c+3}{4N_c}a_1 b_2 m_1 - \frac{N_c+3}{4N_c}a_1^2 m_2 - \frac{3}{2N_c^2}a_1 b_3 m_1 - \frac{3}{2N_c^2}a_1^2 m_3 \Big] \frac{\Delta}{N_c} F_{\mathbf{8}}^{(2)} + \Big[ -\frac{1}{8}a_1^2 m_1 \Big] \frac{\Delta^2}{N_c^2} F_{\mathbf{8}}^{(3)}, \nonumber \\
\end{eqnarray}

\begin{eqnarray}
y_{2} & = & \Big[ \frac{5}{36}a_1^2 m_1 + \frac{N_c+3}{18N_c}a_1 b_2 m_1 + \frac{N_c^2+6N_c+4}{12N_c^2}a_1 b_3 m_1 - \frac{N_c^2+6N_c-1}{3N_c^2}a_1^2 m_3 - \frac{N_c+6}{8N_c}a_1 c_3 m_1 + \frac{N_c+6}{12N_c}a_1^2 m_4 \Big] F_{\mathbf{8}}^{(1)} \nonumber \\
&  & \mbox{} + \Big[ - \frac{1}{24}(N_c^2+6N_c-2)a_1^2 m_1 + \frac{N_c+6}{12N_c}a_1 b_3 m_1 - \frac{11 (N_c+6)}{12N_c}a_1^2 m_3 - \frac{3(N_c+6)}{8N_c}a_1 c_3 m_1 + \frac{N_c+6}{8N_c}a_1^2 m_4 \Big] \frac{\Delta}{N_c} F_{\mathbf{8}}^{(2)}\nonumber \\
&  & \mbox{} + \Big[ -\frac{11}{144}N_c (N_c+6)a_1^2 m_1 \Big] \frac{\Delta^2}{N_c^2} F_{\mathbf{8}}^{(3)},
\end{eqnarray}

\begin{eqnarray}
y_{3} & = & \Big[ \frac{5}{8N_c}a_1 b_2 m_1 + \frac{13}{16N_c}a_1^2 m_2 + \frac{N_c+3}{4N_c^2}a_1 b_3 m_1 - \frac{N_c+3}{N_c^2}a_1^2 m_3 - \frac{3 (N_c+3)}{8N_c^2}a_1 c_3 m_1 + \frac{N_c+3}{4N_c^2}a_1^2 m_4 - \frac{3}{4N_c^3}b_2^2 m_2 \nonumber \\
&  & \mbox{} - \frac{1}{2N_c^3}b_2 b_3 m_1 - \frac{1}{2N_c^3}a_1 b_3 m_2 - \frac{1}{2N_c^3}a_1 b_2 m_3 + \frac{9}{4N_c^3}b_2 c_3 m_1 + \frac{9}{2N_c^3}a_1 c_3 m_2 + \frac{9}{4N_c^3}a_1 b_2 m_4 \Big] F_{\mathbf{8}}^{(1)} \nonumber \\
&  & \mbox{} + \Big[ - \frac{1}{8} (N_c+3)a_1^2 m_1 + \frac{1}{N_c}a_1 b_2 m_1 + \frac{17}{8N_c}a_1^2 m_2 + \frac{N_c+3}{4N_c^2}a_1 b_3 m_1 - \frac{11 (N_c+3)}{4N_c^2}a_1^2 m_3 - \frac{9 (N_c+3)}{8N_c^2}a_1 c_3 m_1 \nonumber \\
&  & \mbox{} + \frac{3 (N_c+3)}{8N_c^2}a_1^2 m_4 \Big] \frac{\Delta}{N_c} F_{\mathbf{8}}^{(2)} + \Big[ - \frac{11}{48} (N_c+3)a_1^2 m_1 + \frac{3}{4N_c}a_1 b_2 m_1 + \frac{9}{4N_c}a_1^2 m_2 \Big] \frac{\Delta^2}{N_c^2} F_{\mathbf{8}}^{(3)}, \nonumber \\
\end{eqnarray}

\begin{eqnarray}
y_{4} & = & \Big[ - \frac{1}{12N_c}a_1 b_2 m_1 - \frac{1}{4N_c}a_1^2 m_2 - \frac{N_c+3}{12N_c^2}b_2^2 m_1 - \frac{N_c+3}{2N_c^2}a_1^2 m_3 - \frac{N_c+3}{2N_c^2}a_1^2 m_4 - \frac{1}{N_c^3}b_2 b_3 m_1 \nonumber \\
&  & \mbox{} - \frac{1}{N_c^3}a_1 b_3 m_2 - \frac{1}{N_c^3}a_1 b_2 m_3 \Big] F_{\mathbf{8}}^{(1)} + \Big[ - \frac{1}{4N_c}a_1 b_2 m_1 - \frac{1}{4N_c}a_1^2 m_2 - \frac{N_c+3}{2N_c^2}a_1 b_3 m_1 \nonumber \\
&  & \mbox{} - \frac{N_c+3}{2N_c^2}a_1^2 m_3 + \frac{3 (N_c+3)}{4N_c^2}a_1 c_3 m_1 - \frac{3 (N_c+3)}{4N_c^2}a_1^2 m_4 \Big] \frac{\Delta}{N_c} F_{\mathbf{8}}^{(2)} + \Big[ \frac{N_c+3}{12}a_1^2 m_1 \Big] \frac{\Delta^2}{N_c^2} F_{\mathbf{8}}^{(3)},
\end{eqnarray}

\begin{eqnarray}
y_{5} & = & \Big[ \frac{1}{4N_c}a_1 b_2 m_1 + \frac{2}{3N_c}a_1^2 m_2 + \frac{N_c+3}{6N_c^2}a_1 b_2 m_2 + \frac{1}{N_c^3}b_2 b_3 m_1 + \frac{2}{N_c^3}a_1 b_3 m_2 + \frac{1}{N_c^3}a_1 b_2 m_3 \Big] F_{\mathbf{8}}^{(1)} \nonumber \\
&  & \mbox{} + \Big[ \frac{1}{4N_c}a_1 b_2 m_1 + \frac{1}{2N_c}a_1^2 m_2 \Big] \frac{\Delta}{N_c} F_{\mathbf{8}}^{(2)},
\end{eqnarray}

\begin{eqnarray}
y_{6} & = & \Big[ \frac{3}{8N_c^2}a_1 b_2 m_2 + \frac{3}{16N_c^2}b_2^2 m_1 + \frac{7}{24N_c^2}a_1 b_3 m_1 + \frac{17}{16N_c^2}a_1^2 m_3 + \frac{1}{3N_c^2}a_1 c_3 m_1 - \frac{1}{6N_c^2}a_1^2 m_4 + \frac{N_c+3}{4N_c^3}b_2 b_3 m_1 \nonumber \\
&  & \mbox{} + \frac{N_c+3}{4N_c^3}a_1 b_3 m_2 + \frac{N_c+3}{4N_c^3}a_1 b_2 m_3 - \frac{N_c+3}{6N_c^3}b_2 c_3 m_1 - \frac{5(N_c+3)}{8N_c^3}a_1 c_3 m_2 - \frac{N_c+3}{6N_c^3}a_1 b_2 m_4 \Big] F_{\mathbf{8}}^{(1)} \nonumber \\
&  & \mbox{} + \Big[ \frac{1}{8}a_1^2 m_1 - \frac{N_c+3}{8N_c}a_1^2 m_2 + \frac{1}{4N_c^2}a_1 b_3 m_1 + \frac{13}{8N_c^2}a_1^2 m_3 + \frac{7}{12N_c^2}a_1 c_3 m_1 - \frac{1}{24N_c^2}a_1^2 m_4 \Big] \frac{\Delta}{N_c} F_{\mathbf{8}}^{(2)} \nonumber \\
&  & \mbox{} + \Big[ \frac{7}{36}a_1^2 m_1 - \frac{N_c+3}{18N_c}a_1 b_2 m_1 - \frac{5 (N_c+3)}{16N_c}a_1^2 m_2 \Big] \frac{\Delta^2}{N_c^2} F_{\mathbf{8}}^{(3)},
\end{eqnarray}

\begin{eqnarray}
y_{7} & = & \Big[ - \frac{1}{2N_c^2}a_1 b_2 m_2 + \frac{5}{8N_c^2}b_2^2 m_1 - \frac{5}{12N_c^2}a_1 b_3 m_1 - \frac{1}{3N_c^2}a_1^2 m_3 + \frac{7}{12N_c^2}a_1 c_3 m_1 + \frac{43}{48N_c^2}a_1^2 m_4 + \frac{2 (N_c+3)}{3N_c^3}b_2 b_3 m_1 \nonumber \\
&  & \mbox{} - \frac{N_c+3}{2N_c^3}a_1 b_3 m_2 - \frac{N_c+3}{3N_c^3}a_1 b_2 m_3 - \frac{N_c+3}{6N_c^3}b_2 c_3 m_1 - \frac{N_c+3}{4N_c^3}a_1 c_3 m_2 + \frac{N_c+3}{12N_c^3}a_1 b_2 m_4 \Big] F_{\mathbf{8}}^{(1)} \nonumber \\
&  & \mbox{} + \Big[ \frac{1}{4}a_1^2 m_1 - \frac{5}{4N_c^2}a_1 b_2 m_2 - \frac{5}{6N_c^2}a_1 b_3 m_1 - \frac{5}{6N_c^2}a_1^2 m_3 - \frac{1}{2N_c^2}a_1 c_3 m_1 + \frac{21}{8N_c^2}a_1^2 m_4 \Big] \frac{\Delta}{N_c} F_{\mathbf{8}}^{(2)} \nonumber \\
&  & \mbox{} + \Big[ \frac{23}{72}a_1^2 m_1 - \frac{N_c+3}{18N_c}a_1 b_2 m_1 - \frac{N_c+3}{12N_c}a_1^2 m_2 \Big] \frac{\Delta^2}{N_c^2} F_{\mathbf{8}}^{(3)},
\end{eqnarray}

\begin{eqnarray}
y_{8} & = & \Big[ - \frac{1}{2N_c^2}a_1 b_2 m_2 + \frac{3}{4N_c^2}b_2^2 m_1 - \frac{1}{2N_c^2}a_1 b_3 m_1 + \frac{1}{3N_c^2}a_1^2 m_3 + \frac{1}{6N_c^2}a_1 c_3 m_1 + \frac{7}{3N_c^2}a_1^2 m_4 + \frac{N_c+3}{2N_c^3}b_2 b_3 m_1 \nonumber \\
&  & \mbox{} - \frac{N_c+3}{6N_c^3}a_1 b_2 m_3 - \frac{N_c+3}{12N_c^3}b_2 c_3 m_1 + \frac{N_c+3}{12N_c^3}a_1 b_2 m_4 \Big] F_{\mathbf{8}}^{(1)} \nonumber \\
&  & \mbox{} + \Big[ \frac{1}{4}a_1^2 m_1 - \frac{1}{2N_c^2}a_1 b_2 m_2 + \frac{7}{12N_c^2}a_1 b_3 m_1 + \frac{7}{12N_c^2}a_1^2 m_3 - \frac{21}{8N_c^2}a_1 c_3 m_1 + \frac{39}{8N_c^2}a_1^2 m_4 \Big] \frac{\Delta}{N_c} F_{\mathbf{8}}^{(2)} \nonumber \\
&  & \mbox{} + \Big[ - \frac{1}{36}a_1^2 m_1 - \frac{N_c+3}{36N_c}a_1 b_2 m_1 \Big] \frac{\Delta^2}{N_c^2} F_{\mathbf{8}}^{(3)},
\end{eqnarray}

\begin{eqnarray}
y_{9} & = & \Big[ \frac{1}{2N_c^2}a_1 b_2 m_2 - \frac{1}{4N_c^2}b_2^2 m_1 + \frac{5}{6N_c^2}a_1 b_3 m_1 + \frac{3}{2N_c^2}a_1^2 m_3 - \frac{1}{6N_c^2}a_1 c_3 m_1 - \frac{13}{12N_c^2}a_1^2 m_4 - \frac{N_c+3}{6N_c^3}b_2 b_3 m_1 \nonumber \\
&  & \mbox{} + \frac{N_c+3}{2N_c^3}a_1 b_2 m_3 + \frac{N_c+3}{12N_c^3}b_2 c_3 m_1 - \frac{N_c+3}{12N_c^3}a_1 b_2 m_4 \Big] F_{\mathbf{8}}^{(1)} + \Big[ - \frac{1}{6}a_1^2 m_1 + \frac{N_c+3}{12N_c}a_1 b_2 m_1 + \frac{1}{2N_c^2}a_1 b_2 m_2 \nonumber \\
&  & \mbox{} + \frac{17}{12N_c^2}a_1 b_3 m_1 + \frac{17}{12N_c^2}a_1^2 m_3 + \frac{3}{8N_c^2}a_1 c_3 m_1 - \frac{21}{8N_c^2}a_1^2 m_4 \Big] \frac{\Delta}{N_c} F_{\mathbf{8}}^{(2)} + \Big[
- \frac{13}{72}a_1^2 m_1 + \frac{N_c+3}{36N_c}a_1 b_2 m_1 \Big] \frac{\Delta^2}{N_c^2} F_{\mathbf{8}}^{(3)}, \nonumber \\
\end{eqnarray}

\begin{eqnarray}
y_{10} & = & \Big[ \frac{1}{12N_c^2}a_1 b_2 m_2 + \frac{1}{8N_c^2}a_1 b_3 m_1 - \frac{1}{2N_c^2}a_1^2 m_3 - \frac{3}{16N_c^2}a_1 c_3 m_1 + \frac{1}{8N_c^2}a_1^2 m_4 \nonumber \\
&  & \mbox{} - \frac{N_c+3}{24N_c^3}b_2^2 m_2 + \frac{N_c+3}{4N_c^3}a_1 b_3 m_2 - \frac{3(N_c+3)}{8N_c^3}a_1 c_3 m_2 \Big] F_{\mathbf{8}}^{(1)} \nonumber \\
&  & \mbox{} + \Big[ - \frac{1}{16}a_1^2 m_1 - \frac{N_c+3}{8N_c}a_1^2 m_2 + \frac{1}{8N_c^2}a_1 b_3 m_1 - \frac{11}{8N_c^2}a_1^2 m_3 - \frac{9}{16N_c^2}a_1 c_3 m_1 + \frac{3}{16N_c^2}a_1^2 m_4 \Big] \frac{\Delta}{N_c} F_{\mathbf{8}}^{(2)} \nonumber \\
&  & \mbox{} + \Big[ - \frac{11}{96}a_1^2 m_1 - \frac{3 (N_c+3)}{16N_c}a_1^2 m_2 \Big] \frac{\Delta^2}{N_c^2} F_{\mathbf{8}}^{(3)},
\end{eqnarray}

\begin{eqnarray}
y_{11} & = & \Big[ - \frac{1}{2N_c^2}a_1 b_3 m_1 + \frac{5}{2N_c^2}a_1^2 m_3 + \frac{7}{12N_c^2}a_1 c_3 m_1 - \frac{11}{12N_c^2}a_1^2 m_4 - \frac{N_c+3}{6N_c^3}b_2 c_3 m_1 - \frac{N_c+3}{6N_c^3}a_1 b_2 m_4 \Big] F_{\mathbf{8}}^{(1)} \nonumber \\
&  & \mbox{} + \Big[ \frac{1}{6}a_1^2 m_1 - \frac{N_c+3}{12N_c}a_1 b_2 m_1 - \frac{1}{N_c^2}a_1 b_3 m_1 + \frac{7}{N_c^2}a_1^2 m_3 + \frac{31}{12N_c^2}a_1 c_3 m_1 - \frac{17}{12N_c^2}a_1^2 m_4 \Big] \frac{\Delta}{N_c} F_{\mathbf{8}}^{(2)} \nonumber \\
&  & \mbox{} + \Big[ \frac{4}{9}a_1^2 m_1 - \frac{N_c+3}{18N_c}a_1 b_2 m_1 \Big] \frac{\Delta^2}{N_c^2} F_{\mathbf{8}}^{(3)},
\end{eqnarray}

\begin{eqnarray}
y_{12} & = & \Big[ - \frac{1}{18N_c^2}a_1 b_3 m_1 + \frac{5}{6N_c^2}a_1^2 m_3 + \frac{17}{36N_c^2}a_1 c_3 m_1 - \frac{1}{9N_c^2}a_1^2 m_4 + \frac{N_c+3}{18N_c^3}b_2 c_3 m_1  + \frac{N_c+3}{18N_c^3}a_1 b_2 m_4 \Big] F_{\mathbf{8}}^{(1)}\nonumber \\
&  & \mbox{} + \Big[ \frac{1}{12}a_1^2 m_1 - \frac{1}{3N_c^2}a_1 b_3 m_1 - \frac{N_c^2+6N_c-34}{12N_c^2}a_1^2 m_3 - \frac{3N_c^2+18N_c-50}{36N_c^2}a_1 c_3 m_1 - \frac{5}{18N_c^2}a_1^2 m_4 \Big] \frac{\Delta}{N_c} F_{\mathbf{8}}^{(2)} \nonumber \\
&  & \mbox{} + \Big[ \frac{55}{216}a_1^2 m_1 + \frac{N_c+3}{54N_c}a_1 b_2 m_1 \Big] \frac{\Delta^2}{N_c^2} F_{\mathbf{8}}^{(3)},
\end{eqnarray}

\begin{eqnarray}
y_{13} & = & \Big[ \frac{3}{8N_c^3}b_2^2 m_2 + \frac{1}{4N_c^3}b_2 b_3 m_1 + \frac{1}{4N_c^3}a_1 b_3 m_2 + \frac{1}{4N_c^3}a_1 b_2 m_3 + \frac{1}{2N_c^3}b_2 c_3 m_1 + \frac{2}{N_c^3}a_1 c_3 m_2  + \frac{1}{2N_c^3}a_1 b_2 m_4 \Big] F_{\mathbf{8}}^{(1)} \nonumber \\
&  & \mbox{} + \Big[ \frac{1}{4N_c}a_1^2 m_2 - \frac{N_c+3}{4N_c^2}a_1^2 m_3 - \frac{N_c+3}{4N_c^2}a_1 c_3 m_1 \Big] \frac{\Delta}{N_c} F_{\mathbf{8}}^{(2)} + \Big[ \frac{1}{6N_c}a_1 b_2 m_1 + \frac{1}{N_c}a_1^2 m_2 \Big] \frac{\Delta^2}{N_c^2} F_{\mathbf{8}}^{(3)},
\end{eqnarray}

\begin{eqnarray}
y_{14} & = & \Big[ \frac{1}{2N_c^3}b_2^2 m_2 - \frac{5}{2N_c^3}a_1 b_3 m_2 - \frac{1}{N_c^3}a_1 b_2 m_3 + \frac{1}{4N_c^3}b_2 c_3 m_1 + \frac{31}{12N_c^3}a_1 c_3 m_2 + \frac{1}{2N_c^3}a_1 b_2 m_4 \Big] F_{\mathbf{8}}^{(1)} \nonumber \\
&  & \mbox{} + \Big[ \frac{2}{3N_c}a_1^2 m_2 - \frac{N_c+3}{12N_c^2}a_1 b_2 m_2 \Big] \frac{\Delta}{N_c} F_{\mathbf{8}}^{(2)} + \Big[ \frac{1}{12N_c}a_1 b_2 m_1 + \frac{101}{72N_c}a_1^2 m_2 \Big] \frac{\Delta^2}{N_c^2} F_{\mathbf{8}}^{(3)},
\end{eqnarray}

\begin{eqnarray}
y_{15} & = & \Big[ \frac{7}{6N_c^3}b_2 b_3 m_1 - \frac{1}{2N_c^3}a_1 b_3 m_2 - \frac{1}{2N_c^3}a_1 b_2 m_3 - \frac{1}{4N_c^3}b_2 c_3 m_1 - \frac{1}{4N_c^3}a_1 c_3 m_2 + \frac{1}{6N_c^3}a_1 b_2 m_4 \Big] F_{\mathbf{8}}^{(1)} \nonumber \\
&  & \mbox{} + \Big[ - \frac{N_c+3}{2N_c^2}a_1^2 m_4 \Big] \frac{\Delta}{N_c} F_{\mathbf{8}}^{(2)} + \Big[ - \frac{1}{12N_c}a_1 b_2 m_1 - \frac{1}{12N_c}a_1^2 m_2 \Big] \frac{\Delta^2}{N_c^2} F_{\mathbf{8}}^{(3)},
\end{eqnarray}

\begin{eqnarray}
y_{16} & = & \Big[ - \frac{1}{2N_c^3}b_2 b_3 m_1 + \frac{5}{3N_c^3}a_1 b_3 m_2 + \frac{1}{2N_c^3}a_1 b_2 m_3 + \frac{1}{4N_c^3}b_2 c_3 m_1 +\frac{2}{3N_c^3}a_1 c_3 m_2 \Big] F_{\mathbf{8}}^{(1)} \nonumber \\
&  & \mbox{} + \Big[ \frac{1}{12N_c}a_1^2 m_2 + \frac{N_c+3}{12N_c^2}a_1 b_2 m_2 \Big] \frac{\Delta}{N_c} F_{\mathbf{8}}^{(2)} + \Big[ \frac{1}{12N_c}a_1 b_2 m_1 + \frac{2}{9N_c}a_1^2 m_2 \Big] \frac{\Delta^2}{N_c^2} F_{\mathbf{8}}^{(3)}, \nonumber \\
\end{eqnarray}

\begin{eqnarray}
y_{17} & = & \Big[ - \frac{1}{3N_c^3}b_2 b_3 m_1 + \frac{1}{N_c^3}a_1 b_3 m_2 + \frac{4}{3N_c^3}a_1 b_2 m_3 - \frac{1}{4N_c^3}b_2 c_3 m_1 - \frac{1}{N_c^3}a_1 c_3 m_2 - \frac{2}{3N_c^3}a_1 b_2 m_4 \Big] F_{\mathbf{8}}^{(1)} \nonumber \\
&  & \mbox{} + \Big[ - \frac{1}{4N_c}a_1^2 m_2 - \frac{N_c+3}{N_c^2}a_1^2 m_3 + \frac{N_c+3}{2N_c^2}a_1^2 m_4 \Big] \frac{\Delta}{N_c} F_{\mathbf{8}}^{(2)} + \Big[ - \frac{1}{12N_c}a_1 b_2 m_1 - \frac{13}{24N_c}a_1^2 m_2 \Big] \frac{\Delta^2}{N_c^2} F_{\mathbf{8}}^{(3)},
\end{eqnarray}

\begin{equation}
y_{18} = y_{19} = y_{20} = y_{21} = y_{22} = 0,
\end{equation}

\begin{equation}
y_{23} = \Big[ \frac{1}{2N_c^2}a_1 c_3 m_1 + \frac{5}{4N_c^2}a_1^2 m_4 \Big] \frac{\Delta}{N_c} F_{\mathbf{8}}^{(2)},
\end{equation}

\begin{equation}
y_{24} = \Big[ \frac{1}{6N_c^2}a_1 b_3 m_1 + \frac{1}{6N_c^2}a_1^2 m_3 - \frac{5}{12N_c^2}a_1 c_3 m_1 - \frac{1}{4N_c^2}a_1^2 m_4 \Big] \frac{\Delta}{N_c} F_{\mathbf{8}}^{(2)},
\end{equation}

\begin{equation}
y_{25} = \Big[ \frac{1}{4N_c^2}a_1^2 m_3 + \frac{1}{4N_c^2}a_1 c_3 m_1 \Big] \frac{\Delta}{N_c} F_{\mathbf{8}}^{(2)},
\end{equation}

\begin{equation}
y_{26} = \Big[ \frac{1}{2N_c^2}a_1 c_3 m_1 + \frac{1}{4N_c^2}a_1^2 m_4 \Big] \frac{\Delta}{N_c} F_{\mathbf{8}}^{(2)},
\end{equation}

\begin{equation}
y_{27} = \Big[ - \frac{1}{8N_c^2}a_1^2 m_3 - \frac{1}{8N_c^2}a_1 c_3 m_1 \Big] \frac{\Delta}{N_c} F_{\mathbf{8}}^{(2)},
\end{equation}

\begin{equation}
y_{28} = \Big[ - \frac{1}{6N_c^2}a_1 b_3 m_1 + \frac{1}{2N_c^2}a_1^2 m_3 + \frac{5}{12N_c^2}a_1 c_3 m_1 - \frac{1}{12N_c^2}a_1^2 m_4 \Big] \frac{\Delta}{N_c} F_{\mathbf{8}}^{(2)},
\end{equation}

\begin{equation}
y_{29} = \Big[ \frac{11}{12N_c^2}a_1^2 m_3 - \frac{11}{24N_c^2}a_1^2 m_4 \Big] \frac{\Delta}{N_c} F_{\mathbf{8}}^{(2)},
\end{equation}

\begin{equation}
y_{30} = \Big[ \frac{1}{6N_c^2}a_1^2 m_3 + \frac{1}{6N_c^2}a_1 c_3 m_1 \Big] \frac{\Delta}{N_c} F_{\mathbf{8}}^{(2)}.
\end{equation}

\end{widetext}

Finally, for the flavor $\mathbf{27}$ representation, the operator basis is listed in Sec.~IV.B of Ref.~\cite{rfm09}. This operator basis also has to be complemented with the following operators:
\begin{eqnarray}
\begin{array}{lll}
Z_{37}^{kc} = \delta^{c8} \mathcal{O}_5^{k8}, \nonumber
Z_{38}^{kc} = \{G^{kc},\{\{J^i,G^{i8}\},\{J^r,G^{r8}\}\}\}, \nonumber \\
Z_{39}^{kc} = \{\mathcal{D}_2^{kc},\{T^8,\{J^r,G^{r8}\}\}\},  \nonumber \\
Z_{40}^{kc} = \{\{J^r,G^{rc}\},\{G^{k8},\{J^i,G^{i8}\}\}\}, \nonumber \\
Z_{41}^{kc} = \{J^k,\{\{J^i,G^{ic}\},\{G^{r8},G^{r8}\}\}\},  \nonumber \\
Z_{42}^{kc} = \{J^2,\{G^{k8},\{T^c,T^8\}\}\}, \nonumber \\
Z_{43}^{kc} = \{J^2,\{G^{kc},\{G^{r8},G^{r8}\}\}\},  \nonumber \\
Z_{44}^{kc} = d^{c8e} \{\mathcal{D}_3^{k8},\{J^r,G^{re}\}\}, \nonumber \\
Z_{45}^{kc} = d^{c8e} \{J^2,\{J^k,\{G^{re},G^{r8}\}\}\},  \nonumber \\
Z_{46}^{kc} = d^{c8e} \{J^2,\{G^{k8},\{J^r,G^{re}\}\}\}, \nonumber \\
Z_{47}^{kc} = \{J^k,\{\{J^i,G^{i8}\},\{G^{rc},G^{r8}\}\}\}. \nonumber
\end{array}
\end{eqnarray}
The matrix elements are listed in Tables \ref{t:mm2733O}, \ref{t:mm2733T}, and \ref{t:mm2733TO}
for the magnetic moments of octet and decuplet baryons and the transition magnetic moment of decuplet-octet baryons, respectively.

\begin{table*}
\caption{\label{t:mm2733O}nontrivial matrix elements of the operators involved in the magnetic moments of octet baryons: $\mathbf{27}$ representation. The entries correspond to $144\langle Z_{m}^{33}\rangle$ $144\sqrt{3}\langle Z_{m}^{38}\rangle$.}
\begin{ruledtabular}
\begin{tabular}{lccccccccc}
& $\displaystyle n$ & $\displaystyle p$ & $\displaystyle \Sigma^-$ & $\displaystyle \Sigma^0$ & $\displaystyle \Sigma^+$ & $\displaystyle \Xi^-$ & $\displaystyle \Xi^0$ & $\displaystyle \Lambda$ & $\displaystyle \Lambda\Sigma^0$ \\[2mm]
\hline
$\langle Z_{37}^{33}\rangle$ & $0$ & $0$ & $0$ & $0$ & $0$ & $0$ & $0$ & $0$ & $0$ \\
$\langle Z_{38}^{33}\rangle$ & $-45$ & $45$ & $-144$ & $0$ & $144$ & $81$ & $-81$ & $0$ & $72 \sqrt{3}$ \\
$\langle Z_{39}^{33}\rangle$ & $-54$ & $54$ & $0$ & $0$ & $0$ & $-162$ & $162$ & $0$ & $0$ \\
$\langle Z_{40}^{33}\rangle$ & $-45$ & $45$ & $-144$ & $0$ & $144$ & $81$ & $-81$ & $0$ & $72 \sqrt{3}$ \\
$\langle Z_{41}^{33}\rangle$ & $-45$ & $45$ & $-432$ & $0$ & $432$ & $153$ & $-153$ & $0$ & $144 \sqrt{3}$ \\
$\langle Z_{42}^{33}\rangle$ & $-54$ & $54$ & $0$ & $0$ & $0$ & $-162$ & $162$ & $0$ & $0$ \\
$\langle Z_{43}^{33}\rangle$ & $-\frac{45}{2}$ & $\frac{45}{2}$ & $-216$ & $0$ & $216$ & $\frac{153}{2}$ & $-\frac{153}{2}$ & $0$ & $72 \sqrt{3}$ \\
$\langle Z_{44}^{33}\rangle$ & $-90$ & $90$ & $-144$ & $0$ & $144$ & $-54$ & $54$ & $0$ & $0$ \\
$\langle Z_{45}^{33}\rangle$ & $-45$ & $45$ & $-144$ & $0$ & $144$ & $-99$ & $99$ & $0$ & $-36 \sqrt{3}$ \\
$\langle Z_{46}^{33}\rangle$ & $-45$ & $45$ & $-72$ & $0$ & $72$ & $-27$ & $27$ & $0$ & $0$ \\
$\langle Z_{47}^{33}\rangle$ & $-45$ & $45$ & $-288$ & $0$ & $288$ & $297$ & $-297$ & $0$ & $0$ \\
$\langle Z_{37}^{38}\rangle$ & $0$ & $0$ & $0$ & $0$ & $0$ & $0$ & $0$ & $0$ & $0$ \\
$\langle Z_{38}^{38}\rangle$ & $27$ & $27$ & $216$ & $216$ & $216$ & $-729$ & $-729$ & $-216$ & $0$ \\
$\langle Z_{39}^{38}\rangle$ & $162$ & $162$ & $0$ & $0$ & $0$ & $-486$ & $-486$ & $0$ & $0$ \\
$\langle Z_{40}^{38}\rangle$ & $27$ & $27$ & $216$ & $216$ & $216$ & $-729$ & $-729$ & $-216$ & $0$ \\
$\langle Z_{41}^{38}\rangle$ & $27$ & $27$ & $648$ & $648$ & $648$ & $-1377$ & $-1377$ & $-216$ & $0$ \\
$\langle Z_{42}^{38}\rangle$ & $162$ & $162$ & $0$ & $0$ & $0$ & $-486$ & $-486$ & $0$ & $0$ \\
$\langle Z_{43}^{38}\rangle$ & $\frac{27}{2}$ & $\frac{27}{2}$ & $324$ & $324$ & $324$ & $-\frac{1377}{2}$ & $-\frac{1377}{2}$ & $-108$ & $0$ \\
$\langle Z_{44}^{38}\rangle$ & $-54$ & $-54$ & $-216$ & $-216$ & $-216$ & $-486$ & $-486$ & $-216$ & $0$ \\
$\langle Z_{45}^{38}\rangle$ & $-27$ & $-27$ & $-324$ & $-324$ & $-324$ & $-459$ & $-459$ & $-108$ & $0$ \\
$\langle Z_{46}^{38}\rangle$ & $-27$ & $-27$ & $-108$ & $-108$ & $-108$ & $-243$ & $-243$ & $-108$ & $0$ \\
$\langle Z_{47}^{38}\rangle$ & $27$ & $27$ & $648$ & $648$ & $648$ & $-1377$ & $-1377$ & $-216$ & $0$ \\
\end{tabular}
\end{ruledtabular}
\end{table*}

\begin{table*}
\caption{\label{t:mm2733T}nontrivial matrix elements of the operators involved in the magnetic moments of decuplet baryons: $\mathbf{27}$ representation. The entries correspond to $48\langle Z_{m}^{33}\rangle$ and $48\sqrt{3} \langle Z_{m}^{38}\rangle$.}
\begin{ruledtabular}
\begin{tabular}{lcccccccccc}
& $\displaystyle \Delta^{++}$ & $\displaystyle \Delta^+$ & $\displaystyle \Delta^0$ & $\displaystyle \Delta^-$ & $\displaystyle {\Sigma^*}^+$ & $\displaystyle {\Sigma^*}^0$ & $\displaystyle {\Sigma^*}^-$ & $\displaystyle {\Xi^*}^0$ & $\displaystyle {\Xi^*}^-$ & $\displaystyle \Omega^-$ \\[2mm]
\hline
$\langle Z_{37}^{33}\rangle$ & $0$ & $0$ & $0$ & $0$ & $0$ & $0$ & $0$ & $0$ & $0$ & $0$ \\
$\langle Z_{38}^{33}\rangle$ & $675$ & $225$ & $-225$ & $-675$ & $0$ & $0$ & $0$ & $225$ & $-225$ & $0$ \\
$\langle Z_{39}^{33}\rangle$ & $810$ & $270$ & $-270$ & $-810$ & $0$ & $0$ & $0$ & $270$ & $-270$ & $0$ \\
$\langle Z_{40}^{33}\rangle$ & $675$ & $225$ & $-225$ & $-675$ & $0$ & $0$ & $0$ & $225$ & $-225$ & $0$ \\
$\langle Z_{41}^{33}\rangle$ & $675$ & $225$ & $-225$ & $-675$ & $360$ & $0$ & $-360$ & $405$ & $-405$ & $0$ \\
$\langle Z_{42}^{33}\rangle$ & $810$ & $270$ & $-270$ & $-810$ & $0$ & $0$ & $0$ & $270$ & $-270$ & $0$ \\
$\langle Z_{43}^{33}\rangle$ & $\frac{675}{2}$ & $\frac{225}{2}$ & $-\frac{225}{2}$ & $-\frac{675}{2}$ & $180$ & $0$ & $-180$ & $\frac{405}{2}$ & $-\frac{405}{2}$ & $0$ \\
$\langle Z_{44}^{33}\rangle$ & $1350$ & $450$ & $-450$ & $-1350$ & $0$ & $0$ & $0$ & $-450$ & $450$ & $0$ \\
$\langle Z_{45}^{33}\rangle$ & $675$ & $225$ & $-225$ & $-675$ & $180$ & $0$ & $-180$ & $-45$ & $45$ & $0$ \\
$\langle Z_{46}^{33}\rangle$ & $675$ & $225$ & $-225$ & $-675$ & $0$ & $0$ & $0$ & $-225$ & $225$ & $0$ \\
$\langle Z_{47}^{33}\rangle$ & $675$ & $225$ & $-225$ & $-675$ & $0$ & $0$ & $0$ & $45$ & $-45$ & $0$ \\
$\langle Z_{37}^{38}\rangle$ & $0$ & $0$ & $0$ & $0$ & $0$ & $0$ & $0$ & $0$ & $0$ & $0$ \\
$\langle Z_{38}^{38}\rangle$ & $675$ & $675$ & $675$ & $675$ & $0$ & $0$ & $0$ & $-675$ & $-675$ & $-5400$ \\
$\langle Z_{39}^{38}\rangle$ & $810$ & $810$ & $810$ & $810$ & $0$ & $0$ & $0$ & $-810$ & $-810$ & $-6480$ \\
$\langle Z_{40}^{38}\rangle$ & $675$ & $675$ & $675$ & $675$ & $0$ & $0$ & $0$ & $-675$ & $-675$ & $-5400$ \\
$\langle Z_{41}^{38}\rangle$ & $675$ & $675$ & $675$ & $675$ & $0$ & $0$ & $0$ & $-1215$ & $-1215$ & $-5400$ \\
$\langle Z_{42}^{38}\rangle$ & $810$ & $810$ & $810$ & $810$ & $0$ & $0$ & $0$ & $-810$ & $-810$ & $-6480$ \\
$\langle Z_{43}^{38}\rangle$ & $\frac{675}{2}$ & $\frac{675}{2}$ & $\frac{675}{2}$ & $\frac{675}{2}$ & $0$ & $0$ & $0$ & $-\frac{1215}{2}$ & $-\frac{1215}{2}$ & $-2700$ \\
$\langle Z_{44}^{38}\rangle$ & $-1350$ & $-1350$ & $-1350$ & $-1350$ & $0$ & $0$ & $0$ & $-1350$ & $-1350$ & $-5400$ \\
$\langle Z_{45}^{38}\rangle$ & $-675$ & $-675$ & $-675$ & $-675$ & $-540$ & $-540$ & $-540$ & $-1215$ & $-1215$ & $-2700$ \\
$\langle Z_{46}^{38}\rangle$ & $-675$ & $-675$ & $-675$ & $-675$ & $0$ & $0$ & $0$ & $-675$ & $-675$ & $-2700$ \\
$\langle Z_{47}^{38}\rangle$ & $675$ & $675$ & $675$ & $675$ & $0$ & $0$ & $0$ & $-1215$ & $-1215$ & $-5400$ \\
\end{tabular}
\end{ruledtabular}
\end{table*}

\begin{table*}
\caption{\label{t:mm2733TO}nontrivial matrix elements of the operators involved in the decuplet to octet transition magnetic moments: $\mathbf{27}$ representation. The entries correspond to $36\sqrt{2}\langle Z_{m}^{33}\rangle$ and $36\sqrt{6}\langle Z_{m}^{38} \rangle$.}
\begin{ruledtabular}
\begin{tabular}{lcccccccc}
& $\displaystyle \Delta^+p$ & $\displaystyle \Delta^0n$ & $\displaystyle {\Sigma^*}^0\Lambda$ & $\displaystyle {\Sigma^*}^0\Sigma^0$ & $\displaystyle {\Sigma^*}^+\Sigma^+$ & $\displaystyle {\Sigma^*}^-\Sigma^-$ & $\displaystyle {\Xi^*}^0\Xi^0$ & $\displaystyle {\Xi^*}^-\Xi^-$ \\[2mm]
\hline
$\langle Z_{37}^{33}\rangle$ & $0$ & $0$ & $0$ & $0$ & $0$ & $0$ & $0$ & $0$ \\
$\langle Z_{38}^{33}\rangle$ & $234$ & $234$ & $18 \sqrt{3}$ & $0$ & $18$ & $-18$ & $153$ & $-153$ \\
$\langle Z_{39}^{33}\rangle$ & $0$ & $0$ & $0$ & $0$ & $0$ & $0$ & $0$ & $0$ \\
$\langle Z_{40}^{33}\rangle$ & $0$ & $0$ & $-9 \sqrt{3}$ & $0$ & $63$ & $-63$ & $-72$ & $72$ \\
$\langle Z_{41}^{33}\rangle$ & $0$ & $0$ & $0$ & $0$ & $0$ & $0$ & $0$ & $0$ \\
$\langle Z_{42}^{33}\rangle$ & $0$ & $0$ & $0$ & $0$ & $0$ & $0$ & $-162$ & $162$ \\
$\langle Z_{43}^{33}\rangle$ & $81$ & $81$ & $54 \sqrt{3}$ & $0$ & $108$ & $-108$ & $\frac{351}{2}$ & $-\frac{351}{2}$ \\
$\langle Z_{44}^{33}\rangle$ & $0$ & $0$ & $0$ & $0$ & $0$ & $0$ & $0$ & $0$ \\
$\langle Z_{45}^{33}\rangle$ & $0$ & $0$ & $0$ & $0$ & $0$ & $0$ & $0$ & $0$ \\
$\langle Z_{46}^{33}\rangle$ & $162$ & $162$ & $-27 \sqrt{3}$ & $0$ & $27$ & $-27$ & $-108$ & $108$ \\
$\langle Z_{47}^{33}\rangle$ & $0$ & $0$ & $0$ & $0$ & $0$ & $0$ & $0$ & $0$ \\
$\langle Z_{37}^{38}\rangle$ & $0$ & $0$ & $0$ & $729$ & $729$ & $729$ & $729$ & $729$ \\
$\langle Z_{38}^{38}\rangle$ & $0$ & $0$ & $0$ & $54$ & $54$ & $54$ & $459$ & $459$ \\
$\langle Z_{39}^{38}\rangle$ & $0$ & $0$ & $0$ & $0$ & $0$ & $0$ & $0$ & $0$ \\
$\langle Z_{40}^{38}\rangle$ & $0$ & $0$ & $0$ & $27$ & $27$ & $27$ & $432$ & $432$ \\
$\langle Z_{41}^{38}\rangle$ & $0$ & $0$ & $0$ & $0$ & $0$ & $0$ & $0$ & $0$ \\
$\langle Z_{42}^{38}\rangle$ & $0$ & $0$ & $0$ & $0$ & $0$ & $0$ & $486$ & $486$ \\
$\langle Z_{43}^{38}\rangle$ & $0$ & $0$ & $0$ & $324$ & $324$ & $324$ & $\frac{1053}{2}$ & $\frac{1053}{2}$ \\
$\langle Z_{44}^{38}\rangle$ & $0$ & $0$ & $0$ & $0$ & $0$ & $0$ & $0$ & $0$ \\
$\langle Z_{45}^{38}\rangle$ & $0$ & $0$ & $0$ & $0$ & $0$ & $0$ & $0$ & $0$ \\
$\langle Z_{46}^{38}\rangle$ & $0$ & $0$ & $0$ & $-81$ & $-81$ & $-81$ & $324$ & $324$ \\
$\langle Z_{47}^{38}\rangle$ & $0$ & $0$ & $0$ & $0$ & $0$ & $0$ & $0$ & $0$ \\
\end{tabular}
\end{ruledtabular}
\end{table*}

The accompanying coefficients are
\begin{equation}
z_{1} = \Big[ \frac{1}{4}a_1^2 m_1 \Big] F_{\mathbf{27}}^{(1)} + \Big[ \frac{1}{4N_c^2}a_1^2 m_3 + \frac{1}{4N_c^2}a_1 c_3 m_1 \Big] \frac{\Delta}{N_c} F_{\mathbf{27}}^{(2)}, \label{eq:z01}
\end{equation}

\begin{equation}
z_{2} = \Big[ \frac{1}{6}a_1^2 m_1\Big] F_{\mathbf{27}}^{(1)},
\end{equation}

\begin{equation}
z_{3} = \Big[ \frac{1}{12}a_1^2 m_1 \Big] F_{\mathbf{27}}^{(1)} + \Big[ \frac{1}{12N_c^2}a_1^2 m_3 + \frac{1}{12N_c^2}a_1 c_3 m_1 \Big] \frac{\Delta}{N_c} F_{\mathbf{27}}^{(2)},
\end{equation}

\begin{equation}
z_{4} = \Big[ \frac{1}{3N_c}a_1 b_2 m_1 \Big] F_{\mathbf{27}}^{(1)},
\end{equation}

\begin{equation}
z_{5} = \Big[ \frac{1}{6N_c}a_1^2 m_2 \Big] F_{\mathbf{27}}^{(1)},
\end{equation}

\begin{equation}
z_{6} = \Big[ \frac{1}{2N_c}a_1 b_2 m_1 \Big] F_{\mathbf{27}}^{(1)},
\end{equation}

\begin{equation}
z_{7} = \Big[ \frac{1}{4N_c}a_1^2 m_2 \Big] F_{\mathbf{27}}^{(1)},
\end{equation}

\begin{eqnarray}
z_{8} & = & \Big[ \frac{1}{4N_c^2}a_1^2 m_3 + \frac{1}{4N_c^2}a_1 c_3 m_1 \Big] F_{\mathbf{27}}^{(1)} \nonumber \\
&  & \mbox{} + \Big[ \frac{1}{N_c^2}a_1^2 m_3 + \frac{1}{2N_c^2}a_1 c_3 m_1 \Big] \frac{\Delta}{N_c} F_{\mathbf{27}}^{(2)} \nonumber \\
&  & \mbox{} + \Big[ \frac{1}{12}a_1^2 m_1 \Big] \frac{\Delta^2}{N_c^2} F_{\mathbf{27}}^{(3)},
\end{eqnarray}

\begin{eqnarray}
z_{9} & = & \Big[ \frac{1}{2N_c^2}a_1 b_3 m_1 + \frac{1}{4N_c^2}a_1 c_3 m_1 + \frac{1}{4N_c^2}a_1^2 m_4 \Big] F_{\mathbf{27}}^{(1)} \nonumber \\
&  & \mbox{} + \Big[ \frac{1}{4N_c^2}a_1 b_3 m_1 + \frac{1}{4N_c^2}a_1^2 m_3 - \frac{1}{8N_c^2}a_1 c_3 m_1 \nonumber \\
&  & \mbox{} + \frac{3}{8N_c^2}a_1^2 m_4 \Big] \frac{\Delta}{N_c} F_{\mathbf{27}}^{(2)} + \Big[ \frac{1}{12}a_1^2 m_1 \Big] \frac{\Delta^2}{N_c^2} F_{\mathbf{27}}^{(3)},
\end{eqnarray}

\begin{eqnarray}
z_{10} & = & \Big[ \frac{1}{3N_c^2}a_1 b_3 m_1 - \frac{1}{6N_c^2}a_1^2 m_3 + \frac{1}{6N_c^2}a_1^2 m_4 \Big] F_{\mathbf{27}}^{(1)} \nonumber \\
&  & \mbox{} + \Big[ -\frac{2}{3N_c^2}a_1^2 m_3 - \frac{1}{6N_c^2}a_1 c_3 m_1 \nonumber \\
&  & \mbox{} + \frac{1}{6N_c^2}a_1^2 m_4 \Big] \frac{\Delta}{N_c} F_{\mathbf{27}}^{(2)},
\end{eqnarray}

\begin{eqnarray}
z_{11} & = & \Big[ \frac{1}{3N_c^2}a_1 c_3 m_1 + \frac{1}{2N_c^2}a_1^2 m_4 \Big] F_{\mathbf{27}}^{(1)} \nonumber \\
&  & \mbox{} + \Big[ \frac{1}{6}a_1^2 m_1 - \frac{1}{6N_c^2}a_1 b_3 m_1 - \frac{1}{6N_c^2}a_1^2 m_3 \nonumber \\
&  & \mbox{} - \frac{1}{4N_c^2}a_1 c_3 m_1 + \frac{17}{12N_c^2}a_1^2 m_4 \Big] \frac{\Delta}{N_c} F_{\mathbf{27}}^{(2)} \nonumber \\
&  & \mbox{} + \Big[ \frac{1}{9}a_1^2 m_1 \Big] \frac{\Delta^2}{N_c^2} F_{\mathbf{27}}^{(3)},
\end{eqnarray}

\begin{equation}
z_{12} = \Big[ \frac{1}{3N_c^2}a_1^2 m_3 \Big] F_{\mathbf{27}}^{(1)} + \Big[ \frac{2}{3N_c^2}a_1^2 m_3 \Big] \frac{\Delta}{N_c} F_{\mathbf{27}}^{(2)},
\end{equation}

\begin{eqnarray}
z_{13} & = & \Big[ \frac{1}{3N_c^2}a_1^2 m_4 \Big] F_{\mathbf{27}}^{(1)} \nonumber \\
&  & \mbox{} + \Big[ \frac{1}{6N_c^2}a_1 b_3 m_1 + \frac{1}{6N_c^2}a_1^2 m_3 - \frac{5}{12N_c^2}a_1 c_3 m_1 \nonumber \\
&  & \mbox{} + \frac{7}{12N_c^2}a_1^2 m_4 \Big] \frac{\Delta}{N_c} F_{\mathbf{27}}^{(2)},
\end{eqnarray}

\begin{eqnarray}
z_{14} & = & \Big[ \frac{1}{6N_c^2}a_1^2 m_3 + \frac{1}{6N_c^2}a_1 c_3 m_1 \Big] F_{\mathbf{27}}^{(1)} \nonumber \\
&  & \mbox{} + \Big[ \frac{2}{3N_c^2}a_1^2 m_3 + \frac{1}{3N_c^2}a_1 c_3 m_1 \Big] \frac{\Delta}{N_c} F_{\mathbf{27}}^{(2)} \nonumber \\
&  & \mbox{} + \Big[ \frac{1}{18}a_1^2 m_1 \Big] \frac{\Delta^2}{N_c^2} F_{\mathbf{27}}^{(3)},
\end{eqnarray}

\begin{equation}
z_{15} = \Big[ \frac{1}{4N_c^2}b_2^2 m_1 \Big] F_{\mathbf{27}}^{(1)},
\end{equation}

\begin{equation}
z_{16} = \Big[ \frac{1}{2N_c^2}a_1 b_2 m_2 \Big] F_{\mathbf{27}}^{(1)},
\end{equation}

\begin{eqnarray}
z_{17} & = & \Big[ - \frac{1}{N_c^2}a_1^2 m_3 - \frac{1}{2N_c^2}a_1^2 m_4 \Big] F_{\mathbf{27}}^{(1)} \nonumber \\
&  & + \Big[ - \frac{1}{N_c^2}a_1 b_3 m_1 - \frac{1}{N_c^2}a_1^2 m_3 + \frac{1}{N_c^2}a_1 c_3 m_1 \nonumber \\
&  & \mbox{} - \frac{1}{N_c^2}a_1^2 m_4 \Big] \frac{\Delta}{N_c} F_{\mathbf{27}}^{(2)},
\end{eqnarray}

\begin{eqnarray}
z_{18} & = & \Big[ \frac{1}{N_c^2}a_1^2 m_3 - \frac{1}{2N_c^2}a_1^2 m_4 \Big] F_{\mathbf{27}}^{(1)} \nonumber \\
&  & \mbox{} + \Big[ \frac{1}{N_c^2}a_1 b_3 m_1 + \frac{1}{N_c^2}a_1^2 m_3 \Big] \frac{\Delta}{N_c} F_{\mathbf{27}}^{(2)},
\end{eqnarray}

\begin{eqnarray}
z_{19} & = & \Big[ - \frac{3}{2N_c^2}a_1^2 m_3 - \frac{1}{2N_c^2}a_1 c_3 m_1 + \frac{1}{4N_c^2}a_1^2 m_4 \Big] F_{\mathbf{27}}^{(1)} \nonumber \\
&  & \mbox{} + \Big[ - \frac{1}{4}a_1^2 m_1 + \frac{1}{2N_c^2}a_1 b_3 m_1 - \frac{7}{2N_c^2}a_1^2 m_3 \nonumber \\
&  & \mbox{} - \frac{3}{2N_c^2}a_1 c_3 m_1 + \frac{1}{2N_c^2}a_1^2 m_4 \Big] \frac{\Delta}{N_c} F_{\mathbf{27}}^{(2)} \nonumber \\
&  & \mbox{} + \Big[ - \frac16 a_1^2 m_1 \Big] \frac{\Delta^2}{N_c^2} F_{\mathbf{27}}^{(3)},
\end{eqnarray}

\begin{eqnarray}
z_{20} & = & \Big[ \frac{1}{2N_c^2}a_1^2 m_3 - \frac{1}{4N_c^2}a_1^2 m_4 \Big] F_{\mathbf{27}}^{(1)} \nonumber \\
&  & \mbox{} + \Big[ - \frac{1}{2N_c^2}a_1 b_3 m_1 + \frac{3}{2N_c^2}a_1^2 m_3 + \frac{1}{2N_c^2}a_1 c_3 m_1 \nonumber \\
&  & \mbox{} - \frac{1}{2N_c^2}a_1^2 m_4 \Big] \frac{\Delta}{N_c} F_{\mathbf{27}}^{(2)},
\end{eqnarray}

\begin{eqnarray}
z_{21} & = & \Big[ - \frac{1}{2N_c^2}a_1 b_3 m_1 + \frac{1}{4N_c^2}a_1 c_3 m_1 + \frac{1}{2N_c^2}a_1^2 m_4 \Big] F_{\mathbf{27}}^{(1)} \nonumber \\
&  & \mbox{} + \Big[ \frac{1}{4}a_1^2 m_1 - \frac{1}{2N_c^2}a_1 b_3 m_1 - \frac{1}{2N_c^2}a_1^2 m_3 \nonumber \\
&  & \mbox{} - \frac{1}{4N_c^2}a_1 c_3 m_1 + \frac{7}{4N_c^2}a_1^2 m_4 \Big] \frac{\Delta}{N_c} F_{\mathbf{27}}^{(2)} \nonumber \\
&  & \mbox{} + \Big[ \frac{1}{12}a_1^2 m_1 \Big] \frac{\Delta^2}{N_c^2} F_{\mathbf{27}}^{(3)},
\end{eqnarray}

\begin{eqnarray}
z_{22} & = & \Big[ \frac{1}{2N_c^2}a_1^2 m_4 \Big] F_{\mathbf{27}}^{(1)} + \Big[ \frac{1}{4N_c^2}a_1 b_3 m_1 + \frac{1}{4N_c^2}a_1^2 m_3 \nonumber \\
&  & \mbox{} - \frac{5}{8N_c^2}a_1 c_3 m_1 + \frac{7}{8N_c^2}a_1^2 m_4 \Big] \frac{\Delta}{N_c} F_{\mathbf{27}}^{(2)},
\end{eqnarray}

\begin{eqnarray}
z_{23} & = & \Big[ \frac{3}{2N_c^2}a_1 b_3 m_1 + \frac{1}{2N_c^2}a_1^2 m_3 - \frac{1}{4N_c^2}a_1 c_3 m_1 \nonumber \\
&  & \mbox{} -\frac{1}{4N_c^2}a_1^2 m_4 \Big] F_{\mathbf{27}}^{(1)} \nonumber \\
&  & \mbox{} + \Big[ - \frac{1}{4N_c^2}a_1^2 m_3 - \frac{7}{4N_c^2}a_1^2 m_4 \Big] \frac{\Delta}{N_c} F_{\mathbf{27}}^{(2)} \nonumber \\
&  & \mbox{} + \Big[ - \frac{1}{12}a_1^2 m_1 \Big] \frac{\Delta^2}{N_c^2} F_{\mathbf{27}}^{(3)},
\end{eqnarray}

\begin{eqnarray}
z_{24} & = & \Big[ \frac{1}{2N_c^2}a_1^2 m_3 - \frac{1}{4N_c^2}a_1^2 m_4 \Big] F_{\mathbf{27}}^{(1)} \nonumber \\
&  & \mbox{} + \Big[ \frac{1}{4N_c^2}a_1 b_3 m_1 + \frac{1}{4N_c^2}a_1^2 m_3 + \frac{1}{8N_c^2}a_1 c_3 m_1 \nonumber \\
&  & \mbox{} - \frac{3}{8N_c^2}a_1^2 m_4 \Big] \frac{\Delta}{N_c} F_{\mathbf{27}}^{(2)},
\end{eqnarray}

\begin{eqnarray}
z_{25} & = & \Big[ \frac{1}{3N_c^3}b_2 c_3 m_1 + \frac{1}{3N_c^3}a_1 b_2 m_4 \Big] F_{\mathbf{27}}^{(1)} \nonumber \\
&  & \mbox{} + \Big[ -\frac{1}{18}a_1^2 m_1 \Big] \frac{\Delta^2}{N_c^2} F_{\mathbf{27}}^{(3)},
\end{eqnarray}

\begin{eqnarray}
z_{26} & = & \Big[ \frac{1}{3N_c^3}a_1 c_3 m_2 \Big] F_{\mathbf{27}}^{(1)} \nonumber \\
&  & \mbox{} + \Big[ \frac{1}{9N_c}a_1 b_2 m_1 + \frac{1}{6N_c}a_1^2 m_2 \Big] \frac{\Delta^2}{N_c^2} F_{\mathbf{27}}^{(3)},
\end{eqnarray}

\begin{equation}
z_{27} = \Big[ \frac{1}{4N_c^3}b_2^2 m_2 \Big] F_{\mathbf{27}}^{(1)},
\end{equation}

\begin{eqnarray}
z_{28} & = & \Big[ - \frac{1}{N_c^3}a_1 c_3 m_2 \Big] F_{\mathbf{27}}^{(1)} + \Big[ - \frac{1}{2N_c}a_1^2 m_2 \Big] \frac{\Delta}{N_c} F_{\mathbf{27}}^{(2)} \nonumber \\
&  & \mbox{} + \Big[ - \frac{1}{3N_c}a_1 b_2 m_1 - \frac{1}{2N_c}a_1^2 m_2 \Big] \frac{\Delta^2}{N_c^2} F_{\mathbf{27}}^{(3)},
\end{eqnarray}

\begin{eqnarray}
z_{29} & = & \Big[ - \frac{1}{N_c^3}b_2 c_3 m_1 - \frac{1}{N_c^3}a_1 b_2 m_4 \Big] F_{\mathbf{27}}^{(1)} \nonumber \\
&  & \mbox{} + \Big[ - \frac{1}{2N_c}a_1 b_2 m_1 \Big] \frac{\Delta}{N_c} F_{\mathbf{27}}^{(2)} + \Big[ \frac{1}{6}a_1^2 m_1 \Big] \frac{\Delta^2}{N_c^2} F_{\mathbf{27}}^{(3)}, \nonumber \\
\end{eqnarray}

\begin{eqnarray}
z_{30} & = & \Big[ - \frac{1}{2N_c^3}b_2 b_3 m_1 - \frac{1}{2N_c^3}a_1 b_2 m_3 + \frac{1}{4N_c^3}b_2 c_3 m_1 \nonumber \\
&  & \mbox{} + \frac{1}{4N_c^3}a_1 b_2 m_4 \Big] F_{\mathbf{27}}^{(1)} + \Big[ - \frac{1}{24}a_1^2 m_1 \Big] \frac{\Delta^2}{N_c^2} F_{\mathbf{27}}^{(3)}, \nonumber \\
\end{eqnarray}

\begin{eqnarray}
z_{31} & = & \Big[ - \frac{1}{2N_c^3}a_1 b_3 m_2 + \frac{1}{4N_c^3}a_1 c_3 m_2 \Big] F_{\mathbf{27}}^{(1)} \nonumber \\
&  & \mbox{} + \Big[ \frac{1}{12N_c}a_1 b_2 m_1 + \frac{1}{6N_c}a_1^2 m_2 \Big] \frac{\Delta^2}{N_c^2} F_{\mathbf{27}}^{(3)}, \nonumber \\
\end{eqnarray}

\begin{eqnarray}
z_{32} & = & \Big[ \frac{1}{2N_c^3}b_2 b_3 m_1 + \frac{1}{2N_c^3}a_1 b_2 m_3 + \frac{1}{4N_c^3}b_2 c_3 m_1 \nonumber \\
&  & \mbox{} + \frac{1}{4N_c^3}a_1 b_2 m_4 \Big] F_{\mathbf{27}}^{(1)} + \Big[ - \frac{1}{24}a_1^2 m_1 \Big] \frac{\Delta^2}{N_c^2} F_{\mathbf{27}}^{(3)}, \nonumber \\
\end{eqnarray}

\begin{eqnarray}
z_{33} & = & \Big[ \frac{1}{2N_c^3}a_1 b_3 m_2 + \frac{1}{4N_c^3}a_1 c_3 m_2 \Big] F_{\mathbf{27}}^{(1)} \nonumber \\
&  & \mbox{} + \Big[ \frac{1}{12N_c}a_1 b_2 m_1 + \frac{1}{12N_c}a_1^2 m_2 \Big] \frac{\Delta^2}{N_c^2} F_{\mathbf{27}}^{(3)}, \nonumber \\
\end{eqnarray}

\begin{eqnarray}
z_{34} & = & \Big[ - \frac{1}{2N_c^3}b_2 b_3 m_1 + \frac{3}{2N_c^3}a_1 b_2 m_3 + \frac{1}{4N_c^3}b_2 c_3 m_1 \nonumber \\
&  & \mbox{} - \frac{1}{4N_c^3}a_1 b_2 m_4 \Big] F_{\mathbf{27}}^{(1)} + \Big[ \frac{1}{4N_c}a_1 b_2 m_1 \Big] \frac{\Delta}{N_c} F_{\mathbf{27}}^{(2)} \nonumber \\
&  & \mbox{} + \Big[ - \frac{1}{24}a_1^2 m_1 \Big] \frac{\Delta^2}{N_c^2} F_{\mathbf{27}}^{(3)},
\end{eqnarray}

\begin{eqnarray}
z_{35} & = & \Big[ \frac{3}{2N_c^3}b_2 b_3 m_1 -\frac{1}{2N_c^3}a_1 b_2 m_3 - \frac{1}{4N_c^3}b_2 c_3 m_1 \nonumber \\
&  & \mbox{} + \frac{1}{4N_c^3}a_1 b_2 m_4 \Big] F_{\mathbf{27}}^{(1)} + \Big[ \frac{1}{24}a_1^2 m_1 \Big] \frac{\Delta^2}{N_c^2} F_{\mathbf{27}}^{(3)}, \nonumber \\
\end{eqnarray}

\begin{equation}
z_{36} = \Big[ \frac{1}{N_c^3}a_1 b_3 m_2 \Big] F_{\mathbf{27}}^{(1)}  + \Big[ \frac{1}{4N_c}a_1^2 m_2 \Big] \frac{\Delta}{N_c} F_{\mathbf{27}}^{(2)},
\end{equation}

\begin{equation}
z_{37} = \Big[ \frac{1}{3N_c^2}a_1 c_3 m_1 + \frac{1}{6N_c^2}a_1^2 m_4 \Big] \frac{\Delta}{N_c} F_{\mathbf{27}}^{(2)},
\end{equation}

\begin{equation}
z_{38} = \Big[ \frac{1}{2N_c^2}a_1^2 m_4 \Big] \frac{\Delta}{N_c} F_{\mathbf{27}}^{(2)},
\end{equation}

\begin{equation}
z_{39} = \Big[ - \frac{1}{4N_c^2}a_1 b_2 m_2 \Big] \frac{\Delta}{N_c} F_{\mathbf{27}}^{(2)},
\end{equation}

\begin{eqnarray}
z_{40} & = & \Big[ \frac{1}{2N_c^2}a_1 b_3 m_1 + \frac{1}{2N_c^2}a_1^2 m_3 - \frac{1}{4N_c^2}a_1 c_3 m_1 \nonumber \\
&  & \mbox{} - \frac{1}{4N_c^2}a_1^2 m_4 \Big] \frac{\Delta}{N_c} F_{\mathbf{27}}^{(2)},
\end{eqnarray}

\begin{equation}
z_{41} = \Big[ - \frac{1}{N_c^2}a_1^2 m_3 + \frac{1}{2N_c^2}a_1^2 m_4 \Big] \frac{\Delta}{N_c} F_{\mathbf{27}}^{(2)},
\end{equation}

\begin{equation}
z_{42} = \Big[ \frac{1}{4N_c^2}a_1 b_2 m_2 \Big] \frac{\Delta}{N_c} F_{\mathbf{27}}^{(2)},
\end{equation}

\begin{equation}
z_{43} = \Big[ - \frac{1}{N_c^2}a_1^2 m_4 \Big] \frac{\Delta}{N_c} F_{\mathbf{27}}^{(2)},
\end{equation}

\begin{equation}
z_{44} = \Big[ \frac{1}{4N_c^2}a_1^2 m_3 - \frac{1}{8N_c^2}a_1^2 m_4 \Big] \frac{\Delta}{N_c} F_{\mathbf{27}}^{(2)},
\end{equation}

\begin{equation}
z_{45} = \Big[ - \frac{1}{2N_c^2}a_1^2 m_3 - \frac{1}{2N_c^2}a_1 c_3 m_1 \Big] \frac{\Delta}{N_c} F_{\mathbf{27}}^{(2)},
\end{equation}

\begin{equation}
z_{46} = \Big[ \frac{1}{4N_c^2}a_1 c_3 m_1+\frac{1}{8N_c^2}a_1^2 m_4\Big] \frac{\Delta}{N_c} F_{\mathbf{27}}^{(2)},
\end{equation}

\begin{eqnarray}
z_{47} & = & \Big[ - \frac{1}{2N_c^2}a_1 b_3 m_1 + \frac{1}{2N_c^2}a_1^2 m_3 + \frac{1}{4N_c^2}a_1 c_3 m_1 \nonumber \\
&  & \mbox{} - \frac{1}{4N_c^2}a_1^2 m_4\Big] \frac{\Delta}{N_c} F_{\mathbf{27}}^{(2)}. \label{eq:z47}
\end{eqnarray}

Of course, flavor singlet and octet pieces must be subtracted off Eqs.~(\ref{eq:z01})-(\ref{eq:z47}) in order to have a truly $\mathbf{27}$ contribution.


\begin{thebibliography}{99}

\bibitem{cg61}
  S.~R.~Coleman and S.~L.~Glashow,
  Phys.\ Rev.\ Lett.\  {\bf 6}, 423 (1961).

\bibitem{coleman}
S.~R.~Coleman, \textit{Aspects of Symmetry. Selected Erice Lectures} (Cambridge University Press, Cambridge, England, 1985).

\bibitem{jm255}
E.~Jenkins and A.~V.~Manohar,
Phys.\ Lett.\ B {\bf 255}, 558 (1991).

\bibitem{jm259}
  E.~Jenkins and A.~V.~Manohar,
  Phys.\ Lett.\  B {\bf 259}, 353 (1991).

\bibitem{djm94}
R.~F.~Dashen, E.~Jenkins, and A.~V.~Manohar,
Phys.\ Rev.\ D {\bf 49}, 4713 (1994); Phys.\ Rev.\ D {\bf 51}, 2489(E) (1995).

\bibitem{djm95}
R.~F.~Dashen, E.~Jenkins, and A.~V.~Manohar,
Phys.\ Rev.\ D {\bf 51}, 3697 (1995).

\bibitem{jen96}
E.~Jenkins,
Phys.\ Rev.\ D {\bf 53}, 2625 (1996).

\bibitem{caldi}
  D.~G.~Caldi and H.~Pagels,
  Phys.\ Rev.\  D {\bf 10}, 3739 (1974).

\bibitem{fmhjm}
R.~Flores-Mendieta, C.~P.~Hofmann, E.~Jenkins, and A.\ V.\ Manohar,
Phys.\ Rev.\ D {\bf 62}, 034001 (2000).

\bibitem{rfm09}
  R.~Flores-Mendieta,
  Phys.\ Rev.\  D {\bf 80}, 094014 (2009).

\bibitem{jen92}
  E.~Jenkins, M.~E.~Luke, A.~V.~Manohar, and M.~J.~Savage,
  Phys.\ Lett.\ B {\bf 302}, 482 (1993); Phys.\ Lett.\ B {\bf 388}, 866(E) (1996).

\bibitem{ms97}
  U.~-G.~Meissner and S.~Steininger,
  Nucl.\ Phys.\ B {\bf 499}, 349 (1997).

\bibitem{geng}
  L.~S.~Geng, J.~M.~Camalich, L.~Alvarez-Ruso, and M.~J.~V.~Vacas,
Phys.\ Rev.\ Lett.\ {\bf 101}, 222002 (2008).

\bibitem{geng2}
  L.~S.~Geng, J.~Martin Camalich, and M.~J.~Vicente Vacas,
  Phys.\ Rev.\  D {\bf 80}, 034027 (2009).

\bibitem{tib}
  D.~Arndt and B.~C.~Tiburzi,
  Phys.\ Rev.\  D {\bf 69}, 014501 (2004).

\bibitem{part}
  J.~Beringer {\it et al.}  (Particle Data Group Collaboration),
  Phys.\ Rev.\ D {\bf 86}, 010001 (2012).

\bibitem{dai}
J.~Dai, R.~F.~Dashen, E.~Jenkins, and A.~V.~Manohar,
Phys.\ Rev.\ D {\bf 53}, 273 (1996).

\bibitem{rfm12}
  R.~Flores-Mendieta, M.~A.~Hernandez-Ruiz, and C.~P.~Hofmann,
  Phys.\ Rev.\ D {\bf 86}, 094041 (2012).

\bibitem{rfm06}
  R.~Flores-Mendieta and C.~P.~Hofmann,
  Phys.\ Rev.\ D {\bf 74}, 094001 (2006).

\bibitem{lmw95}
  M.~A.~Luty, J.~March-Russell, and M.~J.~White,
  Phys.\ Rev.\ D {\bf 51}, 2332 (1995).  

\bibitem{lb}
R.~F.~Lebed and D.~R.~Martin,
Phys.\ Rev.\ D {\bf 70}, 016008 (2004).

\bibitem{krause}
  A.~Krause,
  Helv.\ Phys.\ Acta {\bf 63}, 3 (1990).  

\bibitem{jen12}
  E.~E.~Jenkins,
  Phys.\ Rev.\ D {\bf 85}, 065007 (2012).

\bibitem{lopez}
  G.~Lopez-Castro and A.~Mariano,
  Phys.\ Lett.\  B {\bf 517}, 339 (2001).

\bibitem{clas1}
  D.~Keller {\it et al.}  (CLAS Collaboration),
  Phys.\ Rev.\ D {\bf 83}, 072004 (2011).

\bibitem{clas2}
  D.~Keller {\it et al.}  (CLAS Collaboration),
  Phys.\ Rev.\ D {\bf 85}, 052004 (2012).

\bibitem{keller}
  D.~Keller and K.~Hicks,
  Eur.\ Phys.\ J.\ A {\bf 49}, 53 (2013).

\end{thebibliography}
\end{document}